\documentclass[dvipsnames]{aa}

\usepackage{graphicx}
\usepackage{txfonts}
\usepackage{xcolor}
\usepackage[colorlinks=true, linktocpage, linkcolor={blue!60!black}, citecolor={blue!60!black}, urlcolor={blue!60!black}]{hyperref}

\newcommand{\ignore}[1]{}

\begin{document} 

\title{Wideband 67-116~GHz receiver development for ALMA Band 2}

\author{P.~Yagoubov\inst{1}\thanks{\email{pyagoubo@eso.org}} \and
{T.~Mroczkowski\inst{1}}\thanks{\email{amroczko@eso.org}} \and
{V.~Belitsky\inst{2}} \and
{D.~Cuadrado-Calle\inst{3,4}} \and
{F.~Cuttaia\inst{5}} \and
{G.~A.~Fuller\inst{3}} \and
{J.-D.~Gallego\inst{6}} \and
{A.~Gonzalez\inst{7}} \and
{K.~Kaneko\inst{7}} \and
{P.~Mena\inst{8}} \and
{R.~Molina\inst{8}} \and
{R.~Nesti\inst{9}} \and
{V.~Tapia\inst{8}} \and
{F.~Villa\inst{5}} \and \\
{M.~Beltr\'an\inst{9}} \and
{F.~Cavaliere}\inst{10} \and
{J.~Ceru}\inst{11} \and 
{G.~E.~Chesmore}\inst{12} \and
{K.~Coughlin}\inst{12} \and
{C.~De Breuck\inst{1}} \and
{M.~Fredrixon\inst{2}} \and
{D.~George\inst{4}} \and
{H. Gibson}\inst{11} \and
{J.~Golec}\inst{12} \and
{A.~Josaitis}\inst{12} \and
{F.~Kemper\inst{1,13}} \and
{M.~Kotiranta\inst{14}} \and
{I.~Lapkin\inst{2}} \and
{I.~L\'opez-Fern\'andez\inst{6}} \and
{G.~Marconi\inst{1}} \and
{S.~Mariotti\inst{15}} \and
{W.~McGenn\inst{3,4}} \and
{J.~McMahon}\inst{12} \and
{A.~Murk\inst{14}} \and
{F.~Pezzotta}\inst{10} \and
{N.~Phillips\inst{1}} \and
{N.~Reyes\inst{8}} \and
{S.~Ricciardi\inst{5}} \and
{M.~Sandri\inst{5}} \and
{M.~Strandberg\inst{2}} \and
{L.~Terenzi\inst{5}} \and
{L.~Testi\inst{1}} \and
{B.~Thomas}\inst{11} \and
{Y.~Uzawa\inst{7}} \and
{D.~Vigan\`o}\inst{10} \and
{N.~Wadefalk}\inst{16}
}

\institute{ 
%1
{European Southern Observatory (ESO), Garching, Germany} \and
%2
{Group for Advanced Receiver Development (GARD), Chalmers University of Technology, Gothenburg, Sweden} \and 
%3
{Jodrell Bank Centre for Astrophysics, School of Physics \& Astronomy, The University of Manchester (UoM), Manchester, UK} \and
%4
{School of Electrical \& Electronic Engineering, The University of Manchester (UoM), Manchester, UK} \and
%5
{Istituto Nazionale di Astrofisica (INAF/OAS), Bologna, Italy} \and
%6 
{Observatorio de Yebes, Guadalajara, Spain} \and
%7
{National Astronomical Observatory of Japan (NAOJ), Mitaka, Tokyo, Japan} \and
% 8
{Universidad de Chile (UdC), Santiago, Chile} \and
%9 
{Istituto Nazionale di Astrofisica (INAF/OAA), Arcetri, Italy} \and
%10
{Dipartimento di Fisica, Universita degli Studi di Milano, Milano, Italy} \and
%11 
{Radiometer Physics GmbH (RPG), Meckenheim, Germany} \and
%12 
{University of Michigan, Department of Physics, Ann Arbor, MI, USA} \and
%13 
{Academia Sinica, Institute of Astronomy \& Astrophysics (ASIAA), Taipei, Taiwan} \and
%14
{Institute of Applied Physics, University of Bern, Switzerland} \and
%15
{Istituto Nazionale di Astrofisica / Istituto di Radioastronomia (INAF/IRA), Bologna, Italy} \and
%16
{Low Noise Factory (LNF), Sweden}
}

   \date{---}

% \abstract{}{}{}{}{} 
% 5 {} token are mandatory
 
  \abstract
   {The Atacama Large Millimeter/submillimeter Array (ALMA) has been in operation since 2011, but it has not yet been populated with the full suite of its planned frequency bands.  In particular, ALMA Band 2 (67-90~GHz) is the final band in the original ALMA band definition to be approved for production.}
   {We aim to produce a wideband, tuneable, sideband-separating receiver with 28~GHz of instantaneous bandwidth per polarisation operating in the sky frequency range of 67-116~GHz.  Our design anticipates new ALMA requirements following the recommendations of the 2030 ALMA Development Roadmap.}
   {The cryogenic cartridge is designed to be compatible with the ALMA Band 2 cartridge slot, where the coldest components --
    the feedhorns, orthomode transducers, and cryogenic low noise amplifiers -- operate at a temperature of 15~K. We use multiple simulation methods and tools to optimise our designs for both the passive optics and the active components.  The cryogenic cartridge is interfaced with a room-temperature (warm) cartridge hosting the local oscillator and the downconverter module.
   This warm cartridge is largely based on GaAs semiconductor technology and is optimised to match the cryogenic receiver bandwidth with the required instantaneous local oscillator frequency tuning range.}
   {Our collaboration has resulted in the design, fabrication, and testing of multiple technical solutions for each of the receiver components, producing a state-of-the-art receiver covering the full ALMA Band 2 and 3 atmospheric window.
   The receiver is suitable for deployment on ALMA in the coming years and it is capable of dual-polarisation, sideband-separating observations in intermediate frequency bands spanning 4-18~GHz for a total of 28~GHz on-sky bandwidth per polarisation channel.}
   {We conclude that the 67-116~GHz wideband implementation for ALMA Band 2 is now feasible and that this receiver provides a compelling instrumental upgrade for ALMA that will enhance observational capabilities and scientific reach.}


   \maketitle
%
%-------------------------------------------------------------------

% useful for outlining the balance of topics:
%\tableofcontents
%--------------------------------------------------------------------

\section{Introduction}\label{sec:intro}

The Atacama Large Millimeter/submillimeter Array\footnote{\url{https://www.almaobservatory.org}} \citep[ALMA; for background and project definition, see][]{2009IEEEP..97.1463W} is a versatile observatory delivering groundbreaking scientific discovery since 2011. It  is not, however, fully equipped with all of the receiver bands that had originally been planned.  In particular, Band 2 (67-90~GHz) has only recently been approved by the ALMA board of directors for the initial phase of construction. Recent technological developments in cryogenic monolithic microwave integrated circuits (MMICs) and optical components, such as wide bandwidth feedhorns, orthomode transducers (OMT), and lens designs -- have opened up the opportunity to extend the originally-planned radio-frequency (RF) bandwidth of this receiver, to cover the 67-116~GHz frequency range on-sky with a single receiver.  This holds the potential for combining ALMA Band 2 (67-90~GHz) with ALMA Band 3 (84-116~GHz), serving as an upgrade that paves the way for wider bandwidth ALMA operations.  As discussed in \citealt{band2_wideband_alma_memo}, this approach offers several operational and scientific advantages.  Furthermore, the aim to cover both wider on sky and instantaneous bandwidth is well aligned with the ALMA 2030 Development Roadmap \citep{alma_roadmap}, a document summarising the highest priority upgrades enabling new ALMA science in the 2030s and beyond.  

At the time when the band ranges and original receiver specifications for the ALMA project were defined, roughly two decades ago, an instantaneous bandwidth of 8~GHz was an ambitious goal.  
However, a number of compelling observational advantages come from delivering wider intermediate frequency (IF)\footnote{The intermediate frequency, or IF, bandwidth is the frequency range to which the sky frequencies are downconverted to allow for digitisation and signal processing.  The IF band determines the instantaneous frequency of the receiver.} and RF bandwidths, including lower noise in continuum observations, the ability to probe large portions of an astronomical spectrum for such phenomena as widely spaced molecular transitions, and the ability to efficiently scan in frequency space to perform, for example, surveys where the redshift of an object is not known a priori.  
Since bandpass and phase calibration are typically performed on sources described by power law continuum spectra, receivers with wider instantaneous bandwidth should be faster to calibrate and with fewer uncertainties across a broader frequency range.
Wider bandwidth also has the potential to improve the sensitivity of broadband mm-wave Very Long Baseline Interferometry (VLBI) measurements, such as those undertaken with the Event Horizon Telescope \citep[EHT;][]{EHT_inst_2019}. 

In this paper, we report the design and performance of the ALMA prototype receiver developed to cover an RF bandwidth spanning 67-116~GHz. 
In fact, several versions of each of the components were developed in parallel, providing viable alternatives for implementation in the final receiver to be installed in ALMA.
The prototype is a sideband-separating (2SB) receiver with an output IF band of 4-18~GHz, yielding a total instantaneous bandwidth of 28~GHz.  This exceeds by far the minimum requirement set by the original -- and, at the time of writing, current {\it } -- ALMA Band 2 specifications of 4-12~GHz in a single sideband (SSB). 
After careful consideration of the alternatives -- SSB and dual-sideband (DSB) receivers --  it is clear that the 2SB approach is preferable, whenever possible, for improving the sensitivity of ALMA \citep[see e.g.][]{Iguchi2005,alma_memo_602}.

The optical system consists of a refractive lens, which also serves as a vacuum window for the cryogenic portion of the receiver, a corrugated feedhorn, and an OMT for a dual-polarisation receiver.
As we show here, the measured beam patterns, aperture efficiencies, and polarisation characteristics of the optics prototypes we developed are compliant with ALMA specifications and in good agreement with the electromagnetic (EM) simulations used in each of their respective designs. For the lens, we use high-density polyethylene (HDPE) as our baseline design material, while investigating ultra-high molecular weight polyethylene (UHMW-PE) and high resistivity float zone silicon as alternatives to reduce transmission loss \citep{Chesmore2018}. 

The receiver system uses cryogenic low noise amplifiers (LNAs) at its input, developed using a state-of-the-art commercial 100~nm process at the Low Noise Factory (LNF) in Sweden, and a 35~nm gate length InP high electron mobility transistor (HEMT) amplifier designed by the University of Manchester (UoM; see \citealt{2017ITMTT..65.1589C} for details) using the fabrication processes offered by Northrop Grumman Corporation. 
The best LNAs tested so far show a noise temperature below 28~K from 70~GHz to 110~GHz for a cryogenic ambient operating temperature of 15~K.
With such developments, cryogenic LNAs have begun to offer noise performance comparable to the more traditional superconductor-insulator-superconductor (SIS) mixer technologies employed in ALMA Bands 3-10 \citep[see e.g.][for a comparison of the two technologies]{CuadradoCalle2017}.  We also note the use of cryogenic LNAs in the ALMA Band 1 receivers, now in production, to achieve state-of-the-art performance \citep{Huang2016,Huang2018}.

A prototype cold cartridge assembly (CCA) -- fully compatible with ALMA's electrical and mechanical interfaces -- has been designed and built to accommodate the optics and the LNAs.  The downconverter module for the receiver is located in the (room-temperature) warm cartridge assembly (WCA). 
The full system has been assembled and its performance has been characterised, demonstrating an uncorrected mean noise temperature $<$32~K across the 70-116~GHz frequency range. 
The receiver was also tested for other key requirements common to ALMA receivers such as amplitude stability and passband power variations.

When it was still under development, a brief overview of the Band 2 receiver was provided in \cite{Yagoubov2018}.  
In the following sections, we provide more extensive details of the specific component designs we developed and tested in order to produce the final ALMA Band 2 receiver prototype.  In Section \ref{sec:science}, we briefly summarise the scientific motivation for the project.
In Section \ref{sec:design}, we discuss the overall design for the receiver. Subsection \ref{sec:optics} presents the designs for the optical components produced by the member teams located at: Universidad de Chile (UdC); the Osservatorio Astrofisico di Arcetri (OAA) and Osservatorio di Astrofisica e Scienza dello Spazio di Bologna (OAS), both parts of the Italian National Institute of Astrophysics (Istituto Nazionale di Astrofisica; INAF); and the National Astronomical Observatory of Japan (NAOJ), while
subsection \ref{sec:lna} presents the cryogenic LNA designs produced by UoM and LNF.
In Section \ref{sec:receiver}, we discuss the cartridge assemblies and performance results.
We provide our conclusions, including the next steps for implementing the final receiver design in ALMA, in Section \ref{sec:conclusions}.
As a convenience to the reader, we list the acronyms we use in Table \ref{tab:jargon}.

%--------------------------------------------------------------------

\section{Key science drivers for a wideband ALMA Band~2}\label{sec:science}

In this section, we provide a brief overview of the scientific motivation for a wideband ALMA Band 2, many of which are identified and summarised in the white papers by \citet{Beltran2015}, and \citet{fuller16}, assuming an instantaneous bandwidth of  8~GHz per polarisation, split over two side bands.
However, as technology advances and mm-wave observatories work to take advantage of these new capabilities, it is clear that over the next decade a much  more ambitious upgrade is feasible. The recent ALMA Memo \#605 \citep{band2_wideband_alma_memo} discusses this possibility.

As detailed in Sections \ref{sec:design} and \ref{sec:wca}, the prototype receiver is sideband separating and has 2 IF bands, each spanning 4-18~GHz, for a total of 28~GHz on-sky with an 8~GHz separation between the upper and the lower side bands.  
One overall goal of the project is to deliver this bandwidth in the production receivers, with a minimum requirement that the IF band should cover 4-16~GHz.
As discussed below, when combined with potential upgrades to the ALMA digital processing hardware and software system, this new wideband receiver will address two of the three key science drivers in the ALMA 2030 Development Roadmap \citep{alma_roadmap}. Namely, the two key science drivers addressed directly by this project   are the {\it the Origins of Galaxies} and  {\it the Origins of Chemical Complexity}, although we also note that improved continuum imaging and the ability to probe the carbon monoxide (CO) snowline will partially address the third key science driver, {\it the Origins of Planets}.

\begin{figure}
\centering
\includegraphics[width=\columnwidth]{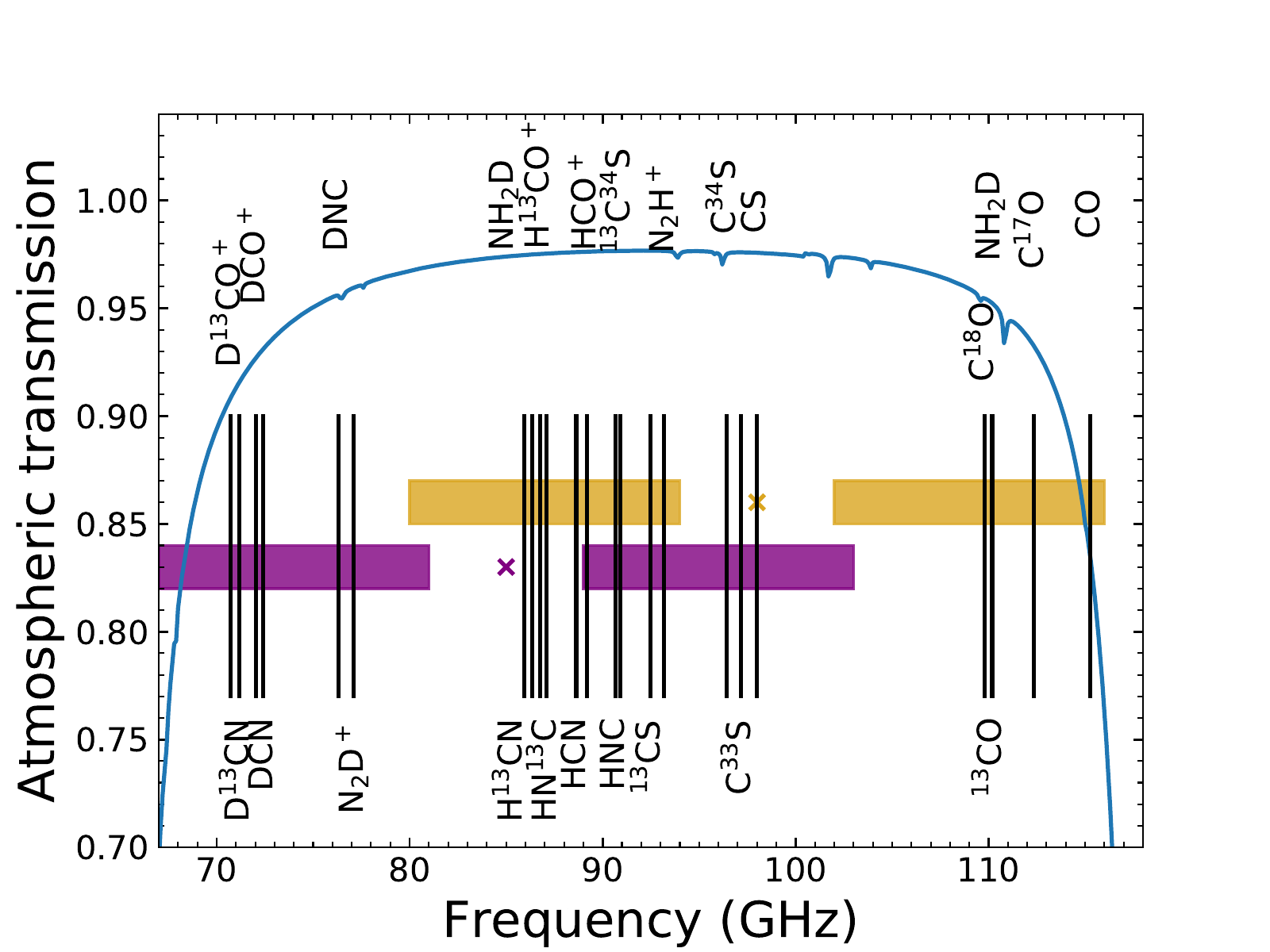}
\caption{Two example spectral setups of the wide RF+IF Band 2 receiver presented here that would cover the entire 67-116~GHz atmospheric window. The x-axis shows the frequency and the horizontal coloured bars show the upper and lower sideband frequency ranges for two distinct LO settings (`x' markers at 85 and 98~GHz on the plot), one shown in purple and the other in gold. The LO frequency is marked by the coloured cross in each case.  The blue curve shows the typical zenith atmospheric transmission assuming the PWV is 2~mm. The vertical black lines denote the frequencies of some molecular transitions in the band.  The full band can be explored in only two spectral setups, assuming the backend digitisation and correlation hardware is upgraded to match the receiver output. Figure adapted from \cite{band2_wideband_alma_memo}.
}
\label{fig:full_B2_lineobs}
\end{figure}

We note that ALMA Band 2 and Band 3 span a single, continuous atmospheric window, the typical zenith atmospheric transmission of which is plotted assuming a precipitable water vapour (PWV) of 2~mm in Fig.~\ref{fig:full_B2_lineobs} (blue curve).
As one of the lowest frequency bands planned for ALMA, Band 2 is a relatively low atmospheric opacity band that is available for $>87.5\%$ of ALMA observing conditions \citep[e.g.][]{otarola2019} and it will contribute substantial observational flexibility to the observatory. 
Many of the science cases we discuss here will benefit from the ability to cover the entire 67--116~GHz window with only two spectral setups, as shown in the example tuning in Fig.~\ref{fig:full_B2_lineobs}.

%--------------------------------------------------------------------
%--------------------------------------------------------------------

In ALMA Memo \#605 \citep{band2_wideband_alma_memo}, we specifically considered the observational and scientific advantages for the deployment of a wideband receiver fully spanning the atmospheric window originally defined as ALMA Bands 2 and 3. 
The memo considered the impacts of a receiver system meeting hypothetical, proposed ALMA system requirements that relax the noise temperature performance modestly in exchange for a substantially wider on-sky and instantaneous bandwidth.  Specifically, it was assumed that up to 10\% on average additional integration time for observations of a single line at a known frequency would be tolerated in exchange for improved frequency coverage.
We note, however, that it is not our goal to relax the ALMA specifications simply to gain more bandwidth; the optimised designs we present in this paper represent the state-of-the-art in mm-wave optics and MMIC HEMT receiver technology, regardless of bandwidth.

We outline below some of the key areas where the wideband Band 2 receiver will have a significant science impact, although we expect it to have a broad impact across the full range of ALMA science. 

%--------------------------------------------------------------------
\begin{figure}
\centering
\includegraphics[width=\columnwidth]{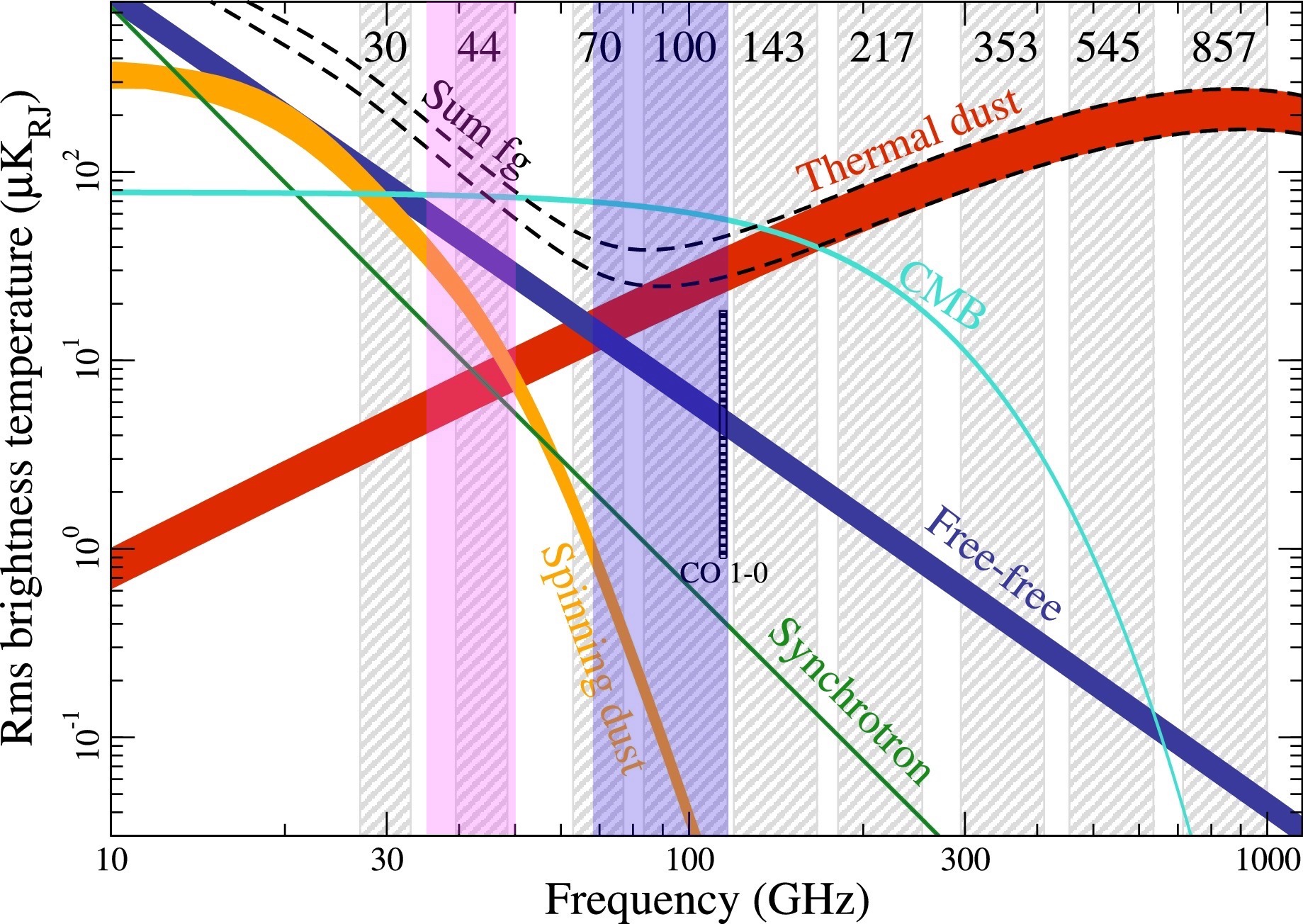}
\caption{ 
Combined millimetre/sub-millimetre spectrum of foreground dust, synchrotron, anomalous microwave emission, etc, plotted with the band centres of the instruments on the {\it Planck} satellite denoted (grey vertical bands). 
The ALMA wide RF Band 2 (transparent dark blue region) will cover most of the {\it Planck} 70 and 100~GHz channels \citep{Tauber2010}, where the sum of the contributions is lowest. We also highlight Band 1 (35-50 GHz; transparent magenta region), which together with Wide RF Band 2 will likely leverage the free-free and spinning dust contributions.
Figure is adapted from \citet{Dickinson2018} to highlight the role of the lower ALMA bands. 
}
\label{fig:ame}
\end{figure}
%--------------------------------------------------------------------
  
\paragraph{The Sunyaev-Zeldovich effect:}   Briefly, the Sunyaev-Zeldovich (SZ; \citealt{Sunyaev1972}) effect is a faint, extended decrement in the Cosmic Microwave Background (CMB) which has a redshift-independent surface brightness that traces the warm and hot gas in large scale structures, such as groups and clusters of galaxies \citep[for a recent review, see][]{Mroczkowski2019}. High-resolution observations of the SZ effect have become nearly routine using ALMA and the Atacama Compact Array (ACA) in Band 3 \citep{Basu2016, Kitayama2016, Ueda2017, DiMascolo2019, DiMascolo2019b, Gobat2019}.
Yet the recent Band 3 studies are all fundamentally limited by the small angular scales accessible with ALMA and the ACA.  Band 2 and Band 1 (35-50~GHz; see e.g.\ \citealt{DiFrancesco2013} for details on the science case) will access larger scales than Band 3, making it better suited for imaging cluster scales (typically several arcminutes at redshifts $z>0.2$), while the improved bandwidth translates to a direct improvement in imaging speed. In addition, Band 2 extends the observations to the frequency range with minimal contamination from free-free and dust emission (see Fig.~\ref{fig:ame}).

%--------------------------------------------------------------------
\begin{figure}
\centering
\includegraphics[width=\columnwidth]{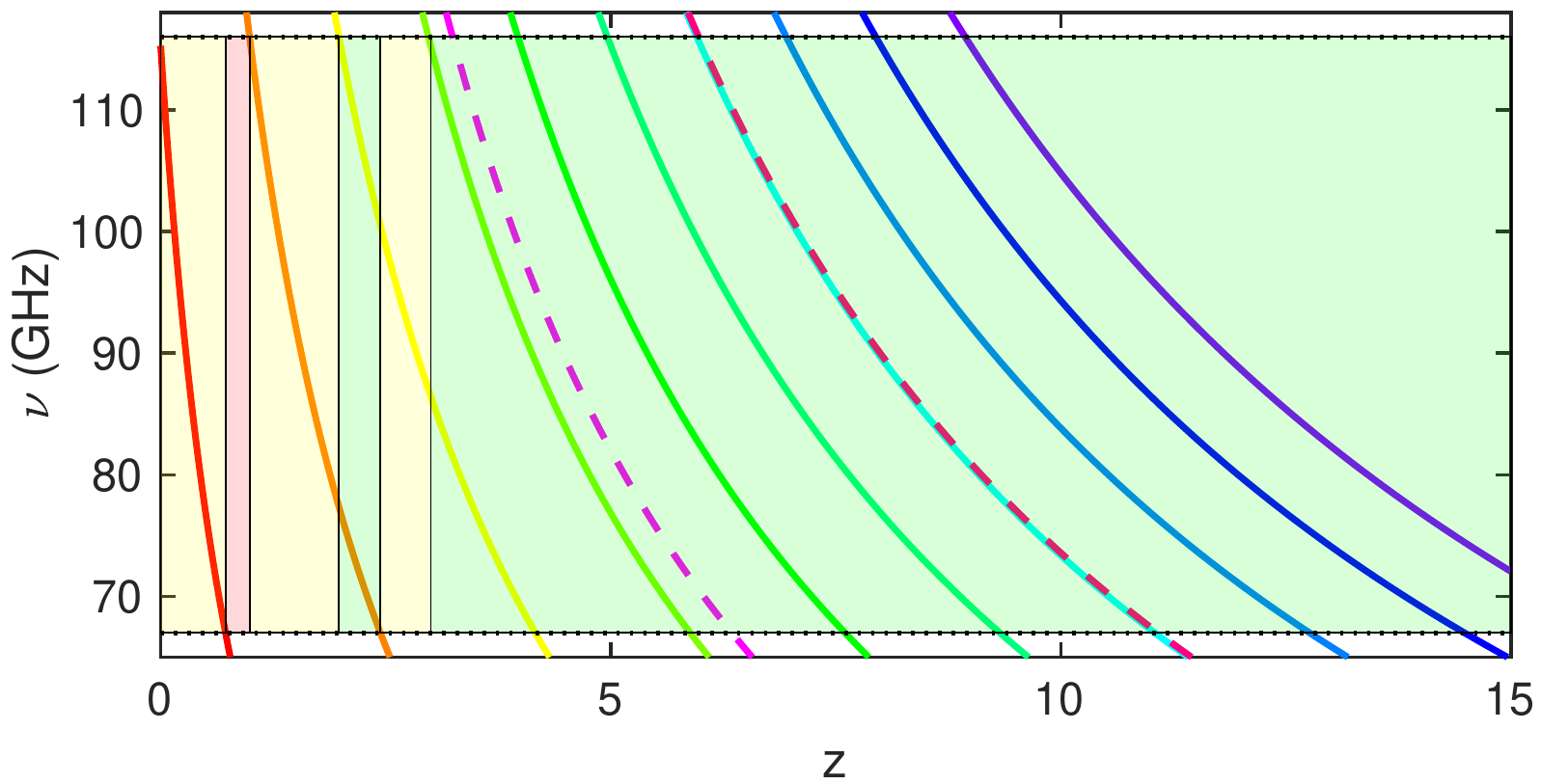}
\includegraphics[width=\columnwidth]{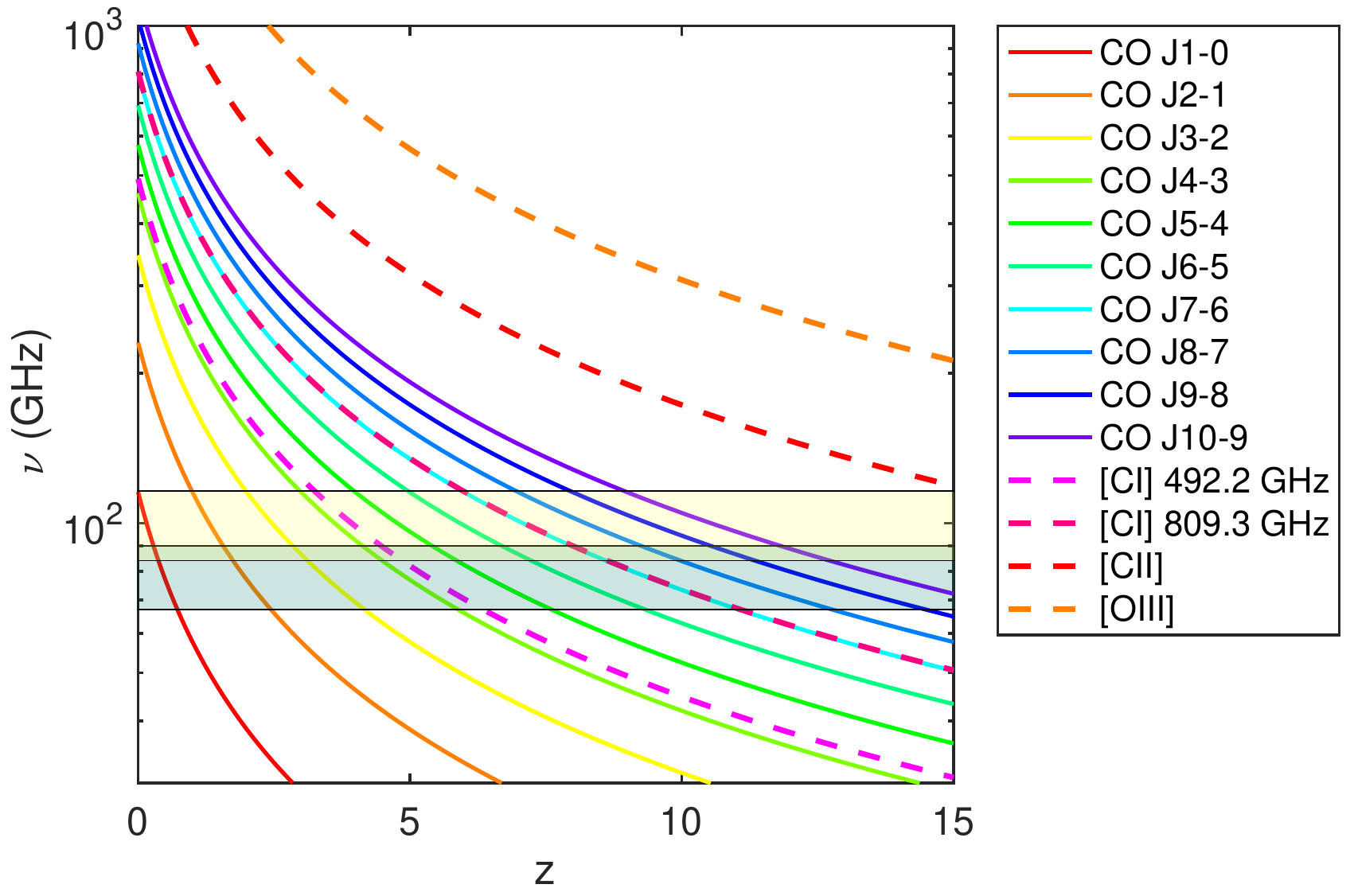}
\caption{{\it Upper panel:}  Curves show bright molecular CO rotational transitions and atomic [CI] lines (neutral carbon at 492 \& 809~GHz in the rest frame) redshifted into the Band 2 \& Band 3 frequency range as a function of redshift.  
For all redshifts other than $0.72<z<0.99$ (pink shaded region) out to formation of the earliest galaxies, at least one CO rotational transition is available (yellow shaded region).
The redshift determination is unambiguous for $1.98<z<2.44$ and $z\geq3$, where two or more lines are available (green shaded regions).
Importantly, the wideband Band 2 receiver system we present here will extend ALMA's capability to probe the [CI] 809.3~GHz rest frame transition at redshifts $z \gtrsim 8.5$.
{\it Lower panel:} Same bright mm/submm lines for all ALMA bands. We note that the full ALMA  frequency range covers the redshifted [CII] and [OIII] lines, although they do not intersect Bands 2\&3 for $z\lesssim15$.
ALMA Bands 2 \& 3, as originally defined, are respectively shaded in green and yellow (horizontal regions).
The wide bandwidth receiver we present here will be unique in its ability to probe simultaneously two or more lines for $z\geq3$.
The legend is common to both panels.  The figures are based on those presented in \citet{fuller16}.
}\label{fig:co_z}
\end{figure}
%--------------------------------------------------------------------

\paragraph{High redshift extragalactic surveys and studies:} As detailed in \cite{band2_wideband_alma_memo,fuller16}, ALMA Band 3 observations have been a powerful tool for redshift determinations, particularly at the high redshifts of $2<z<8.5$, where at least one transition of CO or [CI] line is available.  Wideband Band 2 will be cover 50\% extra frequency in two tunings instead of five (Fig.~\ref{fig:full_B2_lineobs}). This will not only make spectral surveys 2.5 times faster, but it will also increase the redshift identification efficiency (see Fig.~\ref{fig:co_z}) and better leverage their dusty and free-free continuum spectra (see Fig.~\ref{fig:ame}). Combined, these new capabilities directly address the key science driver to probe the origins of galaxies \citep{alma_roadmap}.

\begin{figure}
\centering
\includegraphics[width=\columnwidth]{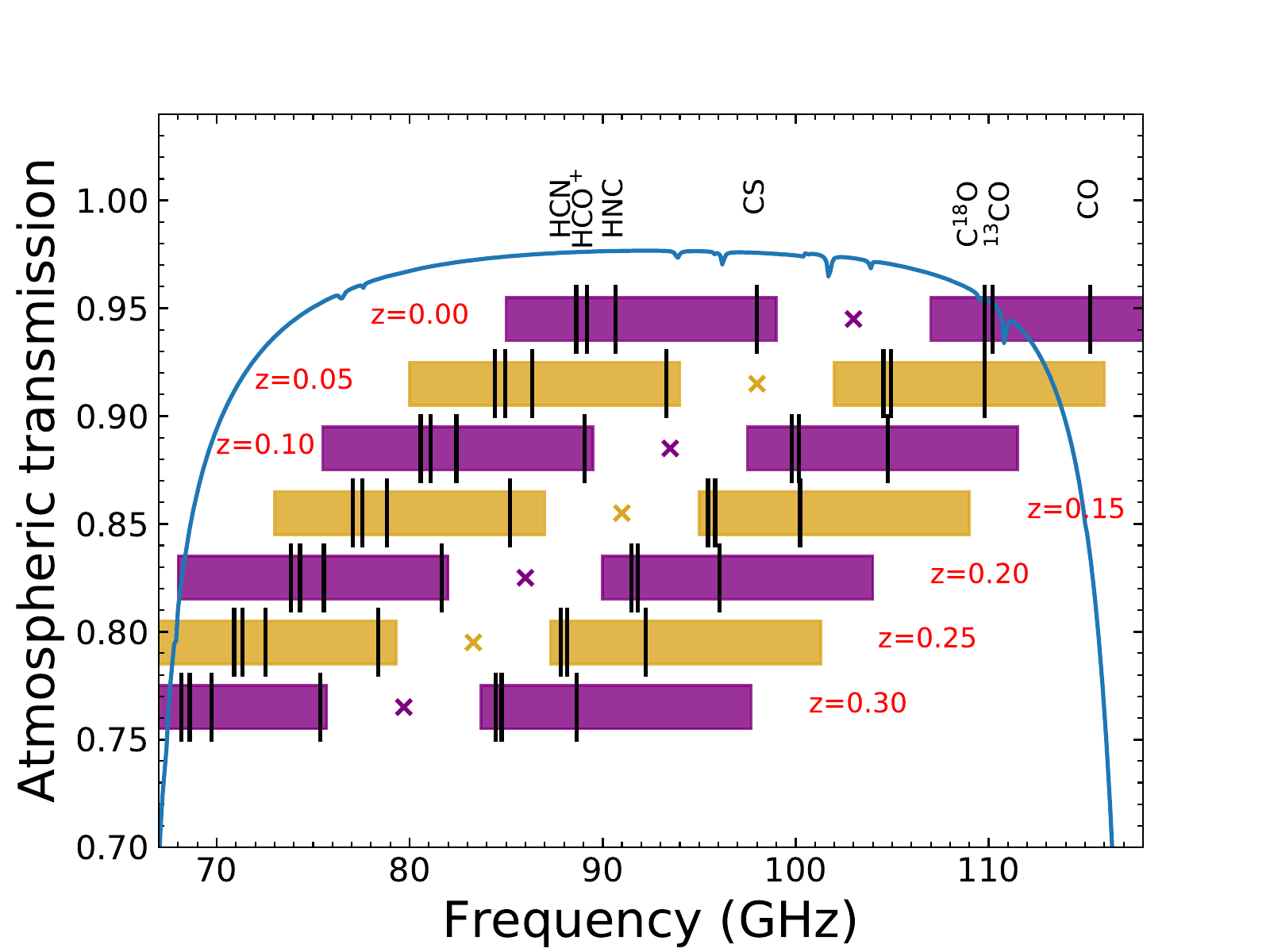}
\caption{ 
Spectral configurations assuming a 4-18\,GHz IF band that allow the simultaneous observation of low density tracers ($^{12}$CO(1-0), $^{13}$CO(1-0), and C$^{18}$O(1-0)),  and tracers of dense gas (CS(2-1), C$^{34}$S(2-1), $^{13}$CS(2-1), N$_{2}$H$^+$(1-0), HNC(1-0), HCO$^+$(1-0), and HCN(1-0)) for a range of redshifts. The vertical black lines indicate the position major spectral lines with each horizontal row corresponding to the redshift ($z$) indicated. The corresponding coloured bars show the sky frequency coverage of the upper and lower sidebands for the LO frequency indicated by the coloured cross. The solid blue curve shows the zenith atmospheric transmission at ALMA for a PWV value of 2~mm.
Figure adapted from \cite{band2_wideband_alma_memo}.
}
\label{fig:zplot}
\end{figure}

\paragraph{Local universe extragalactic science:}  Also detailed in \cite{band2_wideband_alma_memo}, the wideband Band 2 receiver will improve spectral coverage of galaxies in the local universe ($z<0.3$), allowing for observations that probe unique combinations of molecular lines (see Fig.~\ref{fig:zplot}).  At low frequencies, line identification can be less ambiguous due to the ground state transitions being more sparse and it can directly leverage abundances. 

%--------------------------------------------------------------------
%--------------------------------------------------------------------

\paragraph{Molecular/chemical complexity in the Milky Way Galaxy:} The second key science driver proposed in \cite{band2_wideband_alma_memo} is to probe the origins of chemical complexity, one of the key science goals identified in the ALMA 2030 Development Roadmap \citep{alma_roadmap}. As noted in, for example, \cite{Beltran2015, fuller16}, and in Figures \ref{fig:full_B2_lineobs} \& \ref{fig:zplot}, the Band 2 and 3 portion of the spectrum contains numerous molecular and atomic transitions. These include the ground state transitions of deuterated dense gas probing species which trace the cold, quiescent regions at the cusp of star formation and complex organic molecules that can trace building blocks of life (including amino acids such as glycene), which at these lower frequencies are easier to identify due to the lower spectral-crowding. In addition, this frequency range is rich in recombination lines of hydrogen (see \citealt{Murchikova2019} and the discussion in \citealt{band2_wideband_alma_memo}), along with tracers of the chemical evolution of stars and stellar envelopes, to name a few. As with the above case, the molecular lines found in Band 2 are often due to ground state transitions, meaning they are brighter and that they leverage abundances directly.
    
\paragraph{Proto-planetary disks:} 
The large bandwidth of the Band 2 receiver will improve the fidelity of the continuum imaging of proto-planetary disks and the lower frequencies can leverage the tail of the modified blackbody spectrum typically fit to the emission from  dust (see Fig.~\ref{fig:ame}).  Additionally, \cite{Beltran2015} and \cite{fuller16} note that the CO snowline can best be probed in the wideband Band 2 range.
    
\paragraph{Spectral dependence and component separation of the low-frequency continuum:} In addition to thermal emission from  dust, a number of components contribute to the typical Galactic foreground and extragalactic continuum emission (see Fig.~\ref{fig:ame}). 
These include free-free emission and the so-called anomalous microwave emission (AME; see \citealt{Dickinson2018}) such as that coming from spinning dust, which could present a significant foreground source of contamination for such\ studies as that of the polarisation of the primary CMB.  As Fig.~\ref{fig:ame} shows, Band 2 can serve a crucial role in separating the thermal dust emission from the CMB and other emission mechanisms, in combination with Band 1. 

\paragraph{Solar Observations:} The scale height probed by millimetre-wave emission from our Sun is proportional to wavelength \citep{Wedemeyer2016} and this places Band 2 in a good position for probing the upper chromosphere and, potentially, the beginning of the corona.  Wider bandwidth, in turn, allows for a probing of a larger range of scales simultaneously.  Thus, we expect our wideband receiver to transform ALMA's view of the Sun.
    
\paragraph{Solar System physics:} Our own solar system provides a wealth of information about planet formation, potentially informing us of the building blocks of life. As noted in \cite{band2_wideband_alma_memo}, comets may provide key insights into the composition of the early Solar System. Additionally, a number of studies have indicated that low-resolution, next-generation millimetre-wave CMB surveys could probe trans-Neptunian objects, a so-called `Planet X' or `Planet Nine', or even extra-solar Oort clouds, where searches for their thermal emission can be much more effective than for reflected light \citep{Cowan_2016, Gerdes_2017, Baxter2018a, Baxter2018b, Sehgal2019}.  The wideband ALMA Band 2 receiver will be the ideal tool for detailed, high-resolution follow-up, providing more precise locations, imaging, spectroscopy, and kinematic information. With its large instantaneous spectral grasp, the receiver will also enable detailed comparative studies of the temporal variation in spectral emission from rapidly evolving sources such as comets and volcanoes on solar system bodies \citep[see e.g.][]{dePater2019}.

%--------------------------------------------------------------------
\begin{figure*}
\centering
\includegraphics[width=0.95\textwidth]{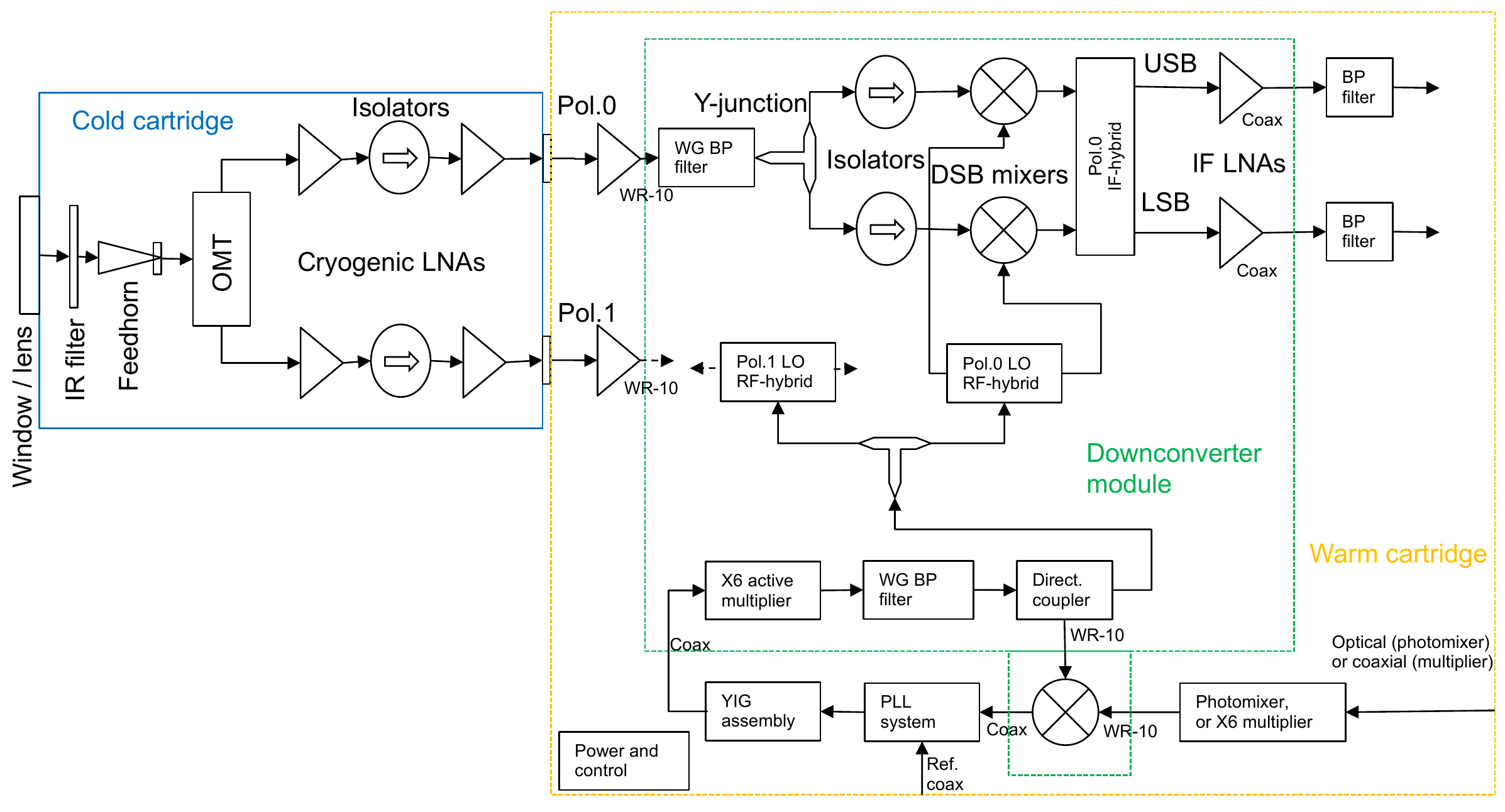}
\caption{67-116~GHz dual-polarisation, 2SB receiver functional block diagram.  The lens and infra-red (IR) filter are installed on the ALMA FE cryostat, the former serving as a vacuum window. Room temperature amplifiers can optionally be used at the input of the downconverter in case the gain in the cold cartridge is insufficient (i.e. if the gain $G < 40$~dB; see Table \ref{tab:lna_specs}).
}\label{fig:b2scheme}
\end{figure*}
%--------------------------------------------------------------------

\section{Design}
\label{sec:design}

\subsection{Receiver Architecture Overview}
\label{sec:design_overview}

The block diagram for the Band 2 receiver is shown in Fig.~\ref{fig:b2scheme}. It consists of two major sub-assemblies: the cold cartridge assembly and the warm cartridge assembly.
The tertiary optics -- defined as those optical components coming after the primary and secondary mirrors of the telescope -- comprise the refractive lens, the cryogenic IR filter(s), the feedhorn, and the OMT. 
The CCA is the portion of the receiver that is installed, using vacuum flanges to make the seals, in the ALMA front end (FE) receiver cryostat.  It consists of the cartridge body support structure, the feedhorn, OMT, two pairs of cryogenic LNAs separated by cryogenic isolators to minimise standing waves, DC bias harnesses, temperature sensors, RF signal transport waveguides (one set for each polarisation), two RF vacuum feedthroughs (one for each polarisation) providing the RF connections to the WCA, and the LNA bias protection boards.
The WCA houses the local oscillator (LO) sub-assembly, downconverter module, bias modules for the cryogenic LNAs, and other power and control units. The LO sub-assembly includes the YIG oscillator, multiplier chain, the digital phase locked loop (PLL), and the monitor and control electronics. The downconverter is a dual polarisation sideband separating receiver. It includes mixers, LO and IF quadrature hybrids, room temperature IF amplifiers, passband filters, and the WCA body. 
The use of cascaded LNAs inside the CCA, as well as the optional room temperature amplifiers in front of the downconverter module, is necessitated by the relatively low gains of the current prototypes. For the final implementation, we anticipate achieving the necessary gain with a single package cryogenic LNA. This will allow to minimise the number of components and improve the overall performance, specifically the noise and the gain flatness.
%--------------------------------------------------------------------

\subsection{Optical Designs}\label{sec:optics}

Here we present the optical designs explored for the lenses, feedhorns, and OMTs. The approaches taken explored differing parallel paths to meet the design challenges presented by the large fractional bandwidth ($>70\%$ of an octave), the requirement of low loss, limited space within the cartridge, and potentially strong truncation by the ALMA FE receiver window. 
The result is that several suitable alternatives, produced mainly by the three groups in parallel within our collaboration -- exist, and each one can readily be implemented in the receiver cartridge.
The designs and performance measurements are presented in the following subsections.

The maximum theoretical aperture efficiency that can be achieved for the designated Band 2 position in the ALMA FE cryostat, under direct illumination from the secondary mirror, can be calculated through Airy-pattern considerations.
\cite{Gonzalez2016b} showed that the maximum aperture efficiency for the Band 2 cartridge slot is below 84\% at the lower end of the range (near 67-70~GHz), whereas it is $\sim$91\% at the upper end (near 110-116~GHz). 
At the lower end of the band, the numbers are very close to the ALMA aperture efficiency specification ($>80\%$), implying that the optimisation of the entire optical chain design is crucial. 

%%%%%%%%%%%%%%%%%%%%%%%%%%%%%%%%%%%%%%%%%%%%%%%%%

\subsubsection{Lens Designs}

The aperture of the Band 2 cartridge contains a lens which serves both as a refractive optical element and as a vacuum window for the cryostat.
Being the first optical component of the cartridge, it is important to minimise the noise contributions from the lens by minimising both the reflection ($S_{11}$) and the dielectric loss.  
For this purpose, we explored three lens materials: HDPE, ultra-high-molecular-weight polyethylene (UHMW-PE), and intrinsic stoichiometry float zone silicon with high purity and resistivity.  For each of these, we developed an anti-reflective (AR) layer, as discussed below.

%-----------------------------------------------------------

\paragraph{High-density polyethylene:}\label{hdpe} Our initial lens design was made of HDPE.  In order to minimise the loss due to reflection at the air-lens and lens-vacuum transitions, we employed AR surfacing techniques that result in meta-material surfaces.  
The full results from five different approaches for the AR are presented in \cite{Tapia:18}, which compares the noise contribution, cross-polarisation leakage, reflection loss, and beam and polarisation efficiencies.

HDPE has served as a standard window and lens material for ALMA and many other mm/submm-wave observatories, but has the drawback of a relatively high dielectric loss \citep[see e.g.][]{Lamb1996, Tapia:18}.  
In order to minimise the total insertion loss in the HDPE lens, we have designed and implemented a Fresnel zoned lens. This geometry puts constraints on the bandwidth, and in practicality it limits the choice of AR geometry that can be manufactured.

%-----------------------------------------------------------
\begin{figure}
\centering
\includegraphics[width=0.95\columnwidth]{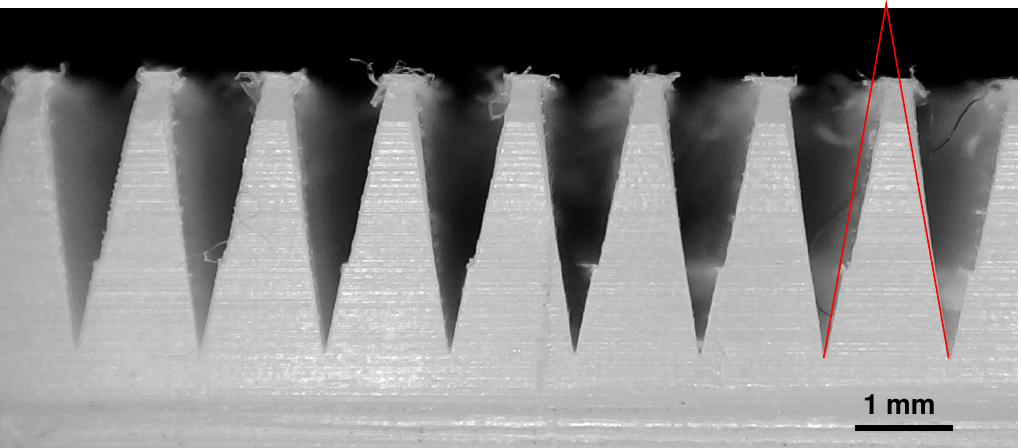}
\includegraphics[width=0.95\columnwidth]{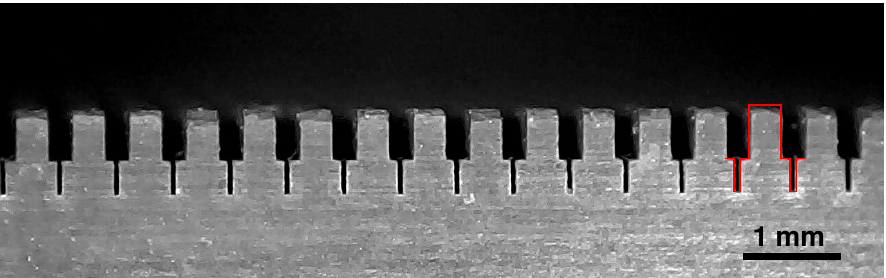}
\caption{Microscope photos of the cross sections of the UHMW-PE and silicon AR profiles implemented on test samples.  The red lines overlaid approximately indicate the target design shape.
A small amount of chipping at the edge of the silicon sample is expected.
}\label{fig:ar_profiles}
\end{figure}
%-----------------------------------------------------------

\paragraph{Ultra-high molecular weight polyethylene:}\label{uhmwpe}

UHMW-PE is a newer form of plastic with potentially lower loss than HDPE.
This may allow one to avoid using Fresnel zone lenses and to implement better performance AR geometry, such as triangular groves.
The upper panel of Fig.~\ref{fig:ar_profiles} shows the groove design implemented to reduce the loss due to reflections, discussed at the end of this section.

%-----------------------------------------------------------
\begin{figure}
\centering
\includegraphics[width=0.49\columnwidth, clip=true, trim=130mm 0mm 250mm 30mm]{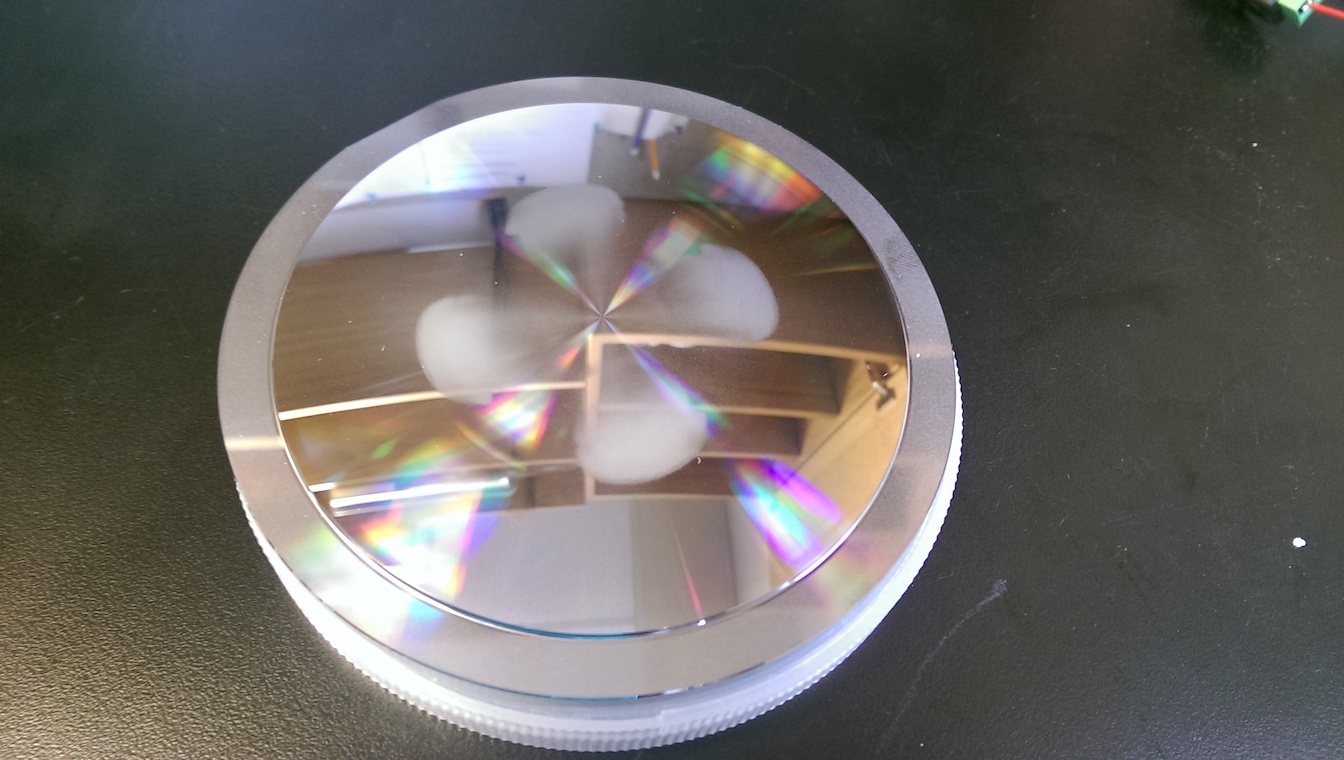}
\includegraphics[width=0.49\columnwidth, clip=true, trim=170mm 20mm 220mm 30mm]{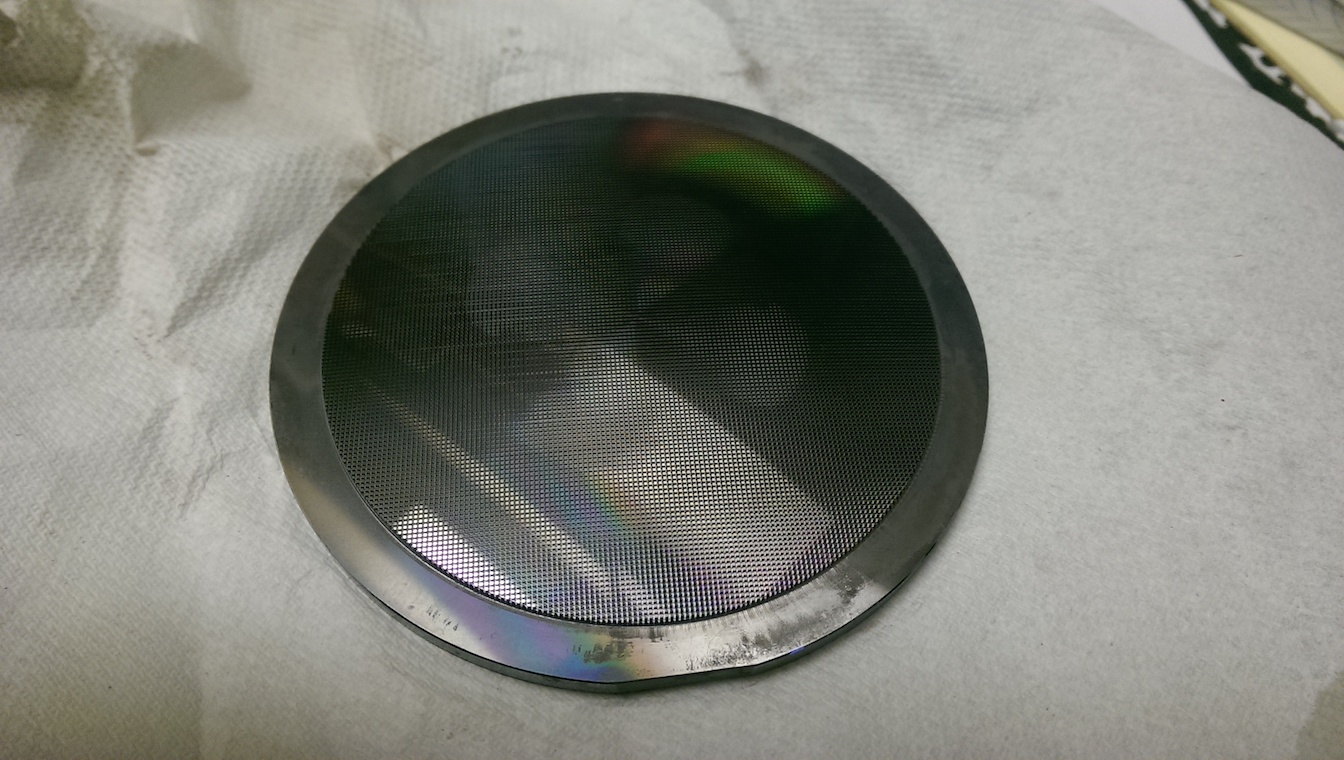}
\caption{Photos of the silicon lens prototype.  The left panel shows the lens blank, which was produced by Nu-Tek Precision Optics using a CNC lathe.  The right panel shows the lens after application of an AR surface layer manufactured using a 5-axis dicing saw.  The outer diameter of the lens, including the outer ring used for mounting, is approximately 11~cm.
}\label{fig:si_lens}
\end{figure}

\paragraph{High resistivity float zone silicon:}

High resistivity float zone (FZ) silicon with high purity and intrinsic stoichiometry has long held the promise of being a very low loss optical material for millimetre and submillimetre-wave applications \citep[see e.g.][]{Parshin1995, Lamb1996, Chesmore2018}.  
Recent advances in machining -- including single point diamond turning on a computer numerical control (CNC) lathe and precision groove cutting on a five-axis CNC dicing saw -- as well as numerical optimisation of meta-material surfaces like those described for the HDPE and UHMW-PE lenses above led us to explore their potential application for wideband ALMA Band 2.
The lower panel of Fig.~\ref{fig:ar_profiles} shows a cross-section of the 2-layer groove design implemented to reduce the reflections at the air-dielectric and dielectric-vacuum interfaces.

The design of the lens and the AR surface (or `coating') was simultaneously optimised using a similar approach as that demonstrated in \cite{Datta2013}. The team at University of Michigan produced the AR surface using a similar approach as that taken for the Simons Observatory receiver design \citep{Galitzki2018}, applying this to a blank produced by Nu-Tek Precision Optics.\footnote{\url{nu-tek-optics.com}}
 The resulting prototype is shown in Fig.~\ref{fig:si_lens}.

%-----------------------------------------------------------
\begin{figure}
\centering
\includegraphics[width=\columnwidth]{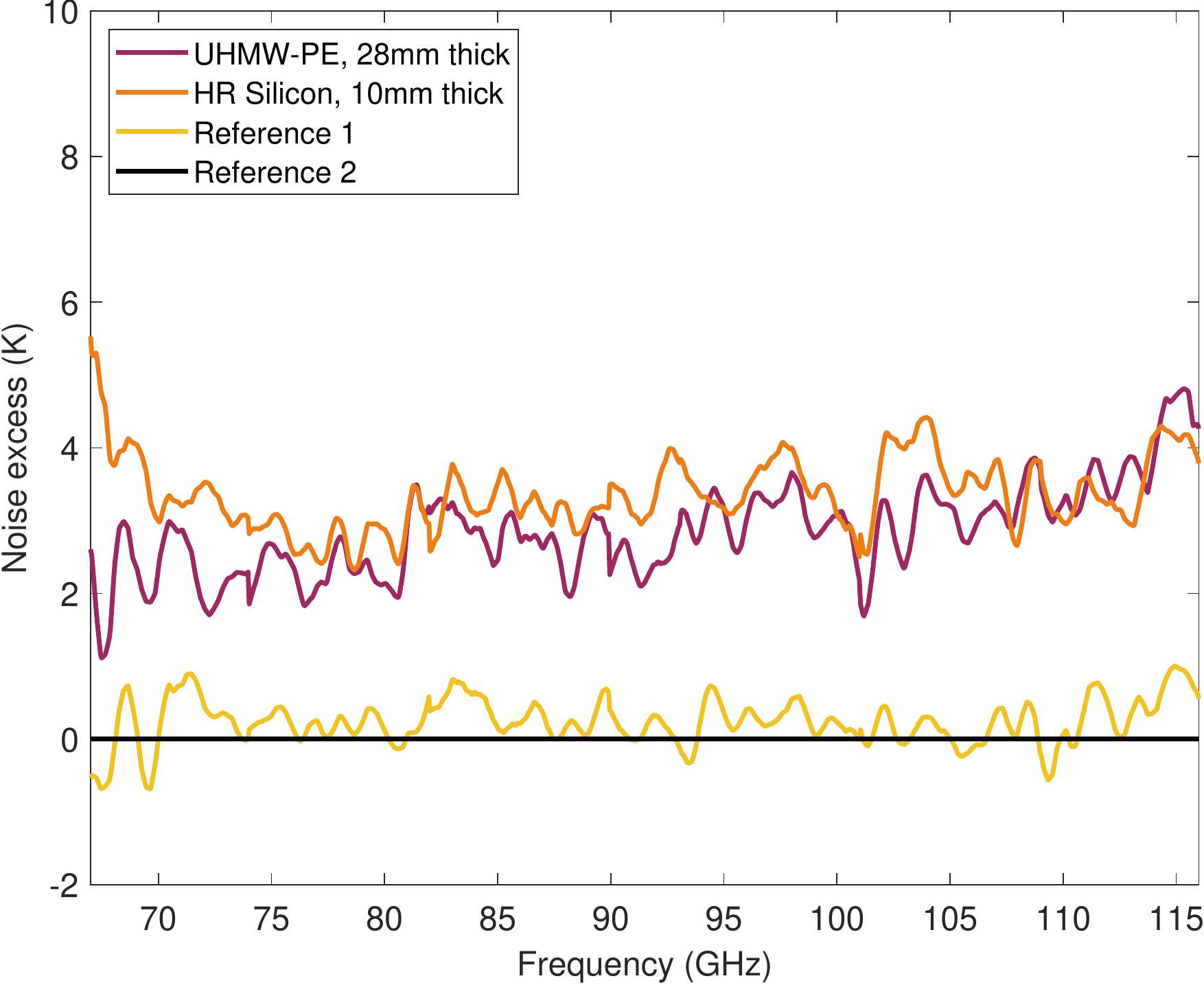}
\caption{Noise temperature (K) contributions of the test samples across the RF band. Reference 2 (black) was used as the baseline noise for comparison.  Reference 1 (yellow) is a second measurement of the reference, and its difference with respect to Reference 2 is indicative of the systematic calibration and measurement uncertainties.
}\label{fig:lens_noise}
\end{figure}
%-----------------------------------------------------------

\paragraph{Comparison of lens materials:}

In order to evaluate and compare the dielectric losses of HDPE, UHMW-PE, and high resistivity silicon, as well as the performances of the AR layers we have designed, we manufactured a number of flat samples from each of these materials, both with and without AR surface layers. The loss tangent characterisation tests are ongoing and include open resonator, transmission and reflection using a vector network analyser (VNA), and radiometric measurements of the total insertion loss.

While most of the above-mentioned measurements are still underway, our preliminary finding is that the UHMW-PE loss tangent is indeed equal or better than that of HDPE. In order to compare against the silicon, we have so far fabricated and tested radiometrically two samples: the 10 mm thick high resistivity silicon with AR layer geometry as in the lower panel of Fig.~\ref{fig:ar_profiles}, and 28 mm thick UHMW-PE with triangular AR layer, in the upper panel of Fig.~\ref{fig:ar_profiles}.
The flattened tips of the UHMW-PE sample, which clearly show burrs, resulted from the non-optimised manufacturing process of the first prototypes.  
More recent samples manufactured by a different workshop are of substantially improved quality and confirm that UHMW-PE is in fact easier to machine than HDPE.

For the tests, the samples are installed in front of the prototype receiver at a distance of a few millimetres from the lens.  The system noise temperature is then measured as described in Appendix \ref{sec:testsystems}. Two reference measurements are performed, one before and one after the samples tests. Figure~\ref{fig:lens_noise} shows the calculated difference in the receiver noise with respect to the second reference measurement. 
The difference between the two reference measurements (yellow and black curves in Fig.~\ref{fig:lens_noise}) thus indicates the level of measurement uncertainty is of the order of $\lesssim 1$~K.

The sample thicknesses were selected for each material to represent the maximum thickness for the 96~mm diameter lens. Thus, the derived excess noise corresponds to an upper limit for the noise contribution expected for a lens made from each material.
The preliminary results from the lens material tests are very encouraging, and despite the nonidealities in the geometry of the AR surface, imply similar total loss for the silicon and UHMW-PE samples.
Further measurements are planned or ongoing, using samples with improved quality fabricated from UHMW-PE and HDPE. These tests will allow us to accurately determine material parameters and AR layer properties individually, the results of which will be reported elsewhere.

%%%%%%%%%%%%%%%%%%%%%%%%%%%%%%%%%%%%%%%%%%%%%%%%%

\subsubsection{Feedhorn Designs}\label{sec:feedhorns}

The feedhorn is located at the 15~K stage of the CCA. All the designs presented here have a corrugated profile and have been optimised for the goals of achieving low cross-polarisation ($< -30$~dB), a reflection loss $S_{11}<-25$~dB, good beam symmetry and appropriate beam size and phase centre location (PCL) as a function of frequency.  The designs for the AR surface vary in their manufacturing approaches, and include modular stacked designs, monolithic machined designs, electroformation, and variations in where the profiling occurs, which can be implemented by varying the inner radius of the horn or the depths of the corrugations.

%----------------------------------------------------------------------------------------------------------

\paragraph{Universidad de Chile feedhorn design:}

\begin{figure}
\centering
\includegraphics[height=4.5cm]{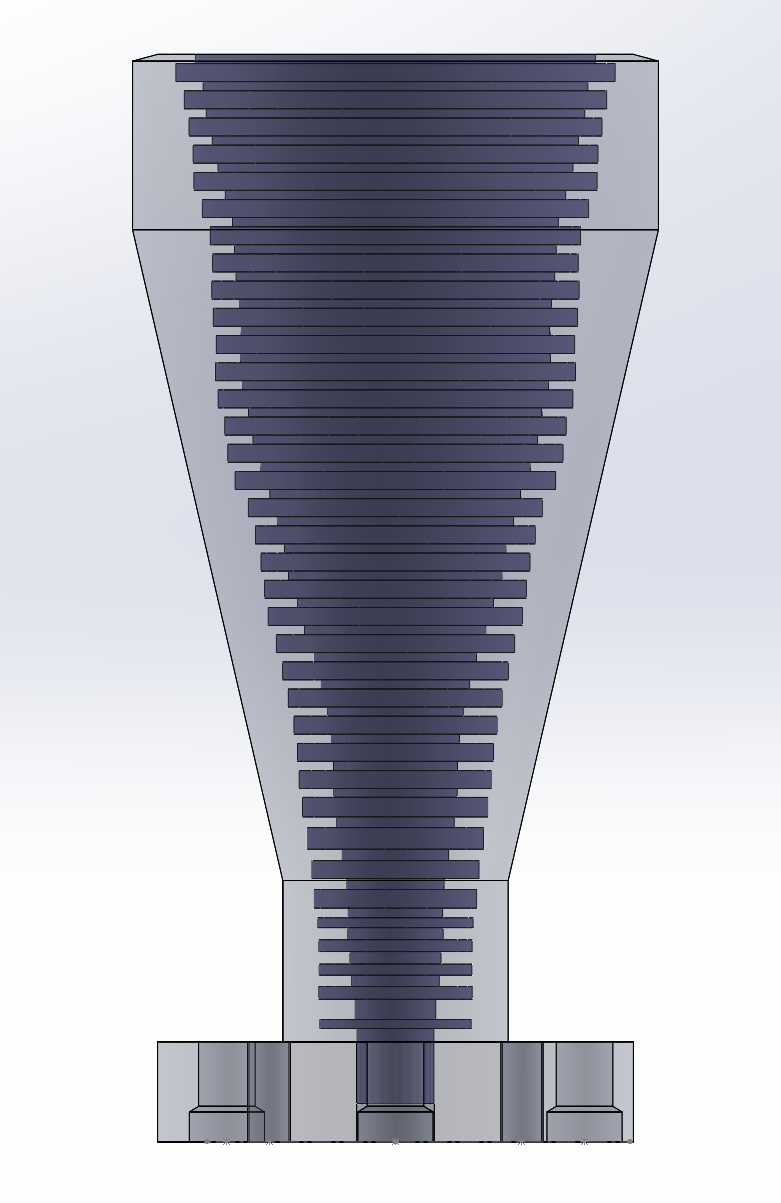}
\includegraphics[height=4.5cm]{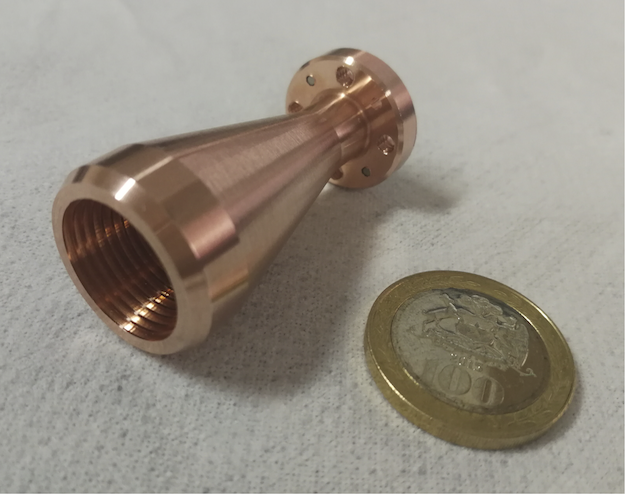}
\caption{Computer model of UdC horn (left) and picture of the final copper version (right). A 100 Chilean peso (CLP) coin is shown for scale (23.5~mm).}\label{fig:feed_udc}
\end{figure}

\begin{figure}
\centering
\includegraphics[width=\columnwidth]{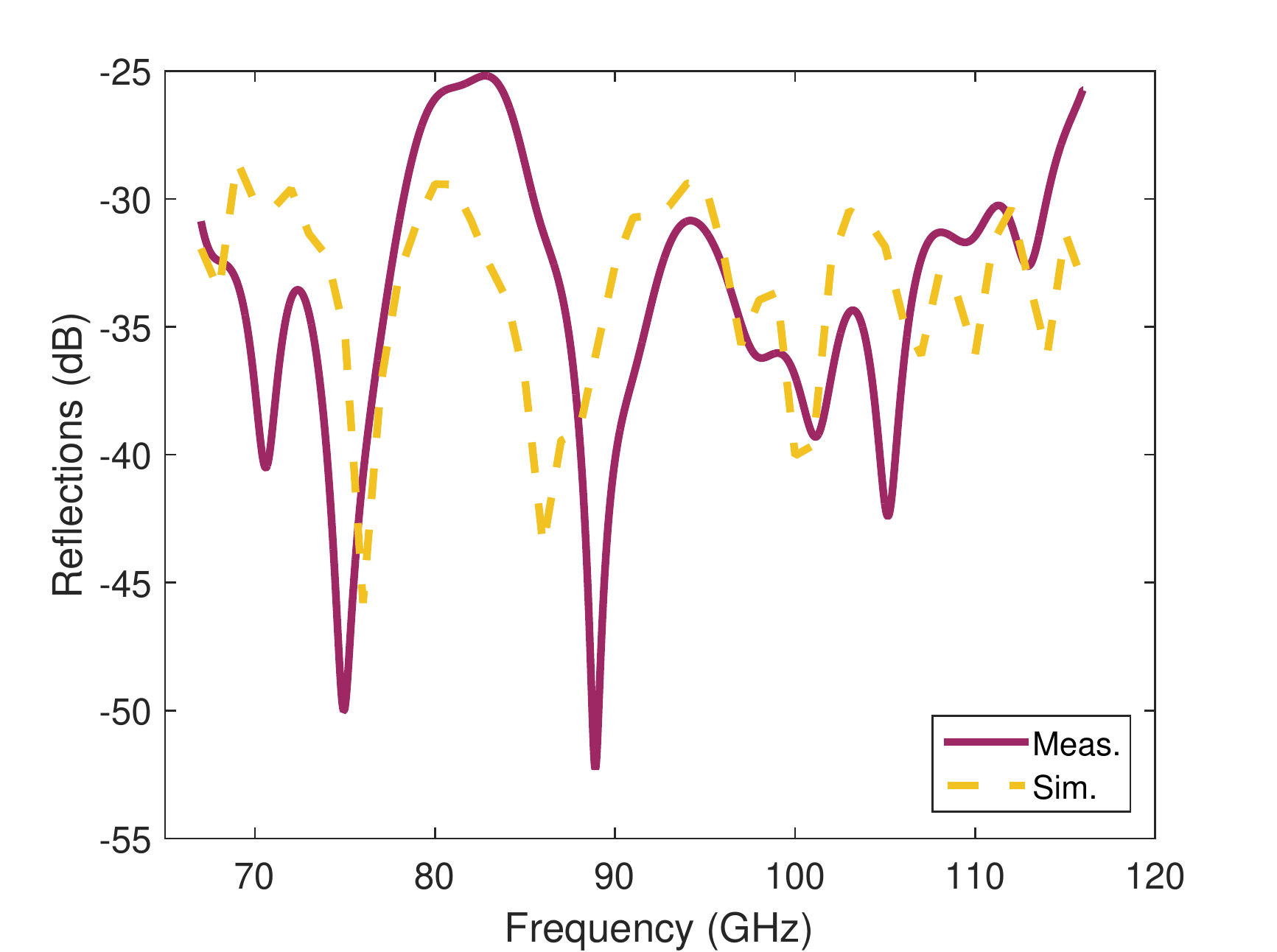}
\includegraphics[width=\columnwidth]{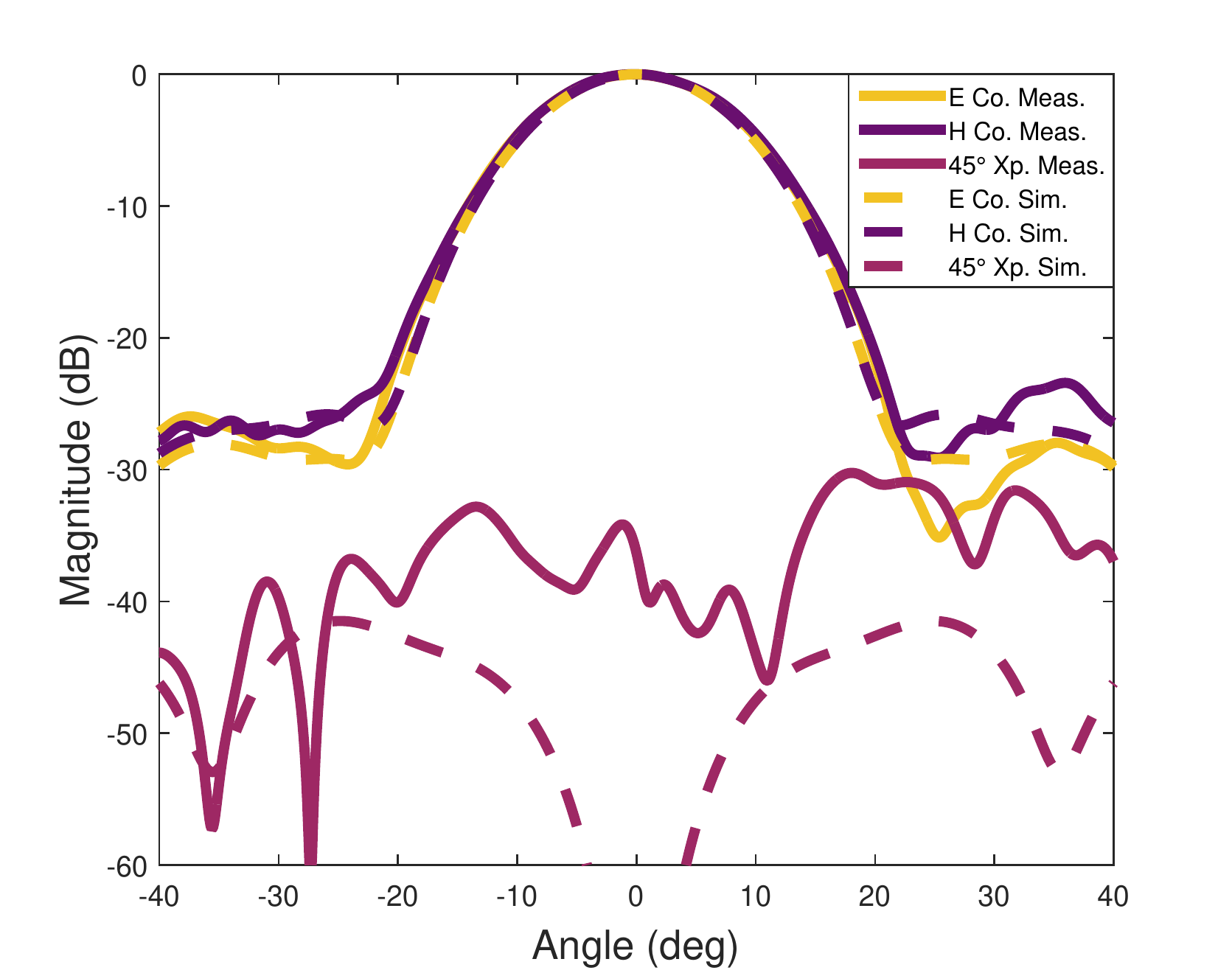}
\caption{{\it Upper:} Measurement and simulation results for the reflected power, $S_{11}$, from the UdC feedhorn.
{\it Lower:} Measured and simulated far-field radiation pattern of the UdC feedhorn at 90~GHz, near the band centre.
Our EM simulations confirm that the beam properties scale smoothly as a function of wavelength and, hence, we only show one representative frequency.}\label{fig:udc_feed_results}
\end{figure}

In addition to being corrugated, the UdC horn has a non-monotonic profile for the feedhorn inner diameter (see Fig.~\ref{fig:feed_udc}, left panel). The number of corrugations, their dimensions, and the profile of the horn were all optimised using the mode-matching method implemented in the commercial software Microwave Wizard. Some mechanical constraints were included to allow the machining of the horn as a single piece on a CNC lathe. 
Particular attention was put in the design of the corrugations corresponding to the converter-mode, since they are critical in the transformation of circular waveguide modes to the hybrid modes propagating into the horn. 
The depth of these corrugations were limited in order to allow the fabrication on a lathe. The resulting design is very compact, with a physical length of 43~mm (corresponding to a 14 wavelengths at the central frequency).  A photograph of the copper feedhorn prototype is shown in the right panel of Fig.~\ref{fig:feed_udc}. 
The reflection loss $S_{11} < -25$~dB (see upper panel of Fig.~\ref{fig:udc_feed_results}), and the maximum cross-polarisation is $<-30$~dB in the centre of the band, as shown in the lower panel of Fig.~\ref{fig:udc_feed_results}. The beam waist and PCL present good Gaussian behaviour with excellent E and H-plane symmetry. The co-polarisation patterns show excellent Gaussicity, symmetry and lower sidelobes as shown in the example presented in Fig.~\ref{fig:udc_feed_results}. The general shape of the measured co-polarisation patterns shows good agreement with simulations.

%----------------------------------------------------------------------------------------------------------

\paragraph{Istituto Nazionale di Astrofisica feedhorn design:}

\begin{figure}
\centering
\includegraphics[angle=0, height=3.9cm]{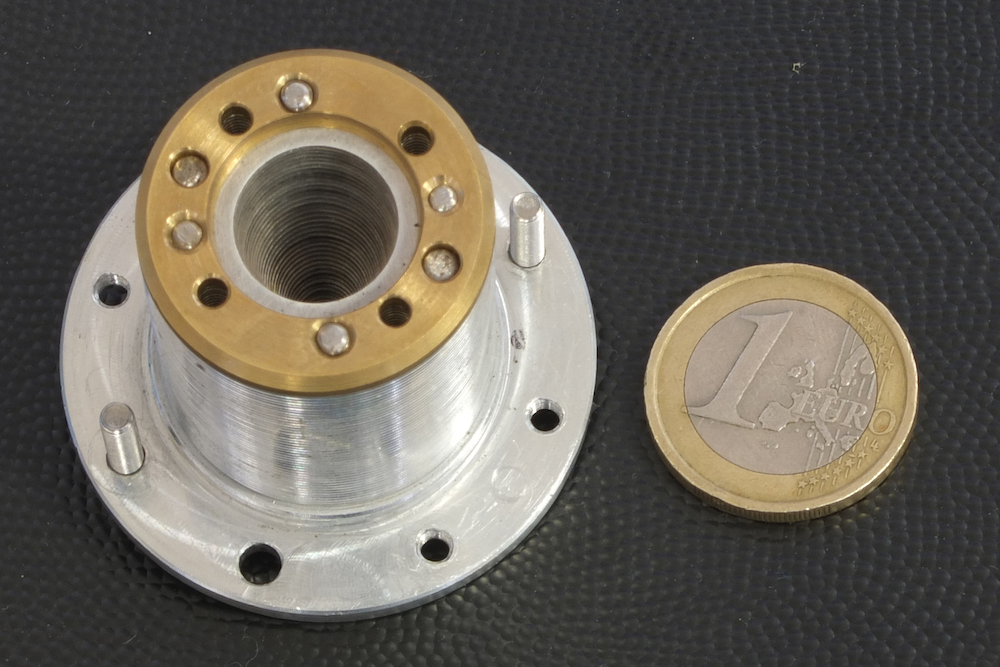}
\includegraphics[angle=0, height=3.9cm]{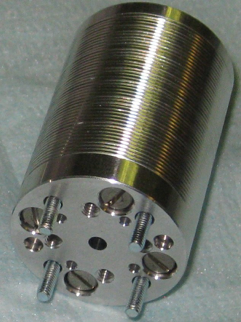}
\caption{The photos show the mouth (left) and waveguide interface end (right) of the platelet feedhorn produced by INAF. 
A 1 euro coin is shown in the left panel for scale (23.25~mm).
The total length of the feedhorn is 41.4~mm, and the outer diameter is 12.92~mm. 
}\label{fig:inaf_feed}
\end{figure}

The INAF feed was designed to obtain a taper of -12~dB at 17$^\circ$ at 91.5~GHz (the centre frequency of the RF band), and a return loss and cross polar performance lower (better) than -30~dB over the full RF band.
To meet such stringent performance requirements over the 53\% fractional bandwidth,
a curved-profile corrugated horn has been designed using a matching step-section before corrugations, as well as very narrow corrugation grooves (0.2~mm wide and 1.54~mm deep) at the throat section. As a consequence, we decided as a baseline to fabricate the horn using a platelet approach (see Fig.~\ref{fig:inaf_feed}). 
The platelet-based prototype was fabricated from aluminium (6082) with the top flange (at the aperture) optimised for cross polarisation performance.
A similar unit, equal in the design and interfaces, was electroformed, and shows similar performance. 

\begin{figure}
\centering
\includegraphics[width=\columnwidth]{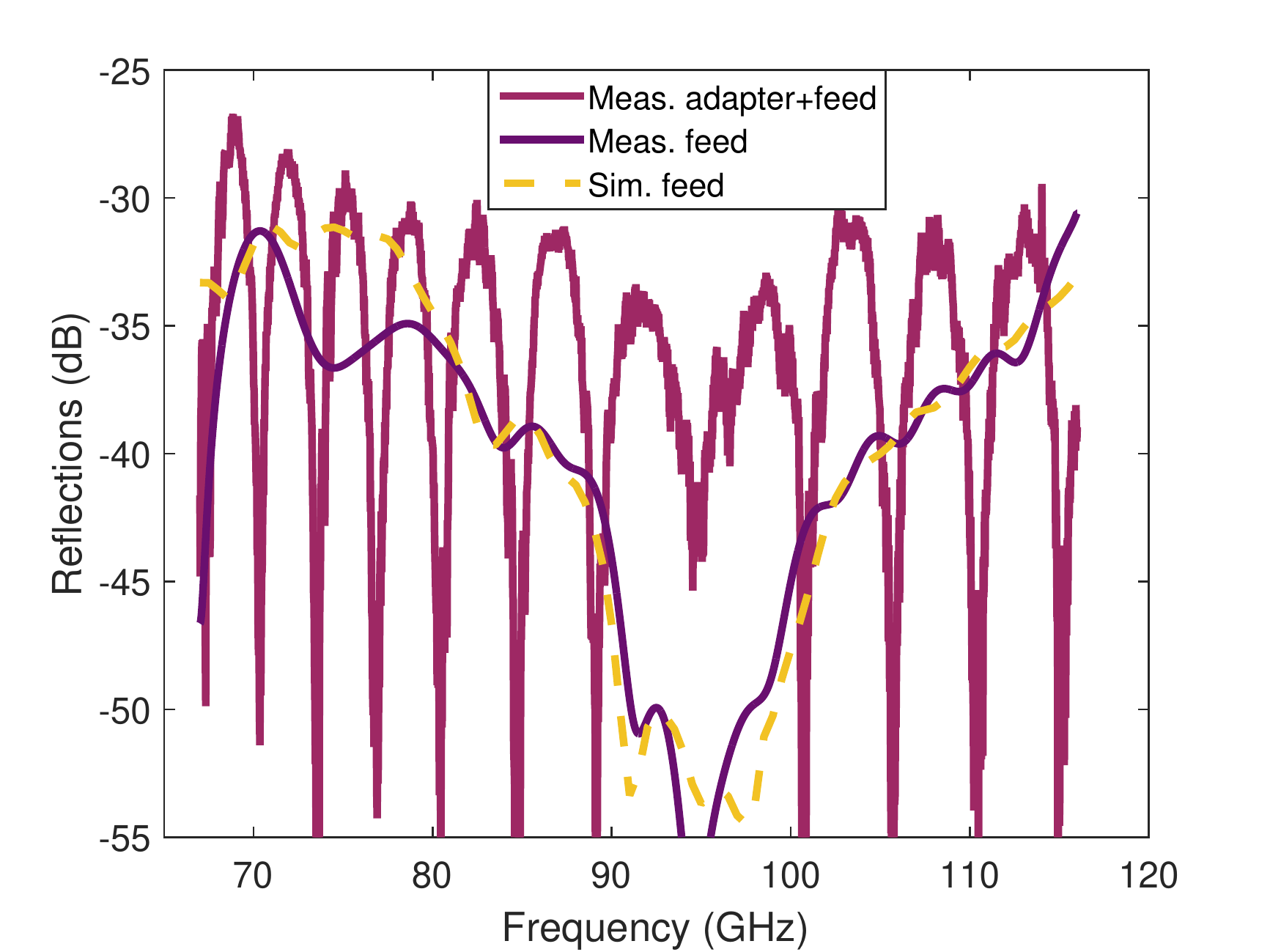}
\includegraphics[width=\columnwidth]{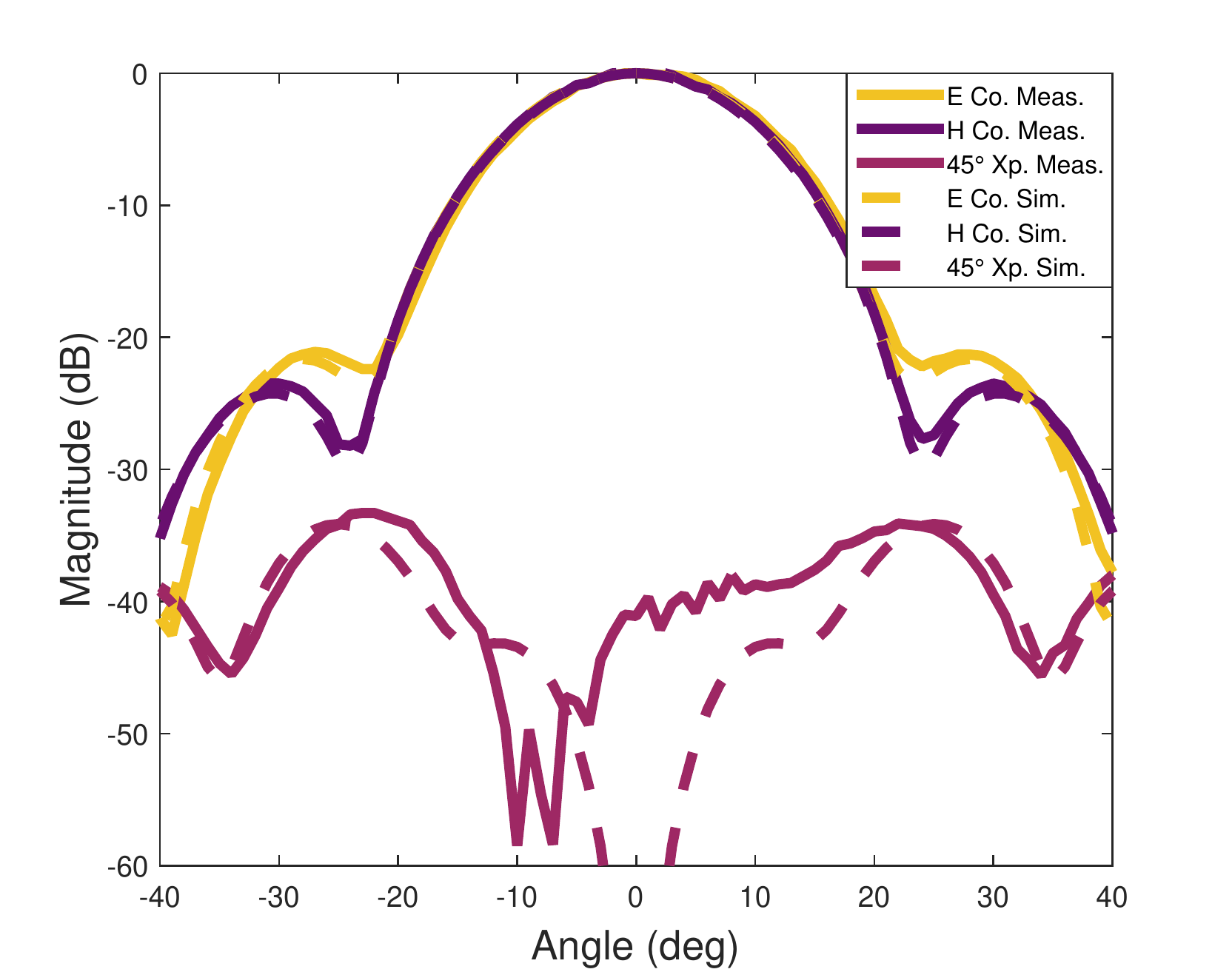}
\caption{{\it Upper:} Reflected power $S_{11}$ for the feedhorn designed at INAF. 
Here, the simulation data are compared with the measurements. The effect of the adaptor used for the measurements is also shown, in lighter purple.
{\it Lower:} INAF feedhorn test results for co-polar (Co) and cross-polar (Xp) beam pattern cuts: E-plane (E), H-plane (H) and 45$^\circ$-plane. The results for the measurements (lines) versus the simulations (dashed) are shown near the band centre, 91.5~GHz.
Our EM simulations confirm that the beam properties scale smoothly as a function of wavelength and, hence, we only show one representative frequency.
}\label{fig:inaf_feed_results}
\end{figure}

A test setup based on the Anritsu Vector Star MS4647B with millimetre wave extension modules has been used for reflection coefficient and beam pattern measurements. The feed reflection coefficient is very good and in agreement with expected simulations once the effects of the circular to rectangular waveguide adaptor, used to connect the feed to the instrument, have been calibrated (upper panel of Fig.~\ref{fig:inaf_feed_results}).
The beam pattern measurements were performed across the RF band in an anechoic chamber environment. Scans have been done in the far-field range on the principal E- and H-planes and the 45$^\circ$-plane, taking both co-polar and cross-polar components. The measurements show very good results also in terms of agreement with simulation data. The lower panel of Fig.~\ref{fig:inaf_feed_results} show  examples of the co-polar and the cross beam pattern measurements at the representative frequency of 91.5~GHz, near the band centre.

%----------------------------------------------------------------------------------------------------------

\paragraph{National Astronomical Observatory of Japan feedhorn design:}

\begin{figure}
\centering
\includegraphics[angle=0, width=0.99\columnwidth]{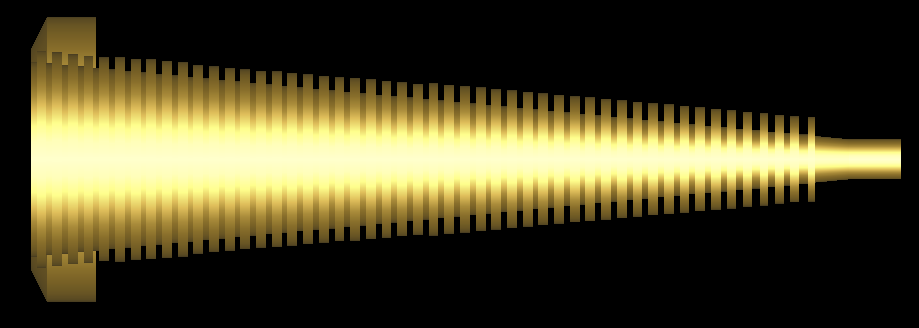}
\caption{Computer model of NAOJ feedhorn, showing the corrugation depth is varied. Figure to appear in Gonzalez et al. (in prep).\ }\label{fig:feed_naoj}
\end{figure}

Wideband corrugated feedhorns traditionally make use of non-linear and often non-monotonic profiles in order to obtain wideband performance, which complicates their fabrication. 
The NAOJ corrugated feedhorn we present here makes use of novel ideas presented in \cite{Gonzalez2017} and obtains good performance over a wide bandwidth by implementing the profile in the corrugation depth, while keeping the horn profile linear (i.e.\ the inner diameter follows a simple flare; see Fig.~\ref{fig:feed_naoj}). This eases the fabrication to a single monolithic block of aluminium by direct machining, since the horn profile can be drilled with a linear taper tool and then corrugations are turned individually on a lathe.
Using this idea, the NAOJ feedhorn is composed of 50 corrugations with constant pitch and optimised depth, which changes every few corrugations. The corrugations in the throat of the horn change both depth and width individually for optimum matching \citep{Zhang1993}. The design has been performed by the hybrid mode-matching/Method-of-Moments code in WaspNET. Two prototypes have been fabricated, each as a single piece, by direct machining of a single block of aluminium.  

\begin{figure}
\centering
\includegraphics[width=\columnwidth]{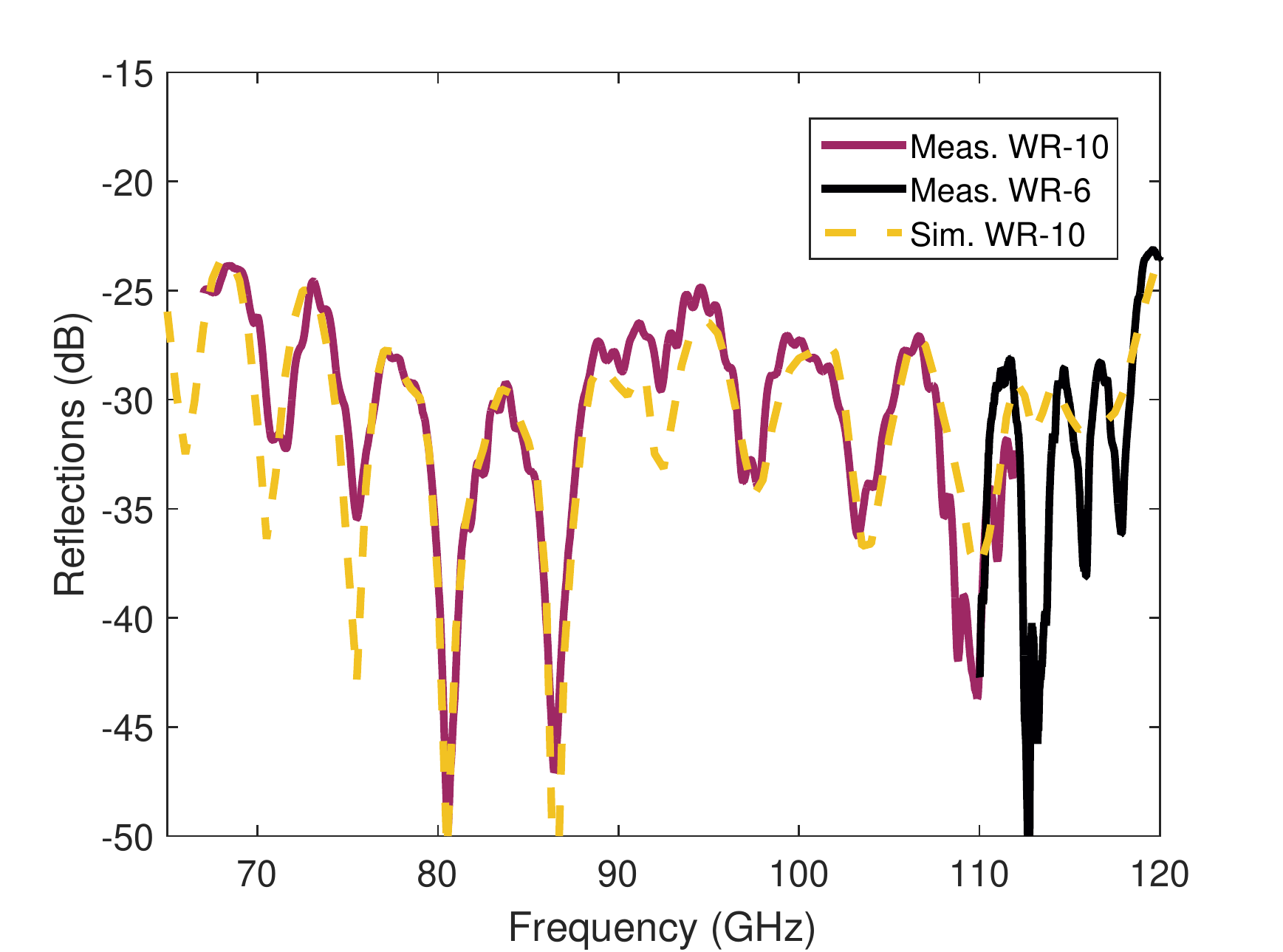}
\includegraphics[width=\columnwidth]{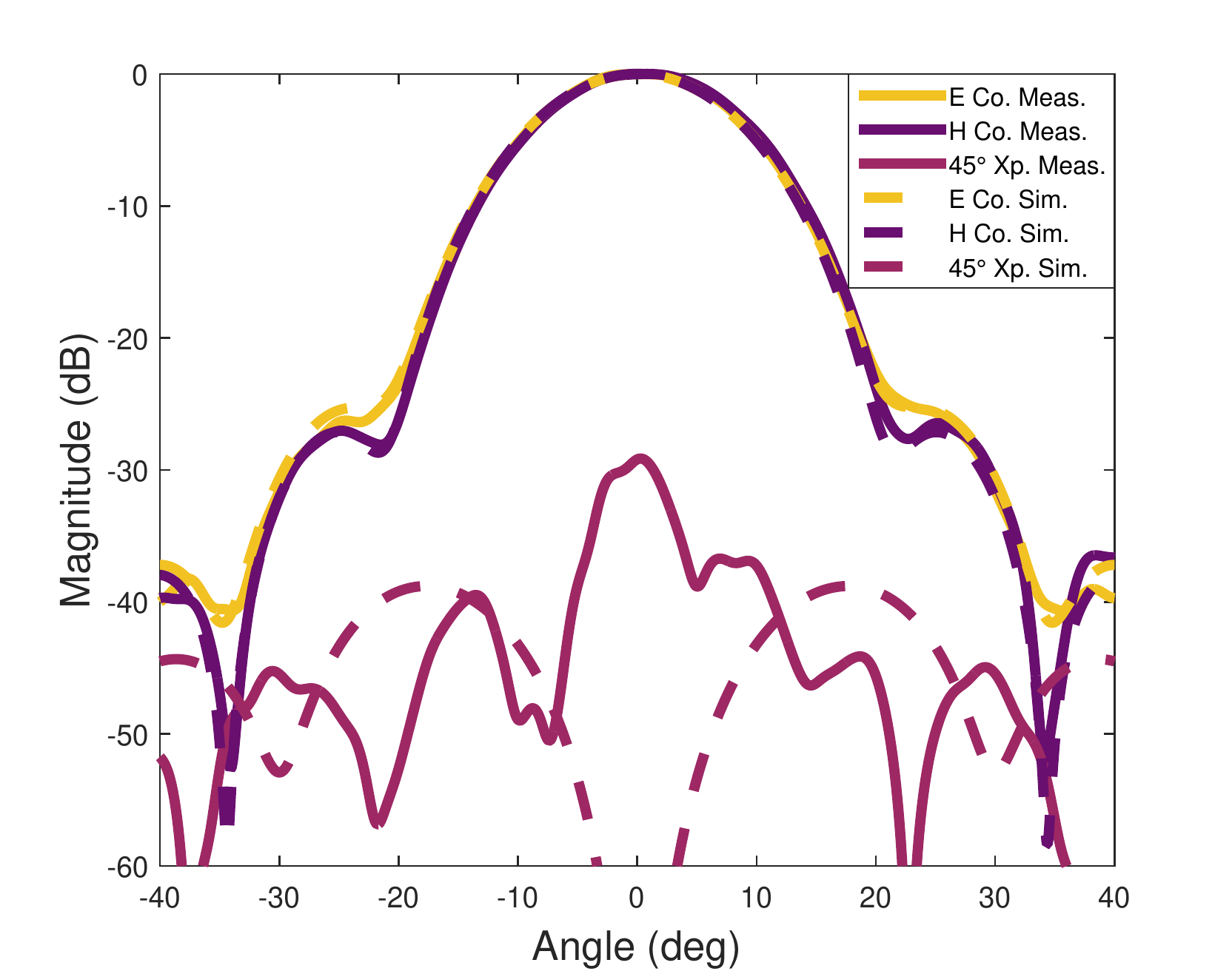}
\caption{{\it Upper:} Measured $S_{11}$ for the NAOJ feedhorn design.  
The purple curve presents data measured directly with WR10 VNA extenders, whereas the black curve ($>110$~GHz) shows data taken with WR6 extenders and waveguide transitions and are affected by the $S_{11}$ of the additional transitions. 
{\it Lower:} Comparison of the NAOJ feedhorn beam pattern from simulations and measurements at 95~GHz, close to the central frequency of 91.5~GHz.  Our EM simulations confirm that the beam properties scale smoothly as a function of wavelength, and hence we only show one representative frequency.
}\label{fig:naoj_horn_results}
\end{figure}

The feedhorns have been characterised using a VNA with mm-wave extenders at NAOJ, and show similar performance (see Fig.~\ref{fig:naoj_horn_results}). Most of the band is measured directly with 67-112~GHz WR10 VNA extenders. However, the upper end of the band, from 110 to 120~GHz, is measured with 110-170~GHz WR6 extenders, which necessitates additional waveguide transitions. 
These transitions have a relatively poor back-to-back $S_{11}\sim -20$~dB, which degrades the measurement results by a few dB. 
In addition, the calibration kit does not have a quarter-wavelength waveguide standard and both the short and open standards are measured through short waveguides with lengths 1 and 1.25 wavelengths.
In addition to these transitions, a circular-to-rectangular (C2R) waveguide transition is necessary for any measurement to connect the circular waveguide output of the horn to the rectangular standard waveguides in the measurement setup. Back-to-back $S_{11}$ values of two C2R transitions are around -30~dB. This degrades the measurement results by a few dB and is indicated in darker purple in the upper panel of Fig.~\ref{fig:naoj_horn_results}.  
The reflection loss is $S_{11} < -24$~dB over the entire 67-116 GHz band and rises sharply for frequencies $\nu \gtrsim 120$~GHz.

Apart from S-parameter characterisation, the NAOJ feedhorn radiation patterns have been measured, together with the OMT described (Section \ref{sec:omt}) and a thin quarter-wavelength-long transition, using a near field beam scanner.  Details of the measurement system can be found in \cite{Gonzalez2016}.
The far field results for the feedhorn are obtained by Fourier transformation of the measured near field data. The comparison between the measured patterns and simulation results using WaspNET (mode-matching + Method-of-Moments) is shown in Fig.~\ref{fig:naoj_horn_results}. We note that, in this comparison, the simulation results are for the horn alone while the measurements were performed with the combination of horn, waveguide transition, and OMT. This is evident in the values of cross-polarisation at the boresight angle of the horn. The measurements and simulation results agree well. The maximum value of the cross-polarisation pattern as a function of frequency is $< -23$~dB over the full range, and is typically $\sim -29$~dB. 
These results are again for the full feedhorn, transition, and OMT assembly. 
The cross-polarisation leakage is less than -23~dB for every frequency in the 67-116~GHz range.

%%%%%%%%%%%%%%%%%%%%%%%%%%%%%%%%%%%%%%%%%%%%%%%%%%%%%%

\subsubsection{Orthomode transducer designs}\label{sec:omt}

\begin{table}
    \centering
        \caption{Requirements for the OMT. Note that the waveguide flange diameters in the designs presented here differ as each OMT design couples to a specific feedhorn design (Sect.~\ref{sec:feedhorns}) and the assembly is jointly optimised. Therefore, different input waveguide diameters were a design choice left to each group's discretion.    }
    \begin{tabular}{lcc}
        \hline\noalign{\smallskip}
        Specification & Value & Unit \\\noalign{\smallskip}
        \hline\noalign{\smallskip}
        Bandwidth  & 67-116 & GHz \\
        Return Loss & $< -15$ ($< -20$ goal) & dB\\
        Insertion loss & $< 0.5$ &      dB \\
        Cross-polarisation maximum & $< -30$ & dB\\
        Isolation & $>30$ & dB\\
        Output waveguides &     WR10 & NA\\
        Input waveguide flange & UG387/U & NA\\
        Input waveguide diameter & 2.93/3.1 & mm\\
        \noalign{\smallskip}
        \hline
    \end{tabular}
    \label{tab:omt_specs}
\end{table}

An OMT is used after the feedhorn to separate the orthogonal linear polarisation components. 
The main performance drivers for the OMT design, summarised in Table \ref{tab:omt_specs}, are to minimise insertion loss and cross-polarisation, and to minimise the return loss (with a goal of $S_{11}< -20$~dB loss) and the isolation between the polarisation channels. 
As with the feedhorn designs (Section \ref{sec:feedhorns}),
in this section we detail the three designs developed in parallel by teams at Universidad de Chile (UdC), Istituto Nazionale di Astrofisica (INAF), and the National Astronomical Observatory of Japan (NAOJ).

\paragraph{Universidad de Chile OMT design:}

\begin{figure}
\centering
\includegraphics[width=0.8\columnwidth]{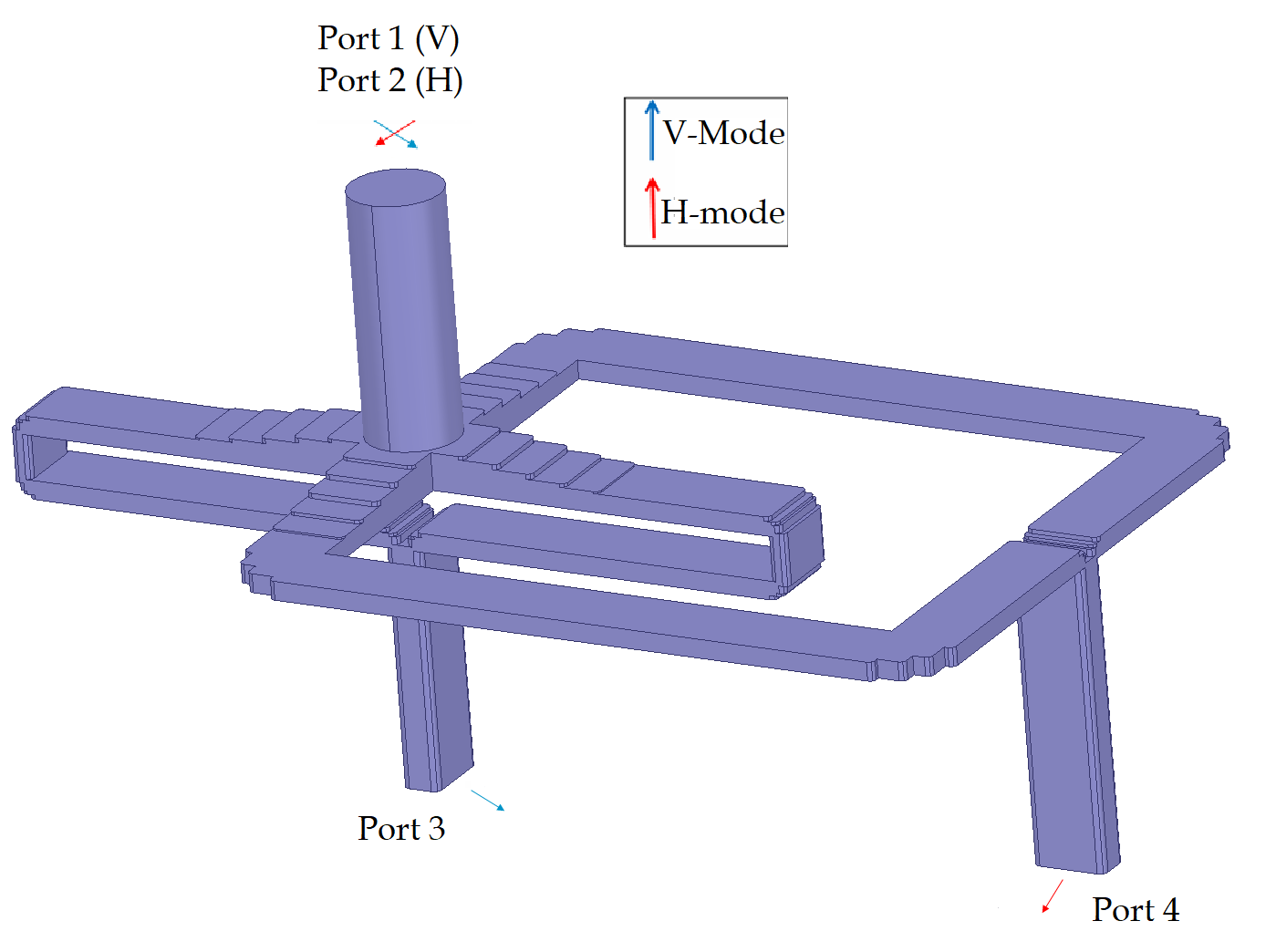}
\includegraphics[width=0.9\columnwidth]{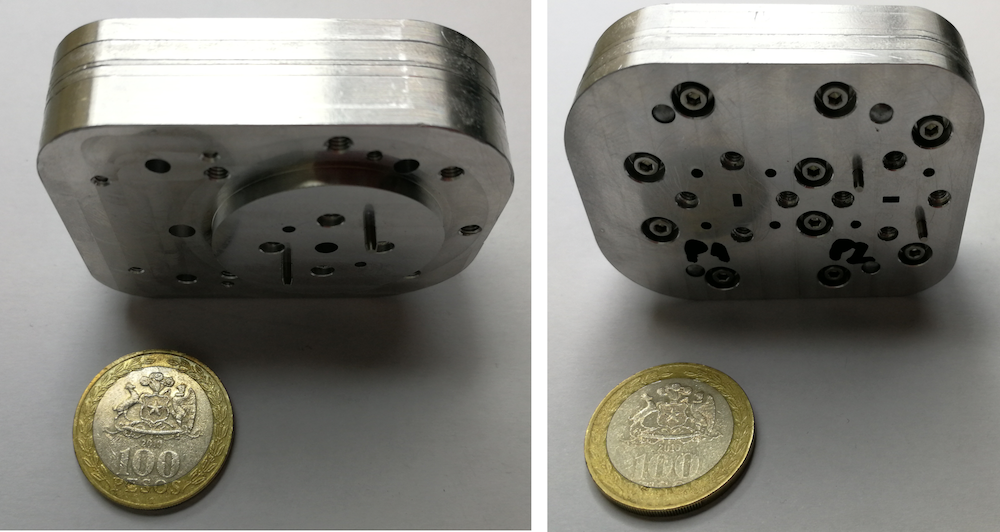}
\caption{3D model of UdC OMT (upper) and a photograph of the final aluminium version (lower). A 100 CLP coin is shown in each photograph for scale (23.5~mm).}\label{fig:udc_omt}
\end{figure}

The OMT designed by UdC is based on a turnstile junction architecture \citep{Henke2014}. It has a circular waveguide input and two WR10 rectangular waveguide outputs (as noted in Table~\ref{tab:omt_specs}). It is composed of the turnstile junction, bends, steps, and power combiners. Each of these elements was optimised separately using finite element simulations. Then, the connection distance between the different elements was optimised to minimise trapped modes. However, there is an inherent trade off in the wave path length. Short paths result in fewer losses, but have a higher probability of trapping undesired modes. The upper panel of Fig.~\ref{fig:udc_omt} shows the final optimised OMT design. The OMT was constructed in five metal plates fabricated on a CNC milling machine, as shown in the photograph in Fig.~\ref{fig:udc_omt} (lower panels).  Aluminium was chosen due to its good electrical conductivity, mechanical robustness, and relative machinability (compared to pure copper). Cuts in the narrow side of the waveguide, which possibly translate in higher loss, were necessary due to the complex and compact OMT’s geometry. 

\begin{figure}
\centering
\includegraphics[width=0.9\columnwidth]{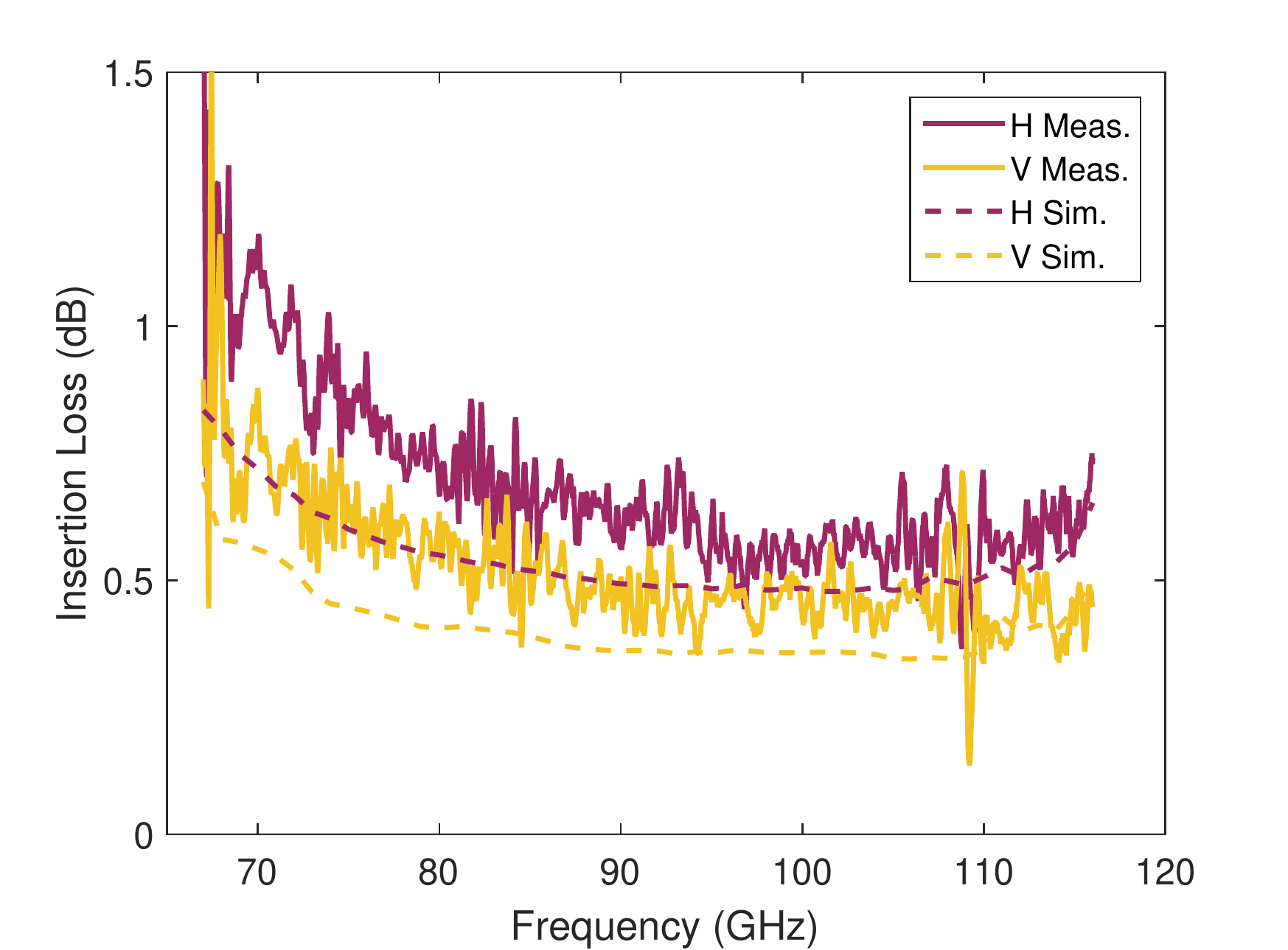}
\includegraphics[width=0.9\columnwidth]{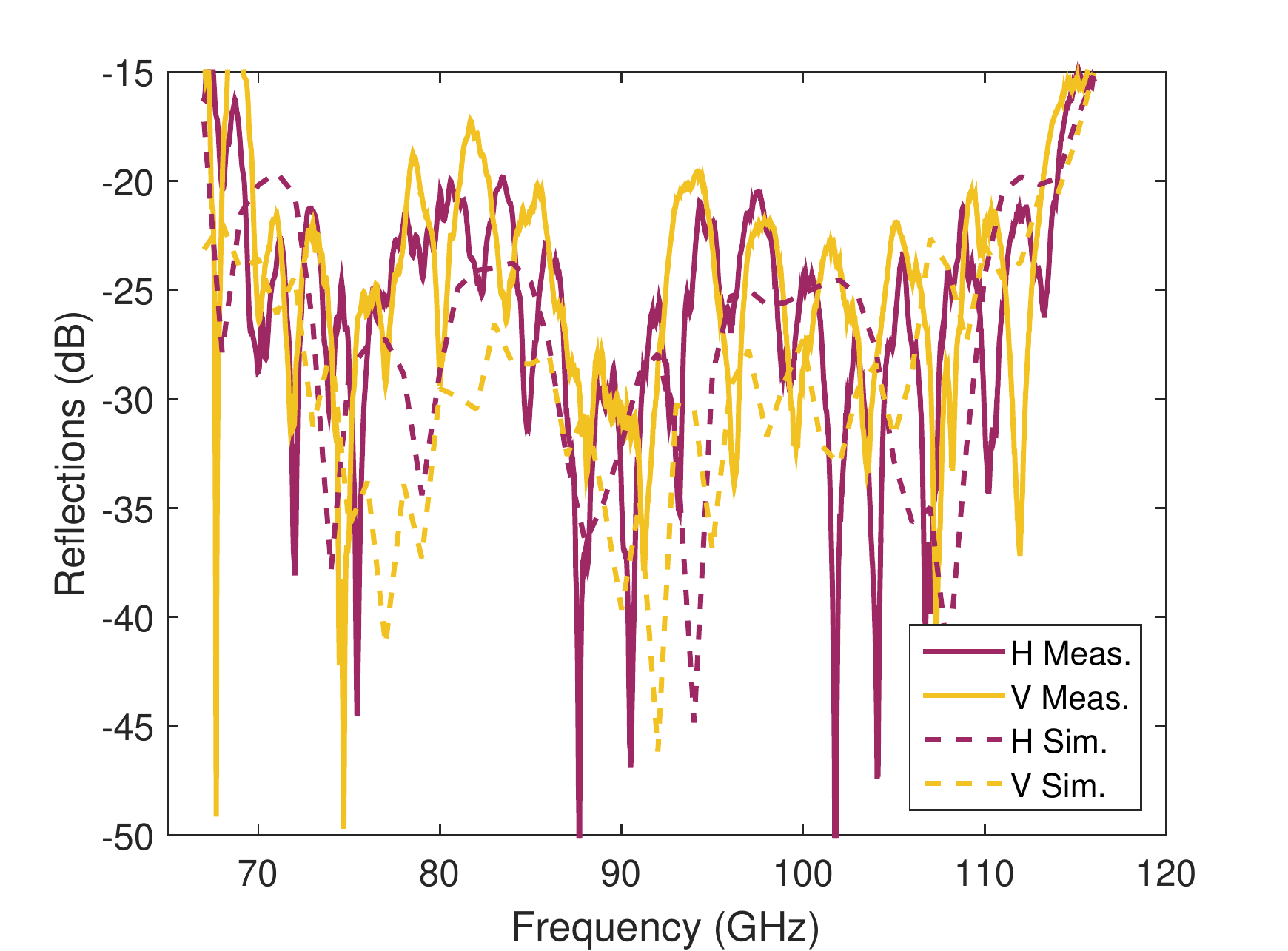}
\includegraphics[width=0.9\columnwidth]{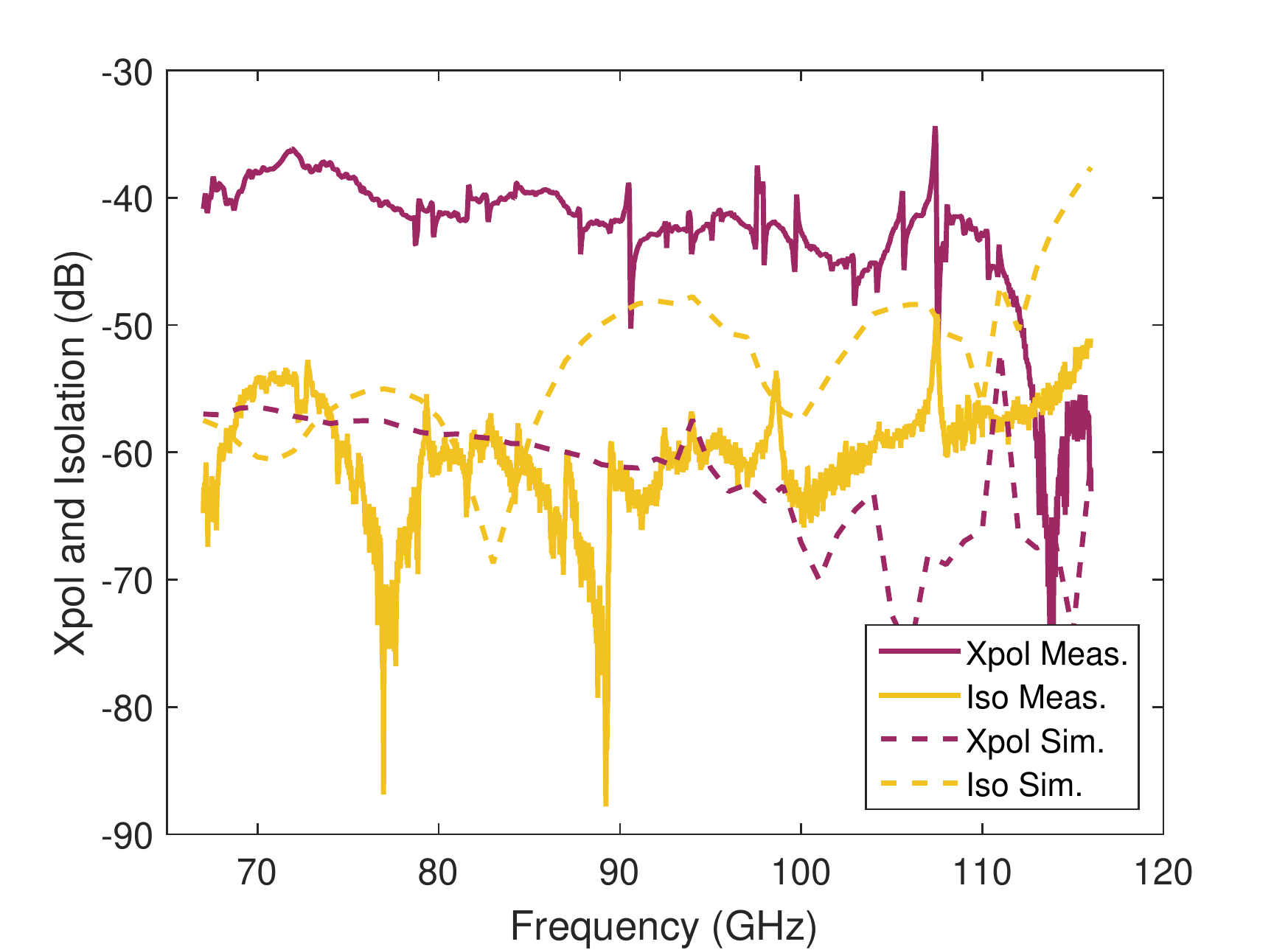}
\caption{Measured (solid lines) and simulated (dashed lines) performance of UdC OMT: insertion losses (upper), reflection (middle) and cross-polar and isolation (lower), using a $1.2\cdot10^{7}$~S/m finite conductivity of aluminium for simulations.}\label{fig:udc_omt_performance}
\end{figure}

Preliminary S-parameter measurements at room temperature were performed using a scalar network analyser built in house, and are shown in Fig.~\ref{fig:udc_omt_performance}. Cross-polarisation and isolation are under $-30$~dB in all bandwidth. Reflection losses are lower than $-15$~dB (except for the range $67-69$~GHz), corresponding to peak differences of about 7~dB with respect to simulations. This suggests that the fabrication procedure could be improved somewhat. Insertion loss is around $0.5$~dB over $85$~GHz and around $1$~dB in the lower part of the frequency band (upper panel of Fig.~\ref{fig:udc_omt_performance}).

\paragraph{Istituto Nazionale di Astrofisica OMT design:}

\begin{figure}
\centering
\includegraphics[height=35mm]{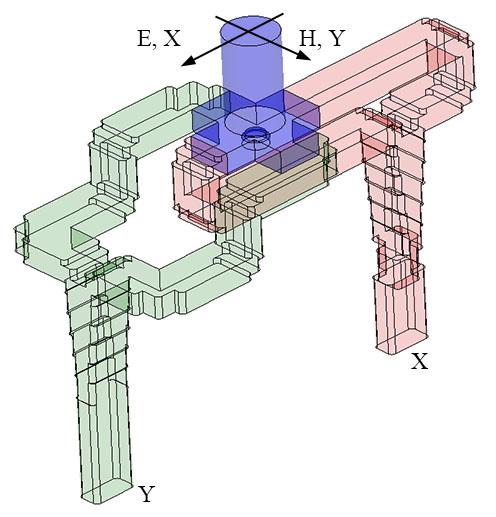}
\includegraphics[clip=true, trim=12mm 8mm 12mm 10mm, height=35mm]{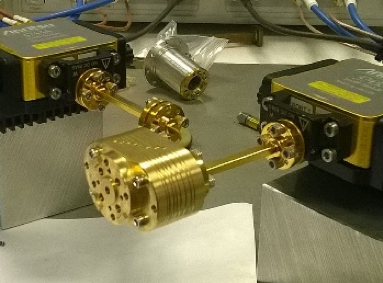}
\caption{{\it Left:} EM model used for simulation of the INAF OMT. 
{\it Right:} Photograph of the INAF OMT connected to the setup used for S-parameter tests in Fig.~\ref{fig:inaf_omt_Sparams}.}\label{fig:inaf_omt}
\end{figure}

The development effort at INAF to produce an OMT design meeting the requirements stated in Table~\ref{tab:omt_specs} resulted in three slightly different designs and models, all of which are based on turnstile junctions. 
Each design was fabricated using platelet technology, where individual plates can be fabricated by standard machining. 
The first OMT model was developed with the primary goal of demonstrating the ability to cover the 53\% of bandwidth. 
It was manufactured using aluminium but it showed high insertion losses (with losses ranging from 0.6-0.8~dB across most of the band to 1.2-1.4~dB at the lower end of the band).
At the time of this first design, matching was presumed to be a more challenging task, due to the very broad bandwidth. 
The second OMT incorporated a revised design to reduce the insertion loss, while maintaining all other performance by minimising the waveguide path lengths and optimising the accuracy and roughness during fabrication. 
For this design we used brass, which easily allows for gold plating the OMT in the future (i.e. to reduce loss by improving the surface conductivity). The resulting unit showed an insertion loss of 0.3-0.4~dB across most of the band, with a degradation up to 0.7-1~dB at lower frequencies. 
Ultimately, a third OMT was developed using a design approach that aimed to reduce the number of plates for manufacturing and to further reduce the insertion loss. 
The core was manufactured using copper tellurium while the external mechanical support flange was stainless steel. The resulting insertion loss performance is excellent, approximately 0.25~dB on average over the whole band. 
However, the OMT performance at cryogenic temperature degrades such that the unit produced by INAF most suitable for use in the wide Band 2 receiver was the second (brass) OMT design. This unit was extensively tested in the CCA during the cryogenic noise measurement campaign together with the platelet INAF feedhorn.

\begin{figure}
\centering
\includegraphics[width=0.9\columnwidth]{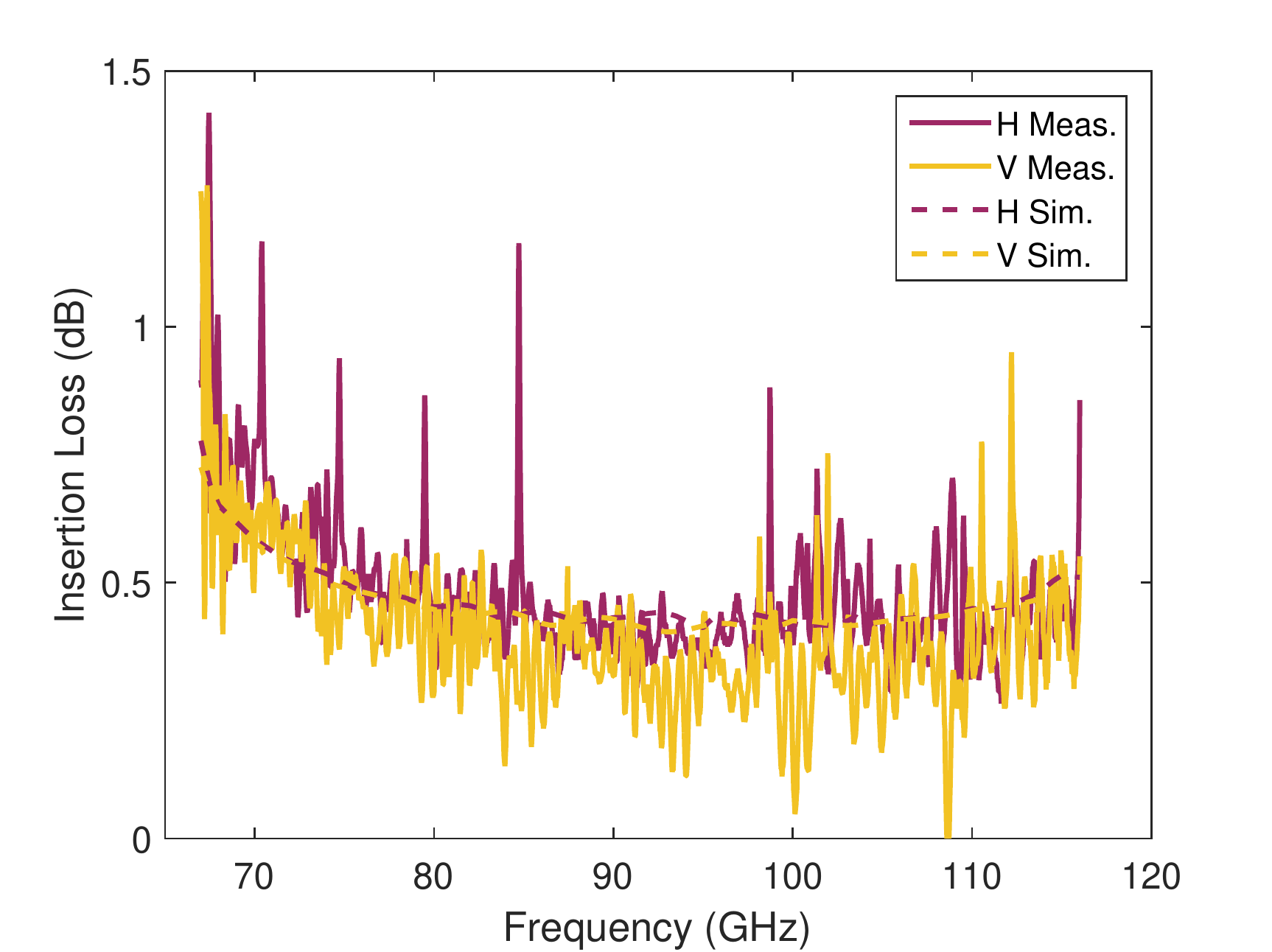}
\includegraphics[width=0.9\columnwidth]{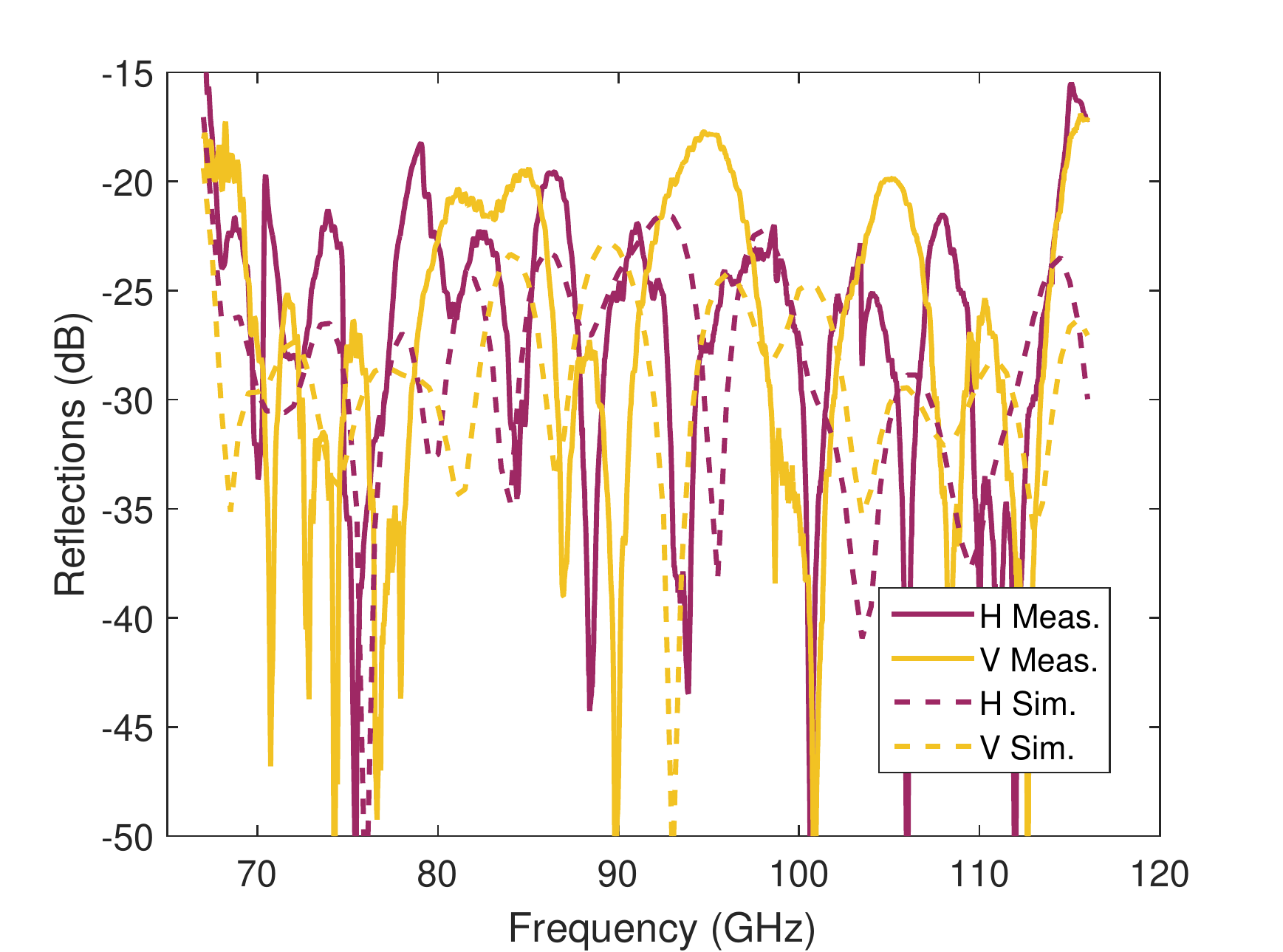}
\includegraphics[width=0.9\columnwidth]{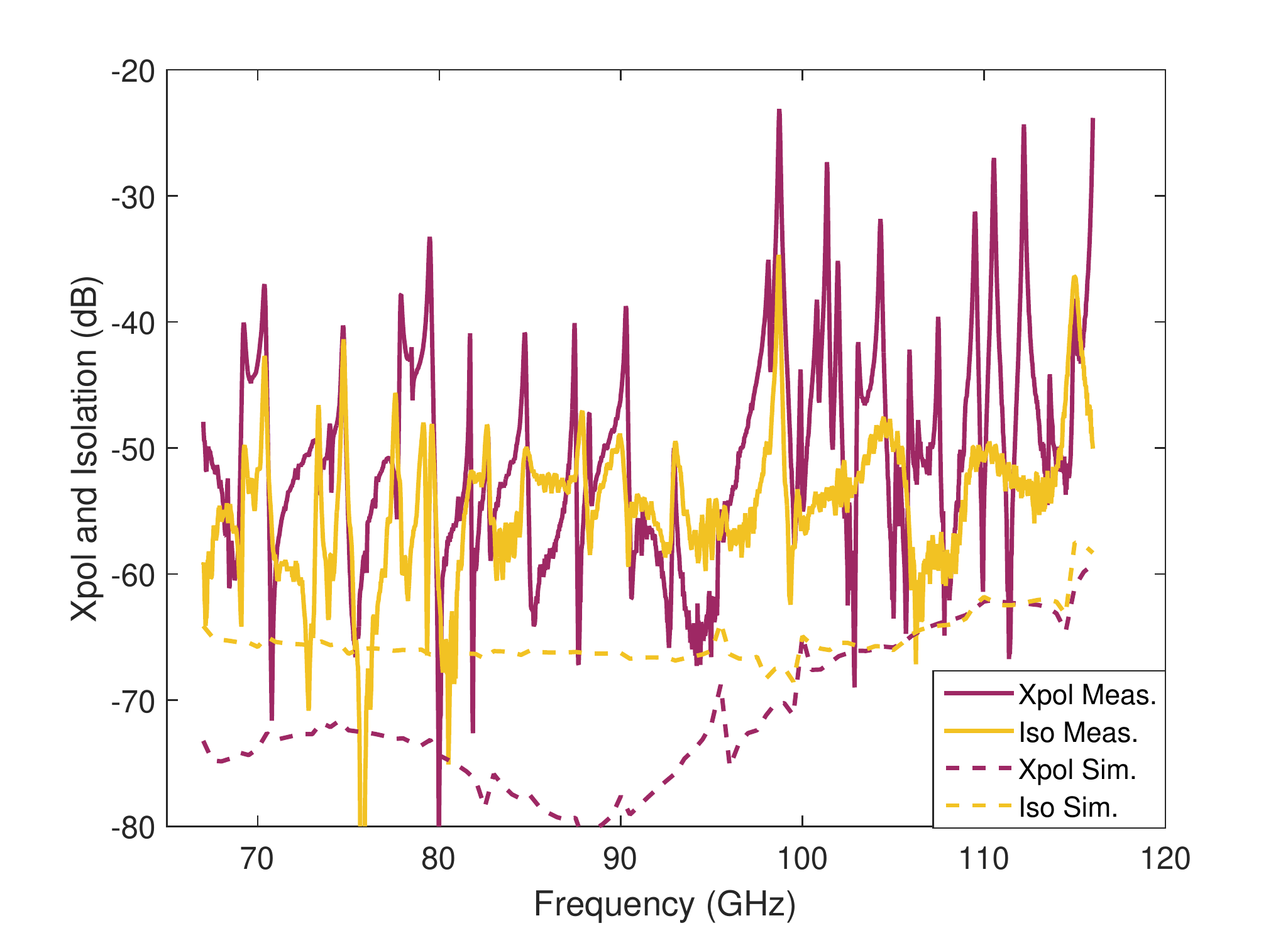}
\caption{Panels show the simulation and measured results for the insertion loss (upper), reflection coefficient (middle), and cross-polarisation and isolation (lower) measurements for the INAF OMT. 
}\label{fig:inaf_omt_Sparams}
\end{figure}

Fig.~\ref{fig:inaf_omt} shows the electromagnetic design (left) and a photograph (right) of the second OMT during S-parameter tests. 
Fig.~\ref{fig:inaf_omt_Sparams} shows the S-parameter test results
using a VNA setup.
As regards return loss, isolation, and cross polarisation, the simulations are in very good agreement with the measurements.
We note that, due to the adopted setup and load, the cross-polarisation measurement is contaminated by effects arising from the polarisation induced by the loading short non-ideality and the resonant peaks of spurious high order mode excitation. Since these effects are very difficult to remove in the post processing, we caution the reader to regard the cross-polarisation measurements as upper limits.

%%%%%%%%%%%%%%%%%%%%%%%%%%%

\paragraph{National Astronomical Observatory of Japan OMT design:}

\begin{figure} 
\centering 
\includegraphics[height=40mm]{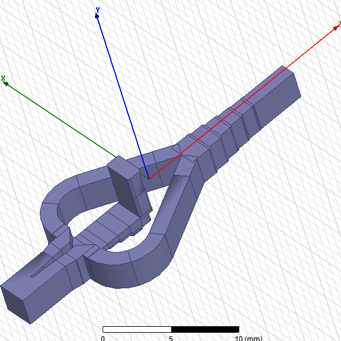} 
\includegraphics[height=40mm]{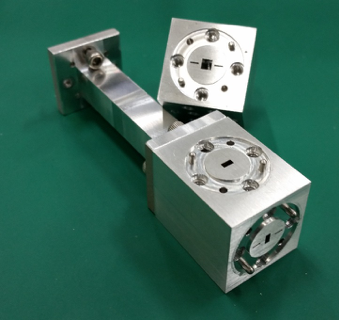} 
\caption{{\it Left:} EM model for the NAOJ OMT, used for simulation and optimisation in HFSS. {\it Right:}  Photograph of the two OMT prototypes, one of them connected to the WR10 to square waveguide transition used for measurements.}\label{fig:naoj_omt} 
\end{figure} 

The NAOJ OMT design is based on a standard dual-ridged waveguide Boifot junction modified for wideband performance \citep{Gonzalez2018}. The design has been performed by the hybrid mode-matching/finite-elements (MMFE) code in WaspNET, and confirmed by FE simulations using the Ansys High-Frequency Structure Simulator (HFSS). 
The EM model for the OMT and a photo of the OMT produced are shown in Fig.~\ref{fig:naoj_omt}. After the initial design was completed, rounded corners with radius 0.2~mm were added in the models and the design was re-optimised. The design is very compact, which translates in very low loss. 
The output waveguide size is standard WR10, consistent with the requirement in Table \ref{tab:omt_specs}. Two prototypes have been fabricated by direct machining as a split-block. These have been characterised by VNA and mm-wave extenders at NAOJ in the same way as for the NAOJ corrugated horn (Section \ref{sec:feedhorns}).

\begin{figure} 
\centering 
\includegraphics[width=0.9\columnwidth]{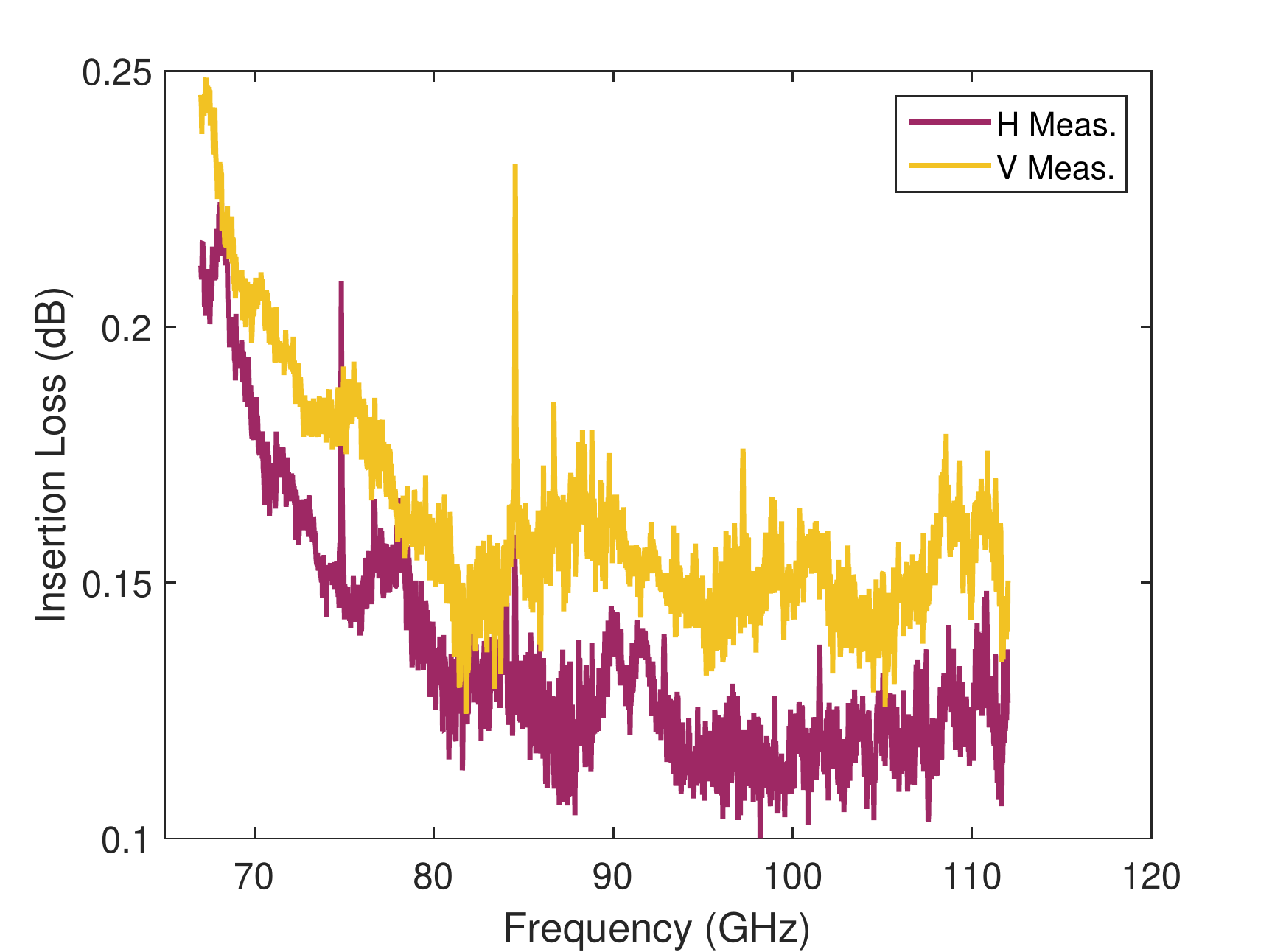}
\includegraphics[width=0.9\columnwidth]{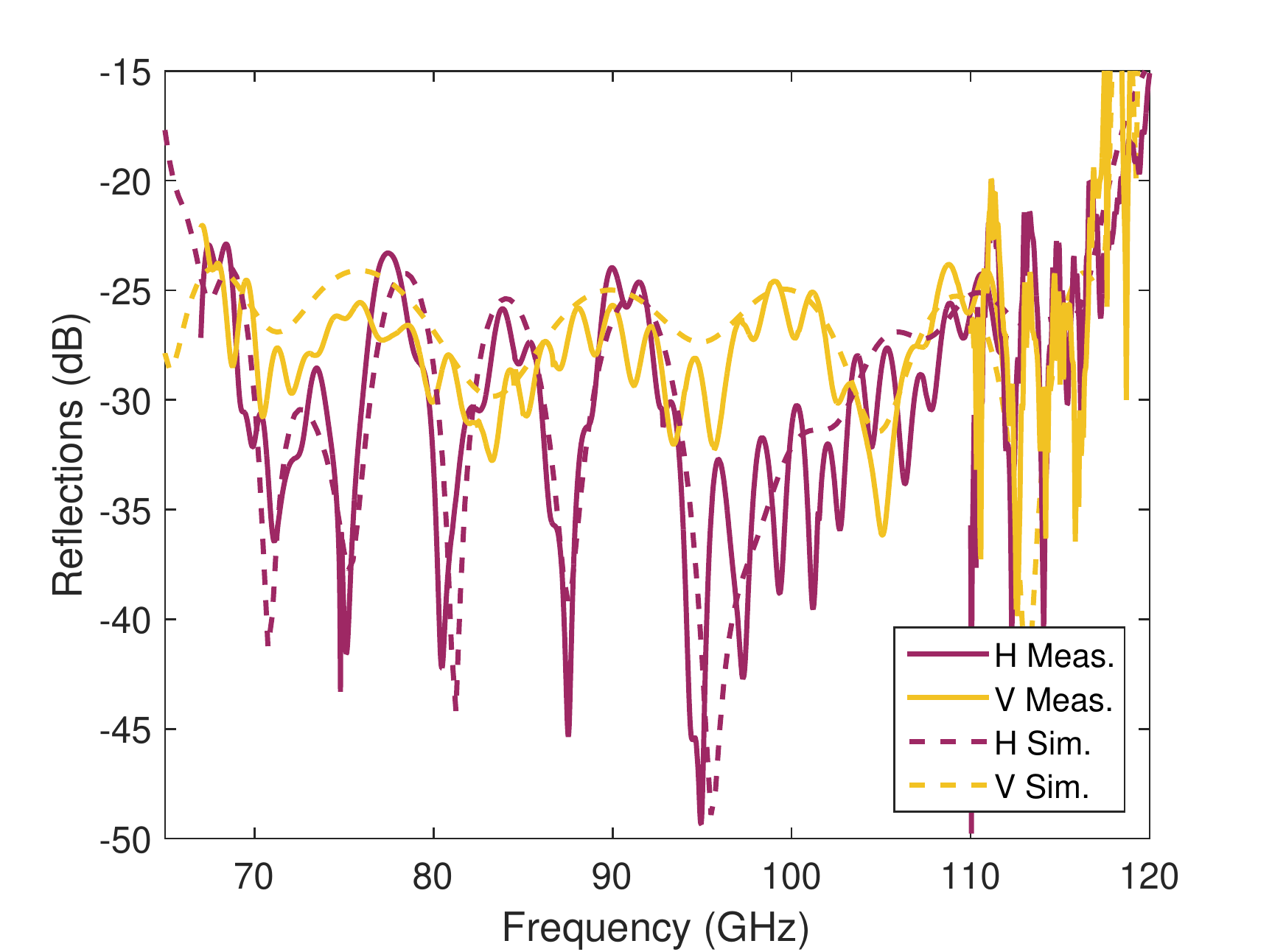}
\includegraphics[width=0.9\columnwidth]{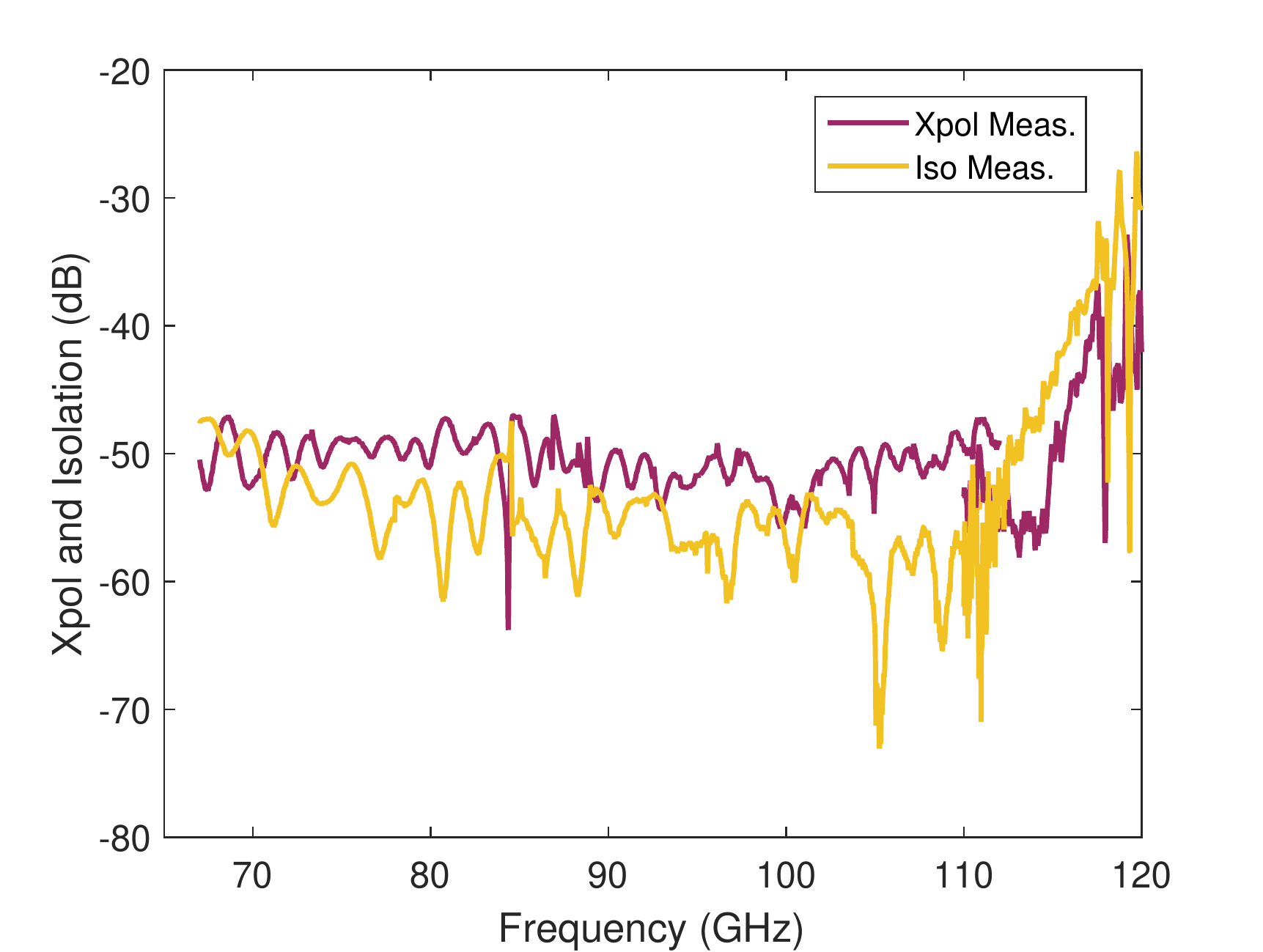}
\caption{The panels show the measured insertion loss (upper), the measured and simulated reflection coefficients (middle), and the cross-polarisation and isolation (lower) of the NAOJ OMT. 
The reflections are shown for the Waspnet simulations, which are in good agreement with the HFSS simulations.  The insertion loss measurement is limited to frequencies $< 112$~GHz by the waveguide extenders used in the measurement setup.  The loss is expected to remain flat or decrease at higher frequency.
}\label{fig:naoj_omt_results} 
\end{figure}

Measurement results for the reflection ($S_{11}$), insertion loss, cross-polarisation, and isolation of the NAOJ OMT are shown in Fig.~\ref{fig:naoj_omt_results}.
The insertion loss at room temperature is less than 0.25~dB across the full band, and $<0.15$~dB at most frequencies. Reflection ($S_{11}$) measurements match simulations and the performance is better (lower) than -23~dB across most of the band, except at 112-116~GHz due to the effect of waveguide transitions. All measured values of cross-polarisation and isolation are below -40~dB.  
Some of the presented measurements exhibit spikes, although they are in general very clean of resonances. These are associated to trapped modes in the connection between the OMT and the transition from square to rectangular waveguide used for 2-port measurements. To verify this, 1-port measurements were performed with a short in the square waveguide of the OMT. Then, the loss can be calculated as half the value of $S_{11}$. In this case, the results do not show resonances at the frequencies at which they appear in the 2-port measurement. This indicates that the resonances do not occur in the internal structure of the OMT. The values of loss in both cases are comparable, although the 1-port measurement shows a stronger standing wave.

%--------------------------------------------------------------------
\subsection{Low noise amplifier designs}\label{sec:lna}

As the first active component in the receiver and based on the assumption that the passive optical components that come before are designed to have low loss, the cryogenic LNA largely determines the overall receiver noise. Here we report on the efforts of two groups to develop LNAs with very low noise, low reflection loss, wide bandwidth ($\sim 53\%$ fractional), high gain, and a relatively flat passband gain.  

\begin{table}
    \centering
    \caption{Proposed key technical requirements for the cryogenic LNAs.
    }
    \begin{tabular}{lcc}
        \hline\noalign{\smallskip}
        Specification & Value & Unit \\\noalign{\smallskip}
        \hline\noalign{\smallskip}
        Bandwidth  & 67-116 & GHz \\
        Noise temperature (67-90~GHz) & $<23$ & K\\
        Noise temperature (90-116~GHz) & $<30$ & K\\
        Input reflection  & $<-6$  & dB\\
        Output reflection & $<-10$ & dB\\
        Gain & $>40$ & dB\\
        1~dB output compression point & $> -10$ & dBm\\
        DC Power Dissipation & $<35$ & mW\\
        Nominal Operating Temperature & 15 & K\\
        \noalign{\smallskip}
        \hline
    \end{tabular}
    \label{tab:lna_specs}
\end{table}

The key performance drivers for the cryogenic LNAs are summarised in Table \ref{tab:lna_specs}. We note that these are the requirements envisaged for the final implementation. In the current development phase, the main focus of optimisation is to meet the most challenging requirements, which are the wide RF bandwidth and the low noise temperature. 

\subsubsection{University of Manchester LNA design and results}

\begin{figure}
\centering
\includegraphics[height=30mm]{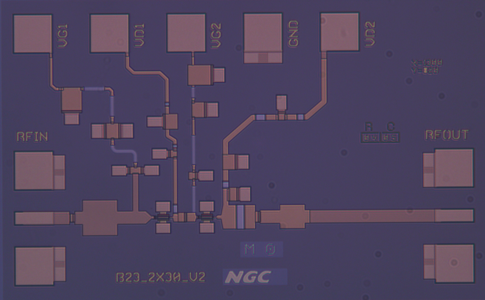}
\includegraphics[height=30mm]{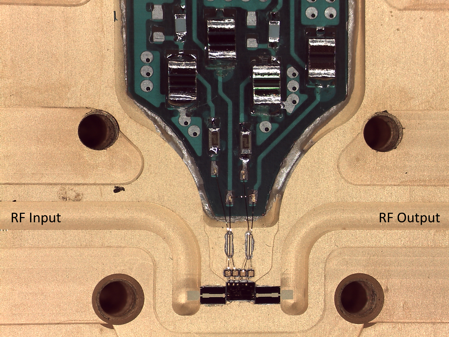}
\caption{{\it Left:} Microscope photograph of one of the fabricated UoM MMICs.  The die size of the UoM MMICs is 1.3~mm $\times$ 0.9~mm.
{\it Right:} Interior of the packaging showing the MMIC, the waveguide to microstrip transitions, WR10 waveguide channels (with the RF input on the left and output on the right), and the off-chip bias protection circuit.  The MMIC die can be seen in the lower middle portion of the photo, below the bias protection circuit.  This figure is based on figures presented in \cite{2017ITMTT..65.1589C}.
  }\label{fig:uom-lna}
\end{figure}

For its MMIC LNA designs covering the full 67-116 GHz bandwidth, the UoM team uses the 35~nm gate length Indium Phosphide (InP) semiconductor process made available by Northrop Grumman Corporation \citep[NGC; ][]{4419013}.  
The MMIC design process was performed with individual simulations of the different matching networks using the EM simulator Momentum, a tool included with the Keysight Advanced Design System (ADS) package. 
The MMIC design consists of two two-finger transistor stages in common source configuration and features the possibility of independently biasing the gate and drain of each transistor stage \citep{2017ITMTT..65.1589C}. 
The size of the fabricated chip is $1300~\mu$m $\times$ 900~$\mu$m and it is shown in Fig.~\ref{fig:uom-lna}.

In order to select and package only the chips with the best noise performance, they were first tested in the cryogenic probe station at the Cahill Radio Astronomy Laboratory  \citep[CRAL;][]{2012RScI...83d4703R}, which is located at the California Institute of Technology (Caltech).  
The MMICs with the best noise performance were selected from those measured on the cryogenic probe station and packaged into highly integrated blocks, as shown in Fig.~\ref{fig:uom-lna}.  
As well as the MMICs, these blocks contain WR10 or WR9.921  waveguide interfaces for the RF signals, gold-on-quartz waveguide-to-microstrip transitions designed specifically to interface to the MMICs and a bias protection circuit consisting of a custom made PCB and off-chip capacitors situated close the MMIC.

The S-parameters and noise performance of the packaged LNAs were first fully characterised at room temperature.  The noise performance was then tested at a cryogenic ambient temperature of 20~K.
The results of the on-wafer testing of the MMICs and the assembled LNAs, along with further information about the LNAs, the MMICs and the other components used in the LNAs, can be found in \cite{2017ITMTT..65.1589C}.
Then the LNAs were installed and tested at a common test facility at Yebes observatory (see Section~\ref{sec:lna-summary}) and, finally, the best LNAs were selected for the integration in the prototype Band 2 receiver.

Efforts to improve the performance of these LNAs to better fit the 67–116~GHz frequency band are underway.  First, we have modified the design of the waveguide elements of the blocks and the waveguide-to-microstrip transitions. A second wafer run with NGC has taken place and updated versions of the MMIC design have been produced.  
These efforts and their results will be presented in detail in a future publication.

\subsubsection{Low Noise Factory LNA design and results}

\begin{figure}
\centering
\includegraphics[width=\columnwidth]{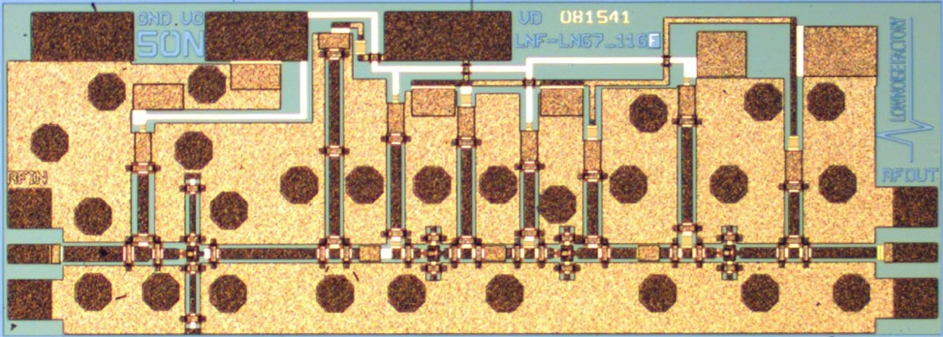}
\caption{Photograph of the W-band MMIC LNA from LNF. The die measures 2.0~mm $\times$ 0.75~mm.}\label{fig:lnf_lna_photo}
\end{figure}

Here we give a short overview of the Low Noise Factory W-band LNA design and performance; more details can be found in \cite{Tang2017}.
The gate length of the InP HEMT process is 0.1 $\mu$m. The epitaxy structure grown on InP wafer, from bottom to top, consists of an In0.52Al0.48As buffer, an In0.65Ga0.35As channel, an In0.52Al0.48As spacer, a Si delta doping, an In0.52Al0.48As Schottky barrier, and a Si-doped In0.53Ga0.47As cap layer. The device selected for this design is 2-finger with total periphery of 40 $\mu$m. 
Biased with a drain voltage $V_d=1~V$ and a drain current of $I_d=10$~mA, device S-parameters are measured at room temperature on a probe station from 0.1--67~GHz.
The small-signal model extracted at room temperature \citep{Dambrine1988} shows that this 2-finger 40 $\mu$m HEMT features an expected transistor cutoff frequency $f_t=220$~GHz, when the short circuit current gain becomes unity, and peak transconductance $g_m = 1400$~mS~mm$^{-1}$ at a drain-source voltage $V_{ds}=1$~V. 
The noise modelling is based on the \citealt{Pospieszalski1989} model, which associates all resistance temperatures to ambient except the channel drain-source resistance $R_{ds}$ that is given a higher noise temperature to model the noise generated by hot electrons inside the channel. In the modelling used for this design, the noise temperature for $R_{ds}$ (operating at room temperature) is assumed to be 3500~K.  This value  is obtained by fitting a noise model to the measured noise factor at room temperature. When the device is cooled to an operating temperature of $\lesssim 16$~K, the noise temperature of $R_{ds}$ is adjusted down to 800~K or lower. The InP substrate is thinned to 75 $\mu$m. 
MIM capacitors, with capacitance densities of 390~pF~mm$^{-2}$, and 50~$\Omega~\square^{-1}$ NiCr thin-film resistors are provided in this process. 

\begin{figure}
\centering
\includegraphics[clip=true, trim=20mm 36mm 35mm 21mm, width=\columnwidth]{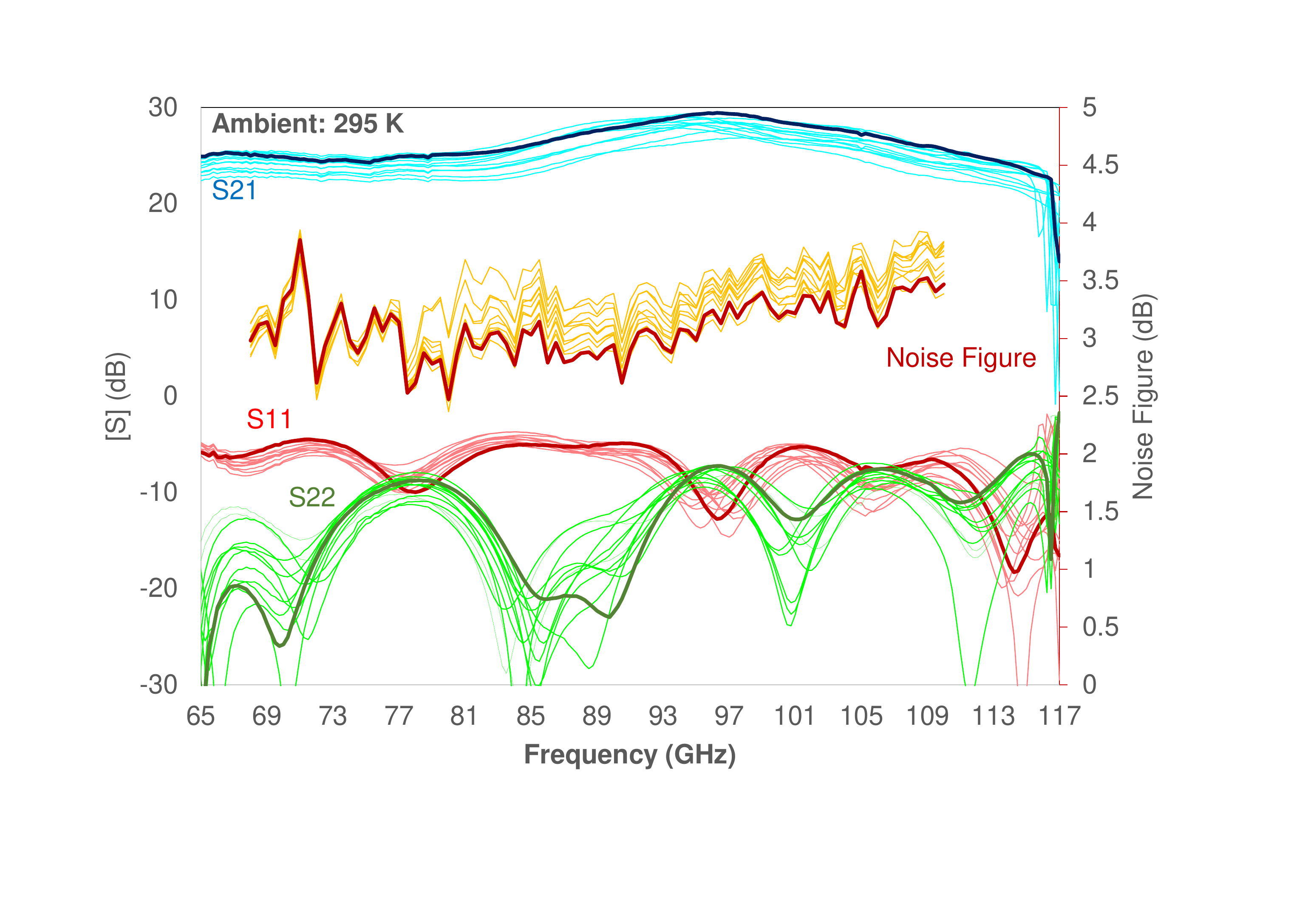}
\includegraphics[clip=true, trim=20mm 20mm 20mm 20mm, width=\columnwidth]{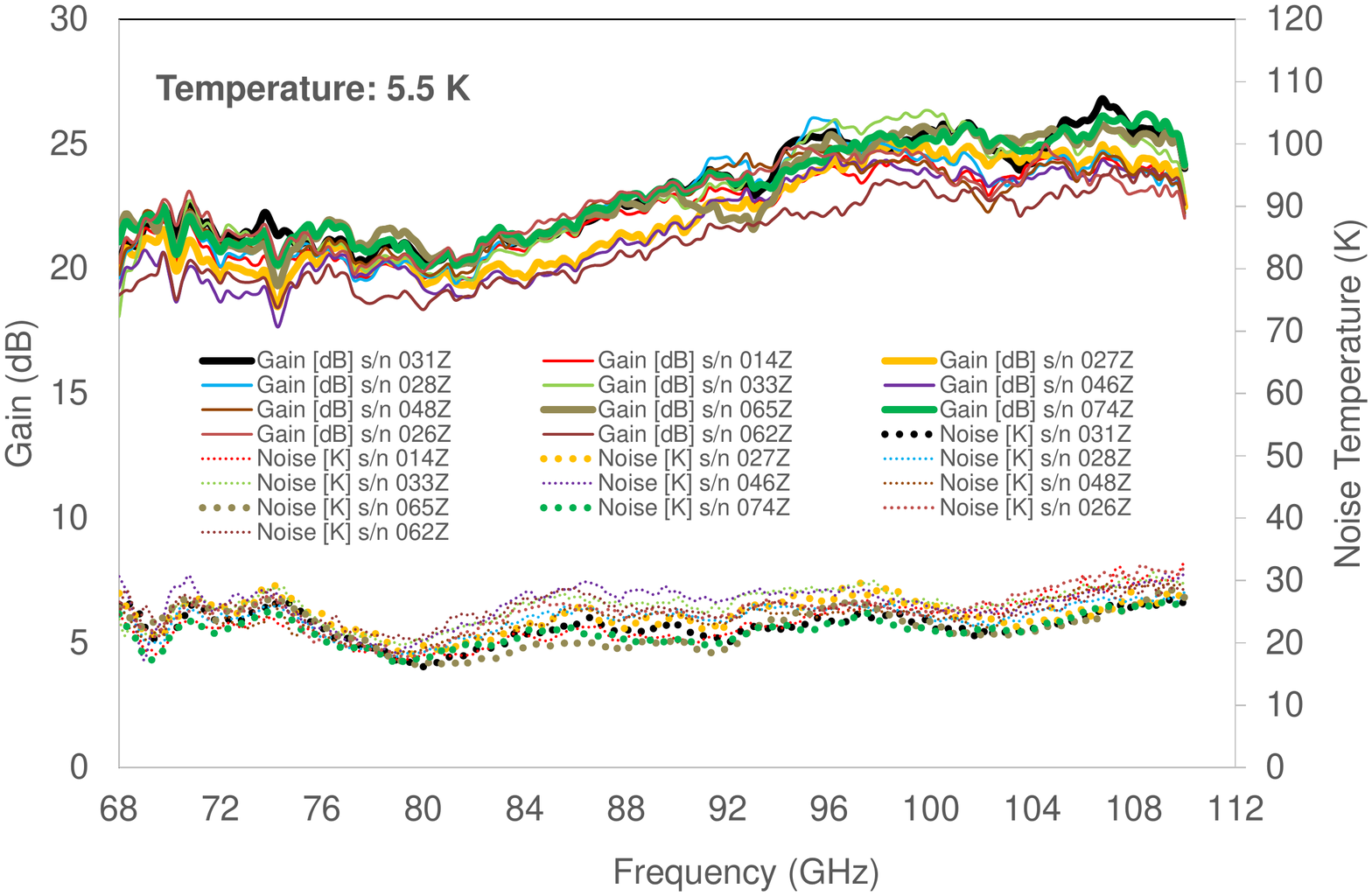}
\caption{{\it Upper:} Measured S-parameters for a batch of thirteen LNA modules from LNF, as well as noise figures for a batch eleven of these modules. 
{\it Lower:} Measured noise temperatures for a batch of eleven LNA modules from LNF.
}\label{fig:lnf_lna_results}
\end{figure}

Fig.~\ref{fig:lnf_lna_photo} shows the die for the LNF LNA MMIC design, which cascades four stages. The 2-finger 40 $\mu$m HEMT is used for each stage but connects to different lengths of shorted stubs at the source terminal. The source inductive stubs function as negative feedback elements that stabilise the device and produce optimal impedances, minimising noise and maximising the available power gain; such inductive feedback decreases the available power gain as well.

The room temperature S-parameters of the LNA were measured at a drain bias of $V_d = 1.5$~V and $I_d = 35$~mA, resulting in 52.5~mW DC power consumption. A batch of thirteen modules was built and characterised. The upper panel of Fig.~\ref{fig:lnf_lna_results} displays the measured S-parameter data for all the modules.
On average, the amplifier gain is greater than 21~dB across the entire band, from 65~GHz to 116~GHz, with an average of 25~dB and $\pm 3$~dB flatness. Input and output return losses are lower (better) than -5~dB and -10~dB, respectively. Excellent reproducibility can be obtained.

The noise temperature is characterised by the $Y$-factor method at both room temperature and cryogenic temperature. Eleven of the modules were tested at a cryogenic operating temperature of 5.5~K. As shown in the lower panel of Fig.~\ref{fig:lnf_lna_results}, repeatable results for the noise temperatures and gains can be observed with the standard deviation less than 3~K across the entire band. All the cryogenic data were taken at 1.0~V drain voltage and 7~mA total LNA current (i.e.\ summed over all stages).

\subsection{Low noise amplifier comparison, summary, and outlook}
\label{sec:lna-summary}

\begin{figure}
\centering
\includegraphics[width=\columnwidth]{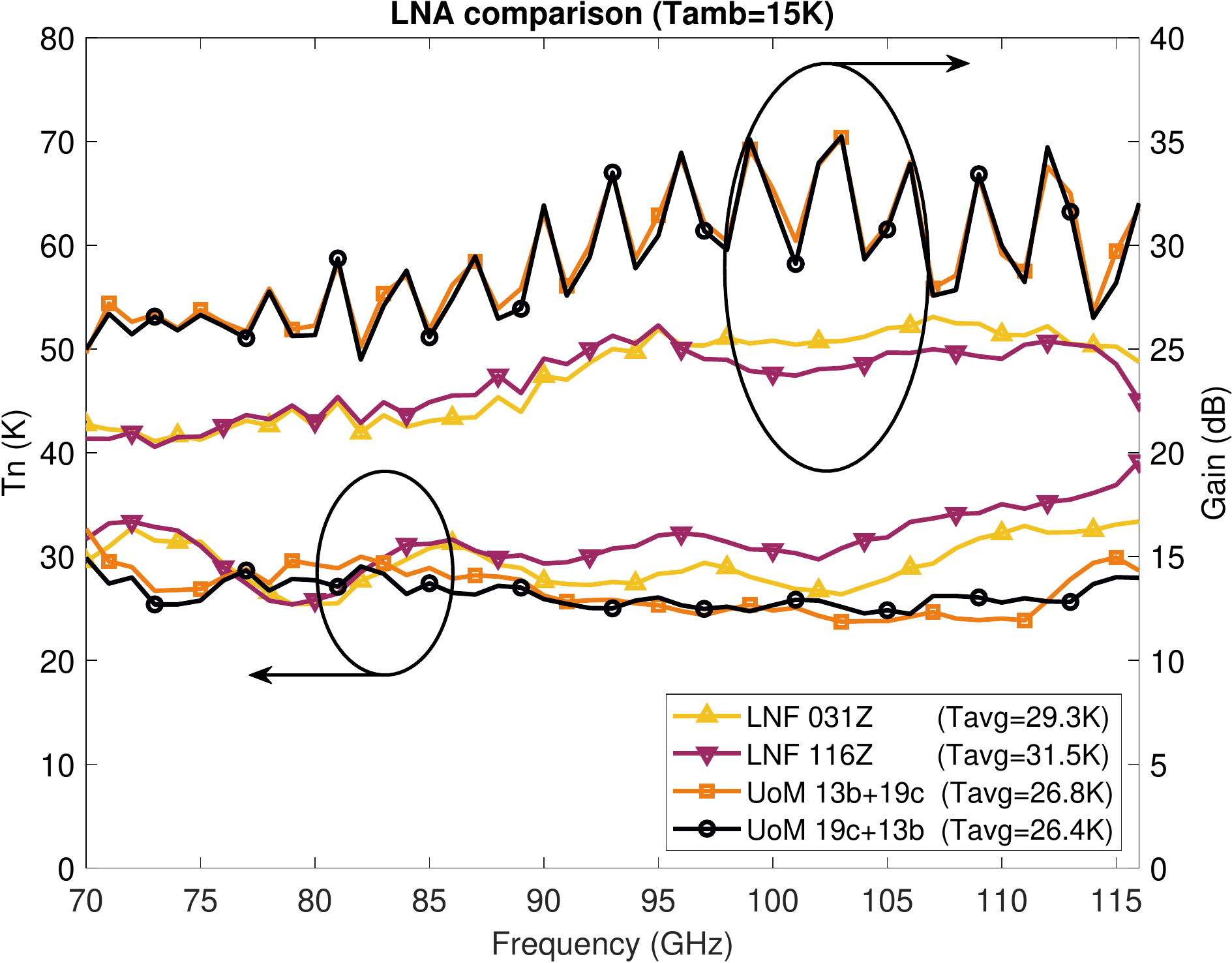}
\caption{Results of the comparison of noise temperature $T_n$ (left y-axis) and gain (right y-axis) of the UoM and LNF amplifiers measured in the Yebes Observatory cryostat with the system described in the text. The gain of the UoM LNA exhibits a significant gain ripple caused by the reflections between the two cascaded units used. This was corrected in the cartridge by inserting a cryogenic isolator between the two units.}
\label{fig:noise_meas}
\end{figure}

\begin{figure*}
\centering
\includegraphics[height=13cm]{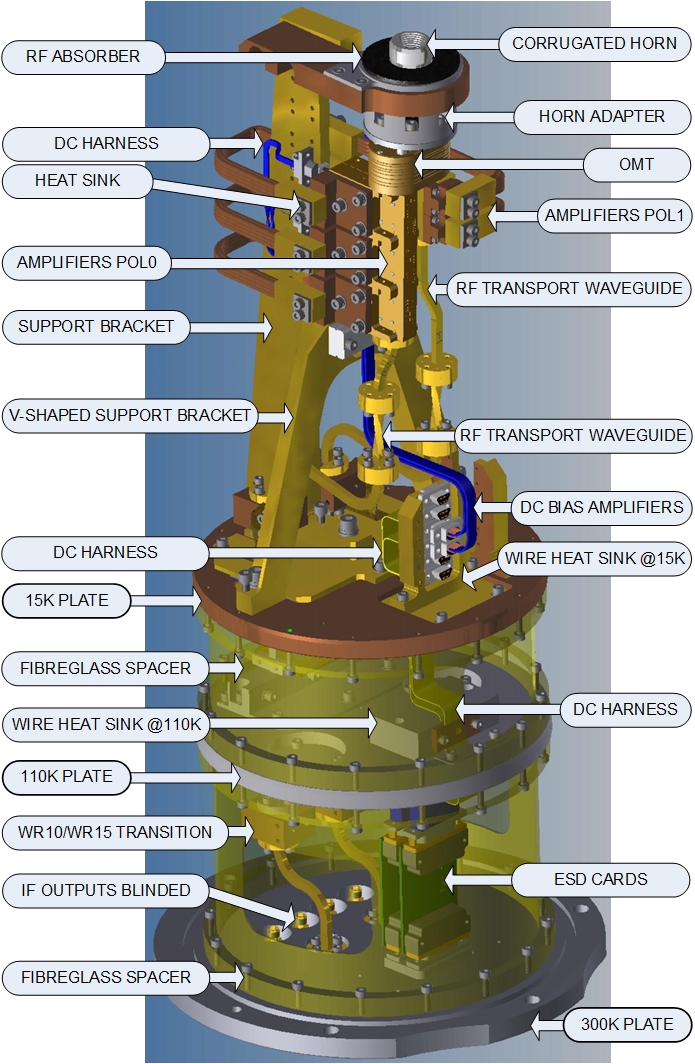} 
\includegraphics[height=13cm]{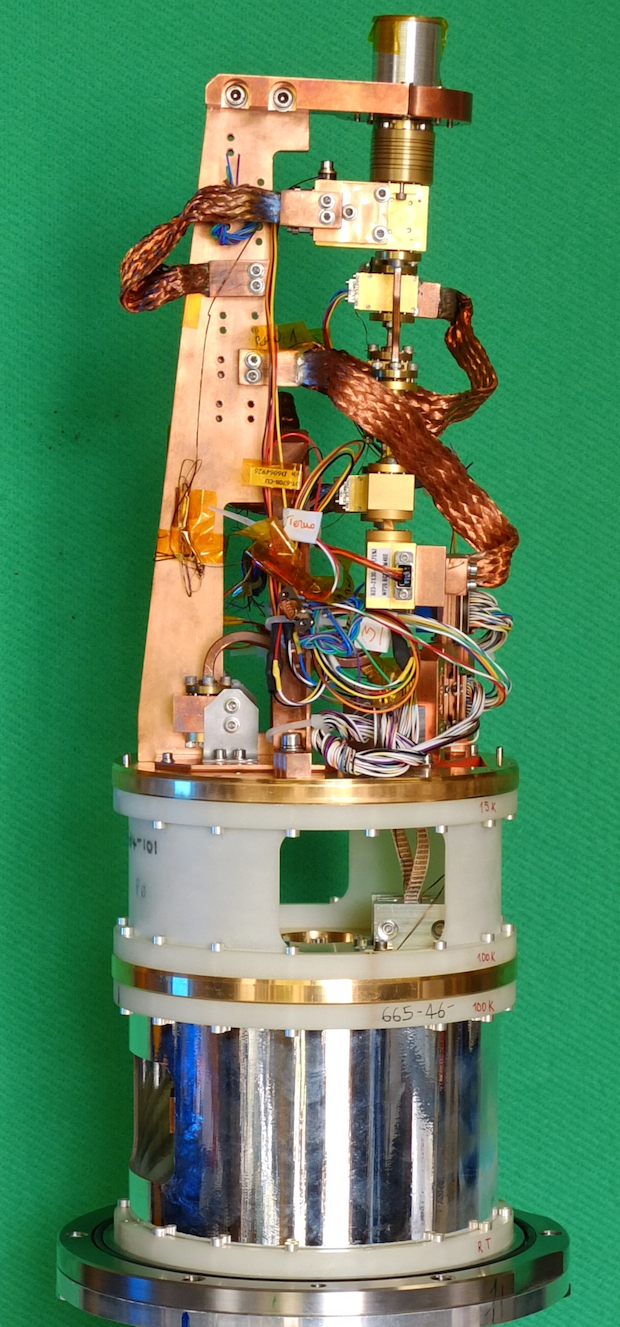}
\caption{{\it Left:} Annotated CAD model for the Band 2 CCA \citep[first presented in][]{Yagoubov2018} is shown here.
{\it Right:} Photograph of the fully-assembled prototype receiver cartridge, integrated at INAF-OAS in Bologna, is shown here.
}\label{fig:b2cartridge}
\end{figure*}

In order to minimise uncertainties due to the calibration of the measurement systems available in different labs and facilities, which can easily lead to erroneous conclusions, we undertook a campaign to compare the performance of the cryogenic LNAs in the same laboratory. The measurement system is described in Appendix \ref{sec:lna-test-setup}.

Fig.~\ref{fig:noise_meas} shows the results obtained in the measurements. The average values obtained in the 70-116 GHz band are shown in the legend of the figure. 
The data for the LNF LNAs correspond to single package LNAs, whereas for the University of Manchester, the data correspond to two units connected in cascade. The order of the units was changed for determining the best option for the first stage, although the results obtained were very similar. The high ripple observed in the gain measurement of these units was caused by the multiple reflections between the two units cascaded. This was solved in the prototype CCA by connecting a cryogenic isolator between the two amplifiers.
We also measured how the LNAs noise temperatures depend on their physical temperature, and discuss the results in Section \ref{sec:cca}.

\section{Prototype Receiver Assembly and Performance}\label{sec:receiver}

\subsection{Cold cartridge assembly}\label{sec:cca}

The Band 2 CCA has been designed to be compatible with the ALMA FE cryogenic system.
The geometry, dimensions, and cryogenic layout of the CCA is largely defined by the electrical, cryogenic, and vaccuum interfaces to the ALMA FE. Cooled by a closed-cycle cooler, each ALMA cryostat contains four temperature levels (nominally at ambient temperature, 110~K, 15~K, and 4~K) to provide the operational temperatures needed for the cold cartridges. The standard ALMA cartridge structure comprises of a stack of four metal plates (corresponding to the cryostat’s four temperature levels), separated by fibreglass spacers that provide mechanical support with a minimum of thermal conduction between the different temperature stages. 
In the left panel of Fig.~\ref{fig:b2cartridge}, we show the annotated 3D computer aided design (CAD) drawing displaying the details of the ALMA Band 2 CCA design; in the right panel of Fig.~\ref{fig:b2cartridge}, we show a photograph of the prototype CCA during the process of assembly.

For the Band 2 CCA position in the ALMA cryostat, only two cryogenic temperature levels are available and they are at 15~K and 110~K, respectively.\footnote{See \url{https://www.cv.nrao.edu/~demerson/almapbk/construc/chap6/chap6.pdf}.} Thus, the cartridge body of Band 2 CCA has three stages, including the ambient stage. An ambient temperature vacuum flange seals the cryostat and is also used to attach the WCAs. The other two levels are used to support the cartridge’s cryogenic opto-mechanical structure and RF components. 

\begin{figure}
\centering
\includegraphics[width=0.95\columnwidth]{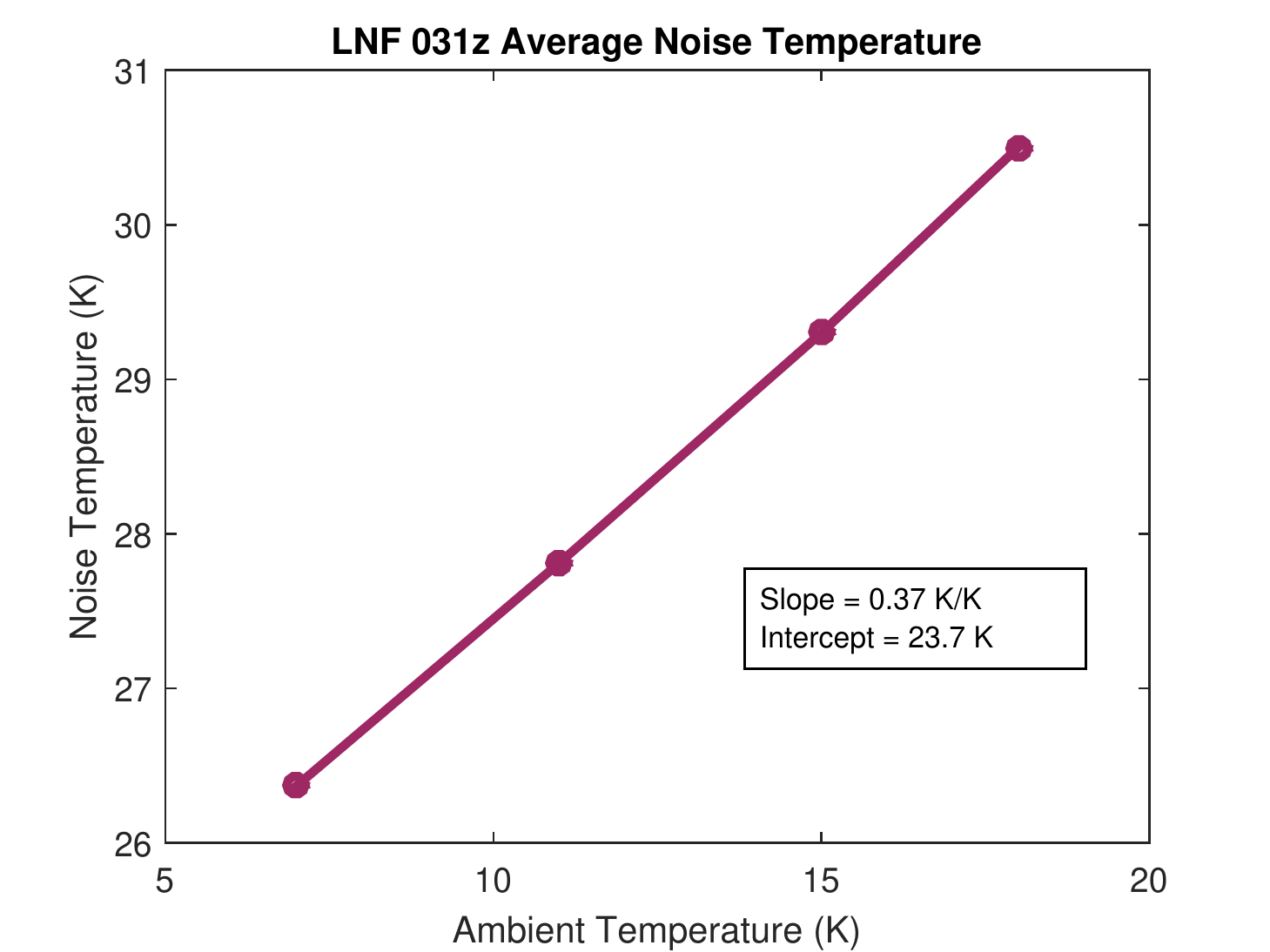}
\caption{Plot shows measurements of the noise temperature of LNF LNA 031z vs.\ ambient temperature in the 7-18~K range. The slope is 0.37~K/K.
}
\label{fig:lna_temp_depend}
\end{figure}

The design of the prototype Band 2 CCA is driven by optimisation for the best performance and versatility to allow testing and integration of different optical solutions and cryogenic low-noise amplifiers (see Sections \ref{sec:optics} \& \ref{sec:lna}).
The horn, OMT, and LNAs are integrated into one unit and should be operated at the lowest achievable temperature. 
This choice is driven by the fact that cryogenic operation of our InP HEMT amplifiers yields the advantage of achieving the low noise temperatures \citep{Bryerton2013b,Pospieszalski2018}.
The standard expectation for Band 2 within the ALMA collaboration has been that a cryogenic stage at 15~K would be suitable since the LNA noise temperature has typically been expected to be a modest function operating temperature for temperatures less than 20~K.  
However, Fig.~\ref{fig:lna_temp_depend} shows measurements of the noise temperature of a typical LNA (in this case, LNF LNA 031z) versus ambient temperature over the 7-18~K range.  This exhibits a slope of 0.37~K per degree change in operating temperature, indicating the dependence is not negligible.
Furthermore, for passive receiver components (e.g. the corrugated feedhorn and OMT) the conduction losses fall approximately linearly with temperature, and their respective RF loss contributions to the receiver noise also drop with physical operating temperature. 
In order to ensure these components operate at the lowest achievable temperature available to the cartridge (15~K), we manufactured the mechanical support for the horn, OMT, and LNAs using oxygen-free copper, which has high thermal conductivity.
The cross-section for the mechanical bracket was calculated to be appropriate for removing the heat dissipated by the LNAs, which is specified to be $< 70$~mW in total for the two polarisation channels (Table~\ref{tab:lna_specs}). All parts were precisely machined to deliver accurate positioning of the optical components with respect to the lens, which serves as a vacuum window for the ALMA FE and is thus fixed to the FE top cover. The entire assembly optics relies on accurate mechanical fabrication and alignment using the dowel pins. The main support arm is used as the primary backbone and heat sink for the receiver active components, as all RF amplifier heat sinks are directly connected to it. The V-shaped support bracket adds stability of the receiver support in the orthogonal direction. The receiver support bracket fixes (with help of adaptor parts) the assembly of the horn and OMT. The mechanical design was modelled using FE simulation software to verify mechanical stress, thermal contraction, and any vibrational eigenfrequencies (the lowest was found to be 79.33~Hz). The total change in length due to contraction as measured from the corrugated horn mouth to the inner plane of the cartridge 300~K plate is estimated to be 1.4~mm; this is accounted for in the receiver's optical design.

In contrast to the higher frequency ALMA CCAs \citep[e.g.\ ALMA Band 5, reported in][]{Belitsky2018}, which contain mixers, the Band 2 CCA only has amplifiers.  This simplifies the CCA, leaving only RF signal transport via waveguides and a DC bias harness. In order to deal with the stress caused by significant contraction of the fibreglass spacers of the cartridge body, we constructed the waveguides transporting the RF signal with substantial flexibility and such that the maximum mechanical stress occurs at room temperature and is reduced to be around zero during cool down. 
At cryogenic temperatures, all materials become much stiffer and more fragile; and by this approach we ensure the integrity of the mechanical parts is not jeopardised by significant mechanical stress under cryogenic temperatures. The WR10 copper waveguides are S-shaped at the interfaces towards the RF amplifiers and the waveguide vacuum feed-through at the cartridge 300~K plate. This provides a desirable level of mechanical flexibility to cope with the stress at 300~K, precluding the development of mechanical stress at both room and cryogenic temperatures.

The RF losses between the cold LNAs and the 300~K RF outputs need to be minimised, with a goal to be below 5 dB. In the design of the RF signal transport waveguide, we have to account for two competing requirements: the RF waveguide should be as short as possible and have high electrical conductivity to minimise RF loss, but at the same time, it should minimise the parasitic heat load. For the latter, one would like to use longer stainless steel waveguide segments, with low thermal conductivity, to limit the thermal flux from 300 K to 110 K and from 110~K to 15~K, keeping the thermal flux below 450~mW and 90~mW at 110 K and 15 K plate temperature busses respectively, in compliance with the ALMA requirements.
Our simulations show that employing WR10 waveguide all way between the cold (15~K) LNAs to the ambient stage (300~K) would result in a level of RF loss exceeding specifications and would exhibit a strong frequency dependence especially at the low frequency end of the RF band.  Using oversized waveguide helps to reduce the RF loss and based on our simulations, we have chosen to use a straight stainless steel WR15 waveguide (serving as thermal isolation between 300~K and 110~K, and that between 110~K and 15~K) connected via WR10-to-WR15 transitions at both ends with S-shaped WR10 gold-plated copper waveguides. 
Where needed, we used WR15-to-WR10-to-WR15 transitions combined with 90-degree waveguide turns made in the WR10 section to avoid generation of higher modes.

An internal harness connects the input DC bulkhead connector via EMI/ESD protecting cards and the terminals at 15~K plate of the cartridge. From the terminals additional cables go to RF amplifiers (DC bias) and temperature sensors. The harnesses are equipped with the thermal links at each respective temperature stage to terminate thermal flux from the higher temperature levels.

\subsection{Warm cartridge assembly}\label{sec:wca}

%-----------------------------------------------------------
\begin{figure}
\centering
\includegraphics[width=\columnwidth]{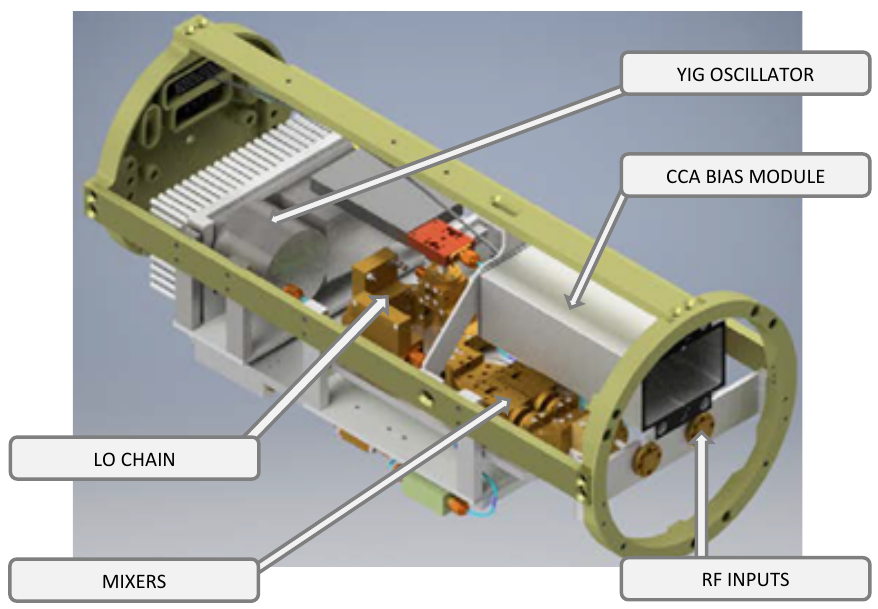}
\includegraphics[width=\columnwidth]{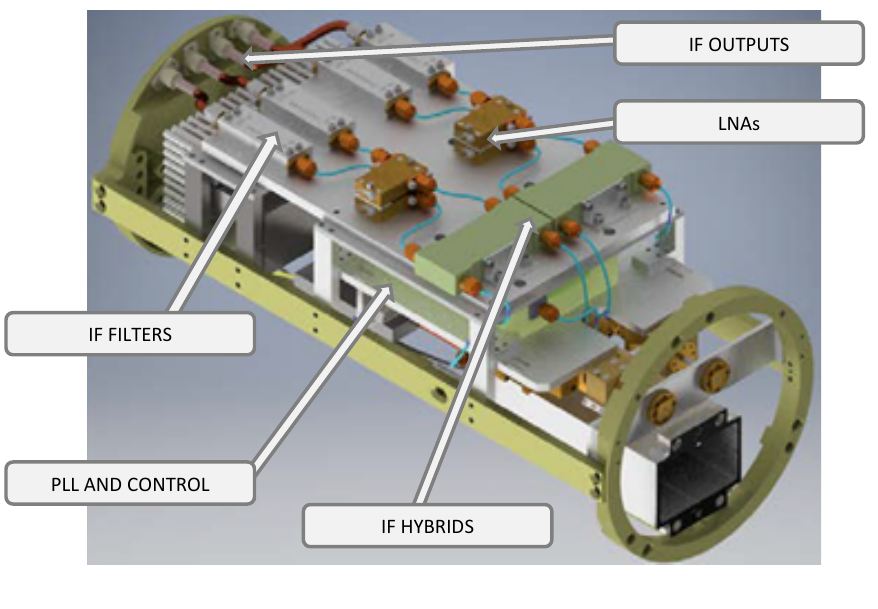}
\includegraphics[clip=true, trim=0mm 0mm 3mm 2mm, width=\columnwidth]{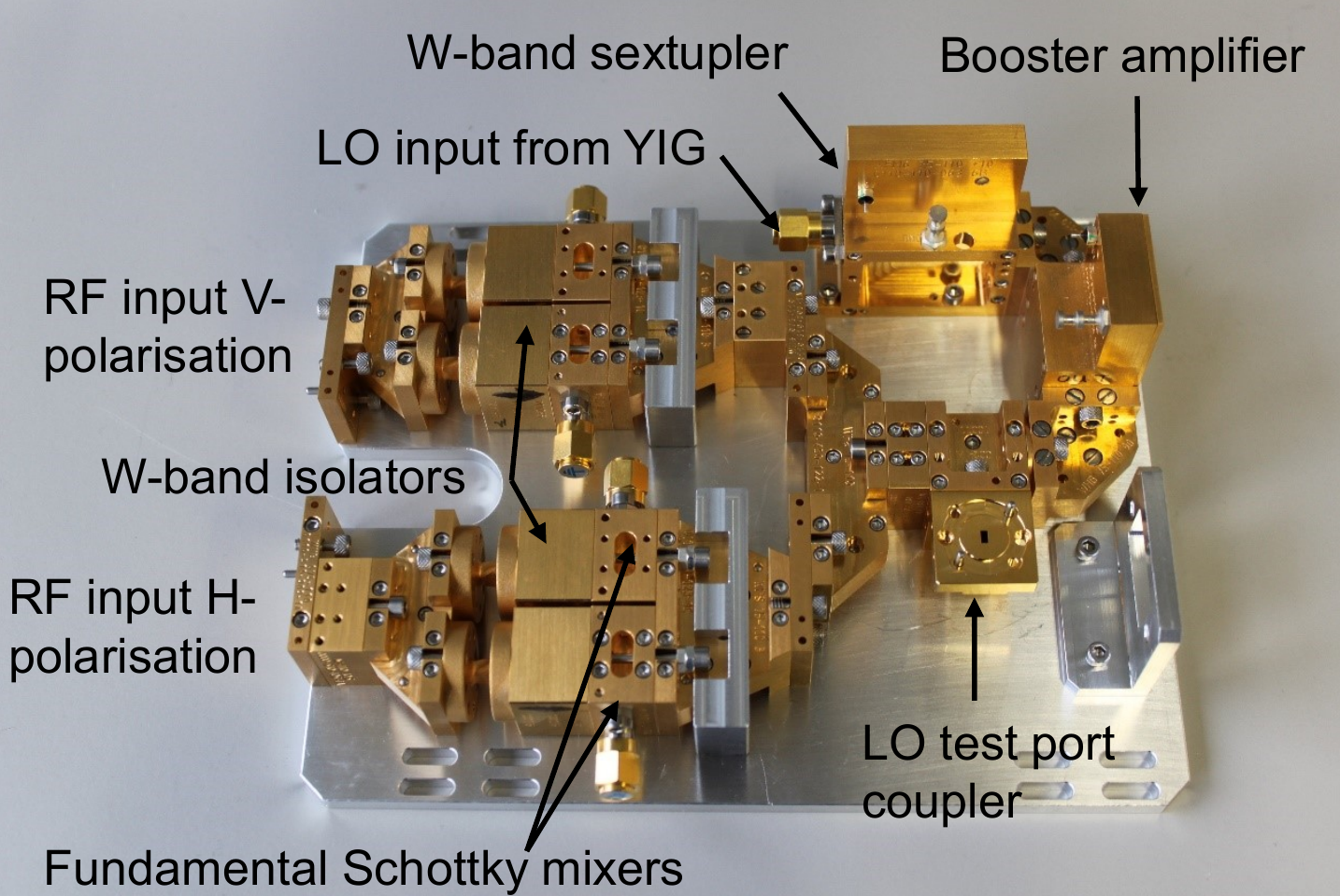}
\caption{
{\it Upper:} Image shows the 3D CAD model for the WCA configuration. The mechanical layout with the downconverter unit visible and placed on its own plate.   
{\it Middle:} Image shows the underside of the WCA, where the IF stage is located.
{\it Lower:} Figure presents a photo of the assembled down-converter unit including all RF parts (i.e. the LO multiplier, amplifier, RF mixers, and isolators), but excluding the IF parts (i.e. the hybrid coupler and IF amplifiers).
}\label{fig:wca}
\end{figure}
%-----------------------------------------------------------

A warm cartridge assembly has been specifically developed for the Band 2 receiver prototype.  The top and bottom views of a 3D CAD model image are shown in the first two panels of Fig.~\ref{fig:wca}, while a photo of the prototype assembly is shown in the lower panel.  The design is based on a dual-polarised, sideband separating receiver concept (2SB) using semiconductor Schottky and MMIC technology working at room temperature \citep{Thomas2018}.\footnote{\url{https://www.radiometer-physics.de/download/space/ISSTT_2018_ESO_ALMA_B23_SSB_Rx_poster.pdf}}

As shown in the design scheme in Fig.~\ref{fig:b2scheme}, each polarisation of the receiver features an 67-116~GHz band-pass filter, a Y-junction splitter, two extended WR10 isolators, and two 67-116 GHz fundamental balanced mixers (FBMs) based on GaAs Schottky diodes pumped by a common LO chain. The LO power required for both polarisation receivers is provided by an active W-band sextupler, followed by a booster 70-105~GHz amplifier split with two 90$^\circ$ hybrid 3~dB couplers \citep{alma_memo_343}.
The LO source is common to both polarisation 2SB channels. 
The LO input is fed by a 13.0-17.5~GHz tunable YIG oscillator with a power level of 10~dBm $\pm$3~dBm. The W-band sextupler outputs +10~dBm of power between 75-110~GHz. The booster W-band amplifier provides between +20~dBm and +22~dBm of output power between 78-104~GHz. A fraction of LO power is split at the LO coupler (`LO test port') and used to pump another mixer as part of the LO PLL system. 
A 78-104 GHz bandpass filter is used to further reject unwanted harmonics from the sextupler and purify the harmonic content of the LO signal. 
The phase shifted RF signal from the mixers is recombined in the IF via a commercially available 90$^\circ$ hybrid coupler 2-18~GHz from Krytar (model 1830). Each sideband of the IF output is then amplified with a (warm) 2-18~GHz LNA from Radiometer Physics GmbH (RPG). The YIG oscillator, PLL system, and photomixer are identical to those of other ALMA receiver bands and are described elsewhere \citep{Bryerton2013}. The unmodified PLL system developed by NRAO for other bands is used to phase lock the YIG. An input to the PLL system is provided by mixing the LO driver with photonic reference in a fundamental mixer (part of the LO driver). 

We note that the instantaneous bandwidth realised in the receiver already exceeds the current ALMA requirements as well as those anticipated for the ALMA 2030 Development Roadmap. It could be further increased by either selecting IF components with wider bandwidths, or by changing the room temperature downconverter architecture. This may not be directly relevant for ALMA, but it could be applicable elsewhere.

The complete dual-polarisation channel 67-116~GHz 2SB receiver has been designed to be located in the WCA mechanical body, following ALMA standards as closely as possible.  The main drivers for the design were to preserve the YIG assembly, bias module and PLL location and interfaces, and locating the IF components on the bottom part of the WCA, where  the IF amplifiers are normally located in other ALMA receivers. 

%--------------------------------------------------------------------
\begin{figure}
\centering
\includegraphics[clip=true, trim=18mm 0mm 18mm 0mm, width=\columnwidth]{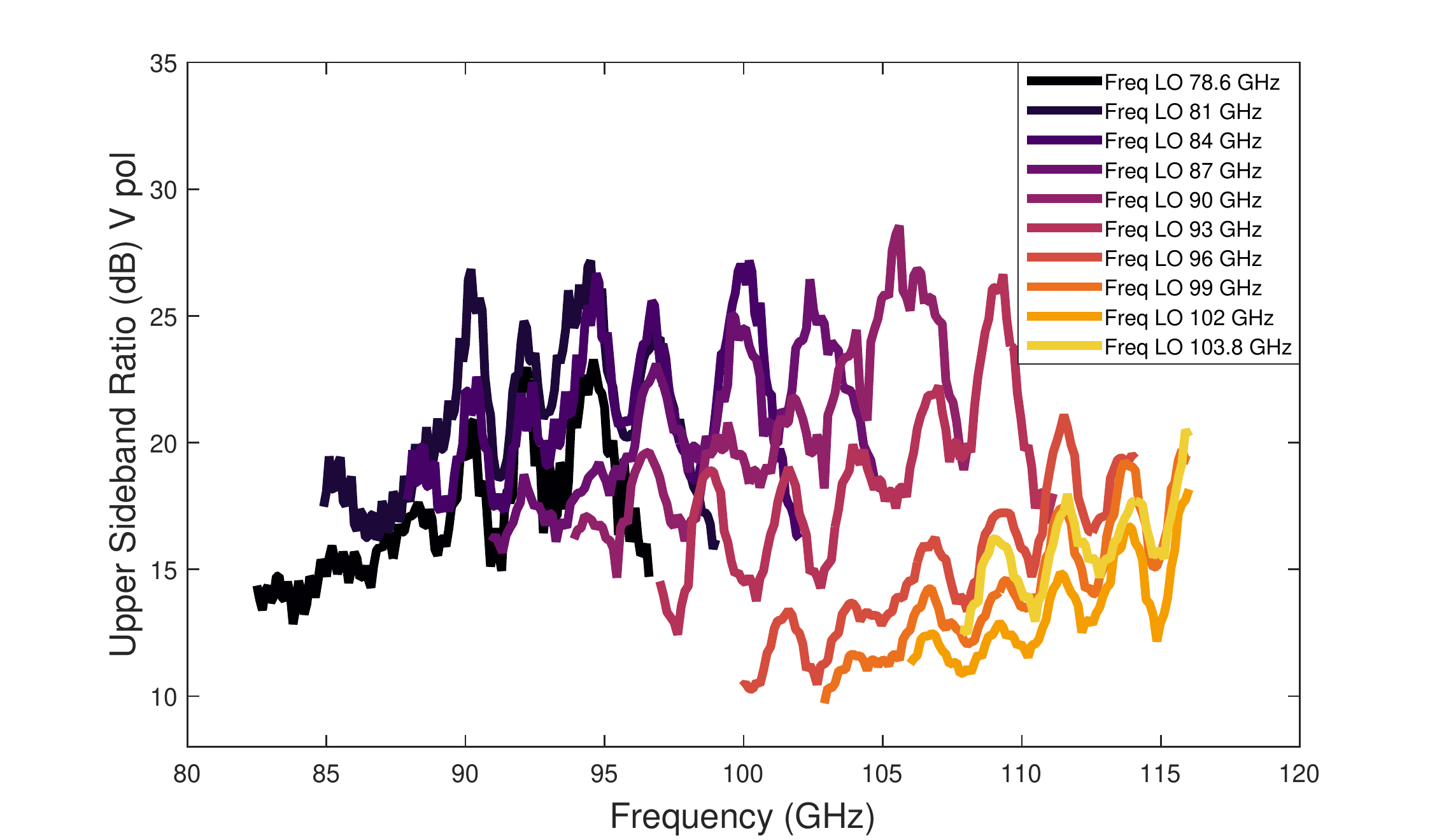}
\includegraphics[clip=true, trim=18mm 0mm 18mm 0mm, width=\columnwidth]{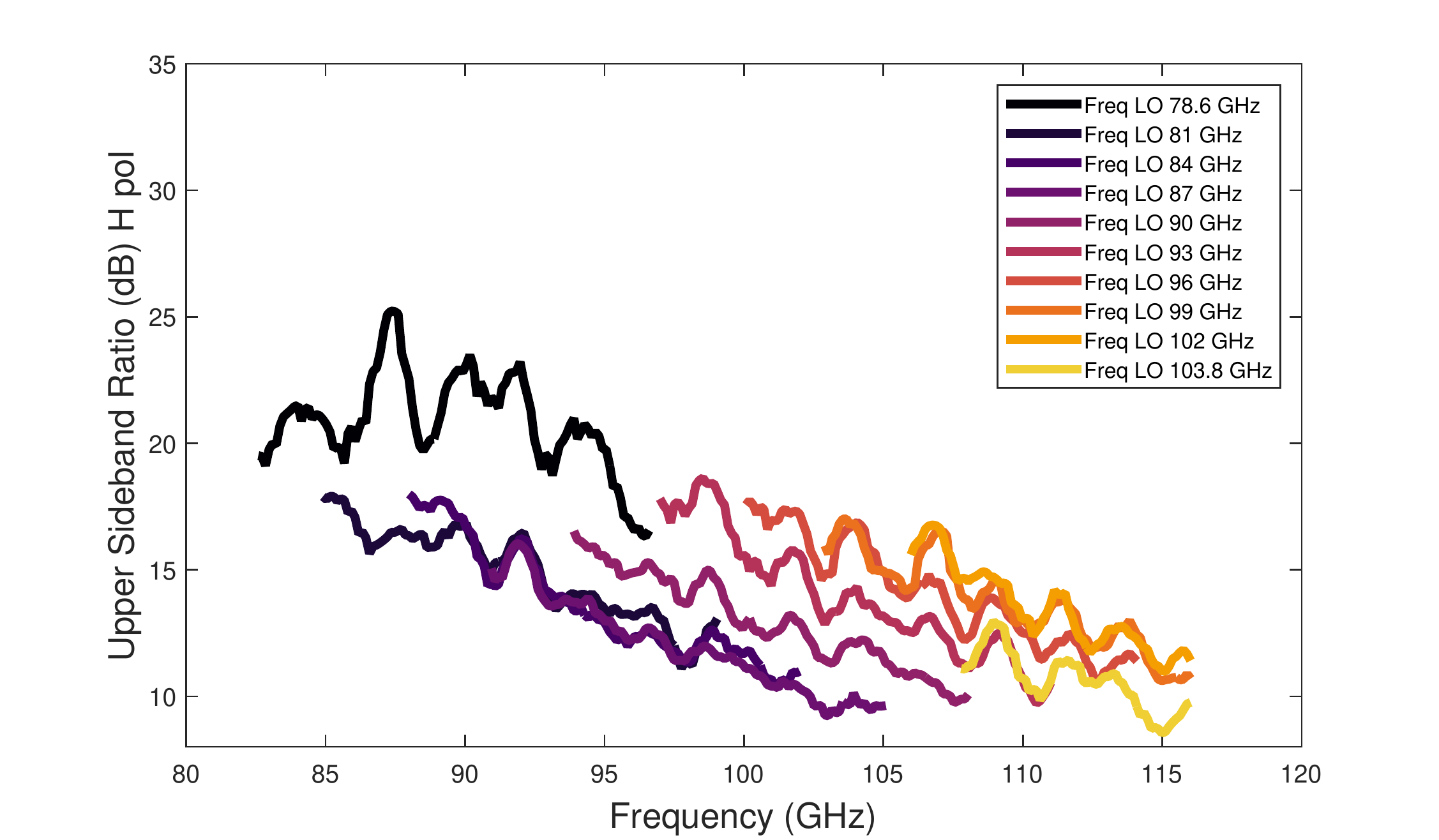}
\caption{Panels show the measured sideband rejection ratio in the upper sideband of the WCA receiver.  The upper panel shows the V-polarisations,  and the lower shows the  H-polarisations,  vs centre RF range. Each curve on the graphs represents the upper sideband ratio for a fixed LO frequency, over an IF range of 4-18~GHz. 
}\label{fig:usb_ratio}
\end{figure}
%--------------------------------------------------------------------

The performance of the WCA has been extensively measured at room temperature. 
In particular, the characterisation of the LO sextupler source together with the booster amplifier and the 79-104~GHz bandpass filter was performed showing the desired 6$^\mathrm{th}$ harmonic of the input signal and the rejection of the 4$^\mathrm{th}$ (-20~dBc), 5$^\mathrm{th}$ (-40~dBc), 7$^\mathrm{th}$ (-30~dBc) and 8$^\mathrm{th}$ (-30~dBc) harmonics, in compliance with the ALMA requirements. The gain slope and sideband ratio of the complete 2SB receiver has been characterised every 3~GHz over the entire 79-104~GHz LO range (63-120~GHz in the RF range) resulting in a gain slope of about 5~dB over 15~GHz bandwidth and a side-band ratio better than 
10~dB and 8~dB respectively for the V-polarisation, and H-polarisation. A summary of the gain and sideband rejection ratio in the upper sideband of the WCA receiver is given in the Fig.~\ref{fig:usb_ratio}. 
The measurement setup used to characterise the sideband ratio includes a swept source with -20~dBm of power approximately over the full band calibrated via a directional coupler with a R\&S power sensor (ref. NRP-Z58). The input signal of this source as well as the IF output of the 2SB receiver is controlled via a R\&S Vector Network Analyser (ref. ZVA-67).

%--------------------------------------------------------------------
\begin{figure}
\centering
\includegraphics[width=\columnwidth]{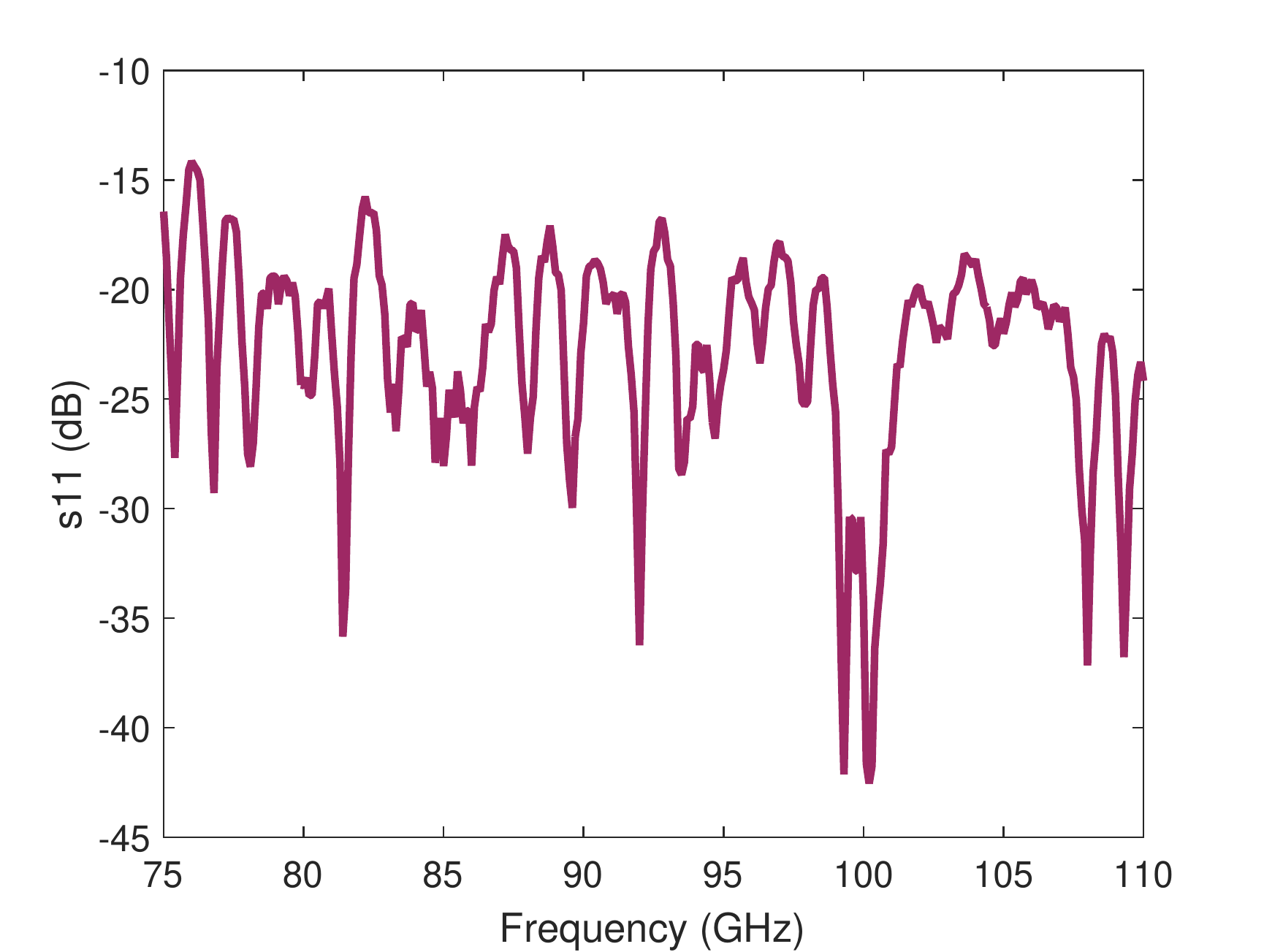}
\caption{Measured RF input return losses of the 2SB receiver, including the input RF bandpass filter in the range of 67-116 GHz. A maximum input return loss of -14~dB is measured, with average values typically better (lower) than -20~dB.
}\label{fig:retloss}
\end{figure}
%--------------------------------------------------------------------

Noise temperature measurements were extensively performed by Y-factor measurements using ambient and liquid nitrogen cooled targets, both on an early prototype single polarisation channel receiver and on the full prototype showing a typical noise temperature between 3000-10000~K across the full RF and IF frequency ranges, contributing less than 1~K to the overall receiver noise temperature, provided there is $>40$~dB total gain attained prior to the downconverter. RF input return losses are characterised between 75-110~GHz, and do not exceed -14~dB, as shown in Fig.~\ref{fig:retloss}. This limits dramatically the standing wave between the CCA and WCA, minimising the overall gain ripple.

\subsection{Receiver cartridge performance}\label{sec:testsresults}

%--------------------------------------------------------------------
\begin{figure}
\centering
\includegraphics[width=0.94\columnwidth]{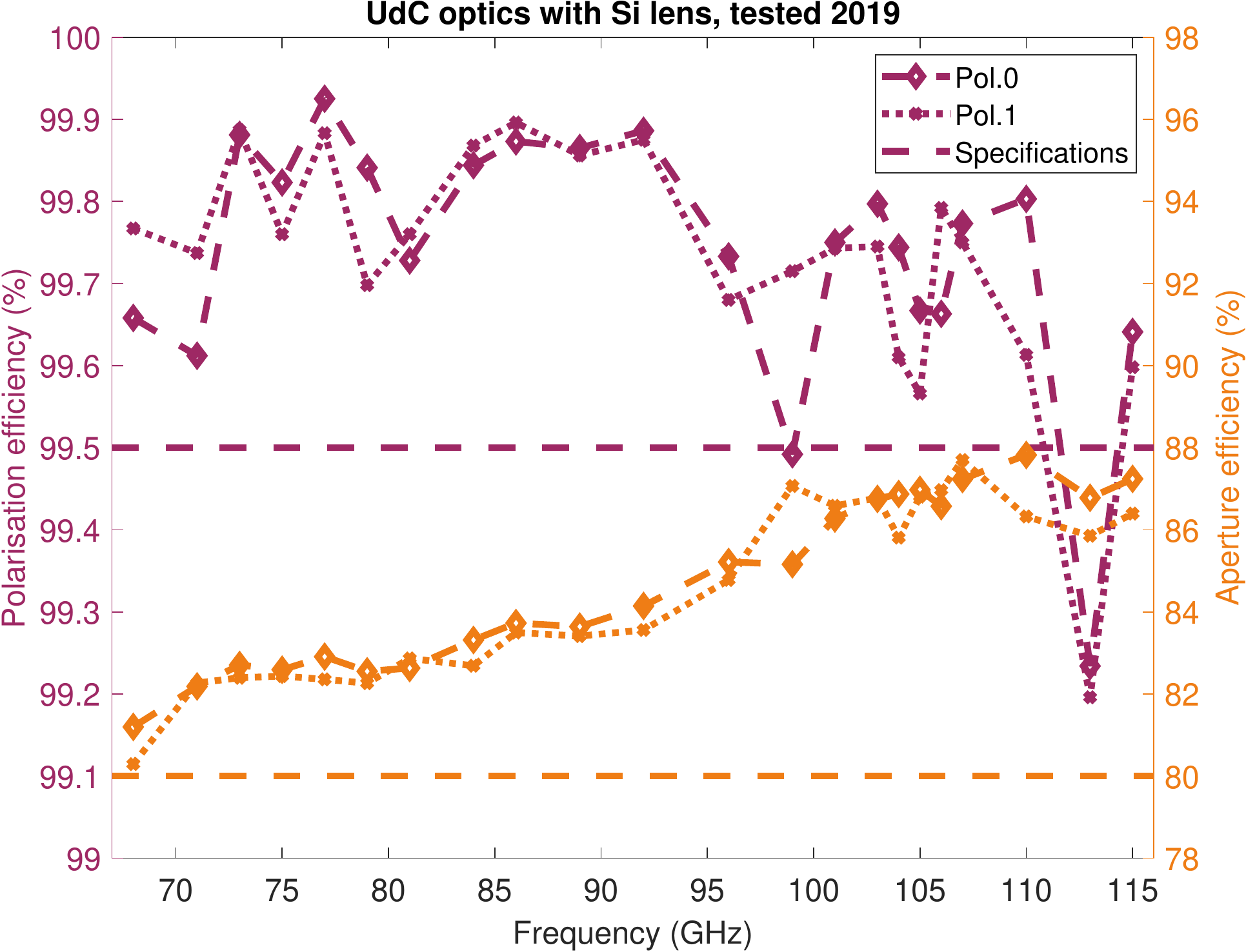}
\includegraphics[width=0.94\columnwidth]{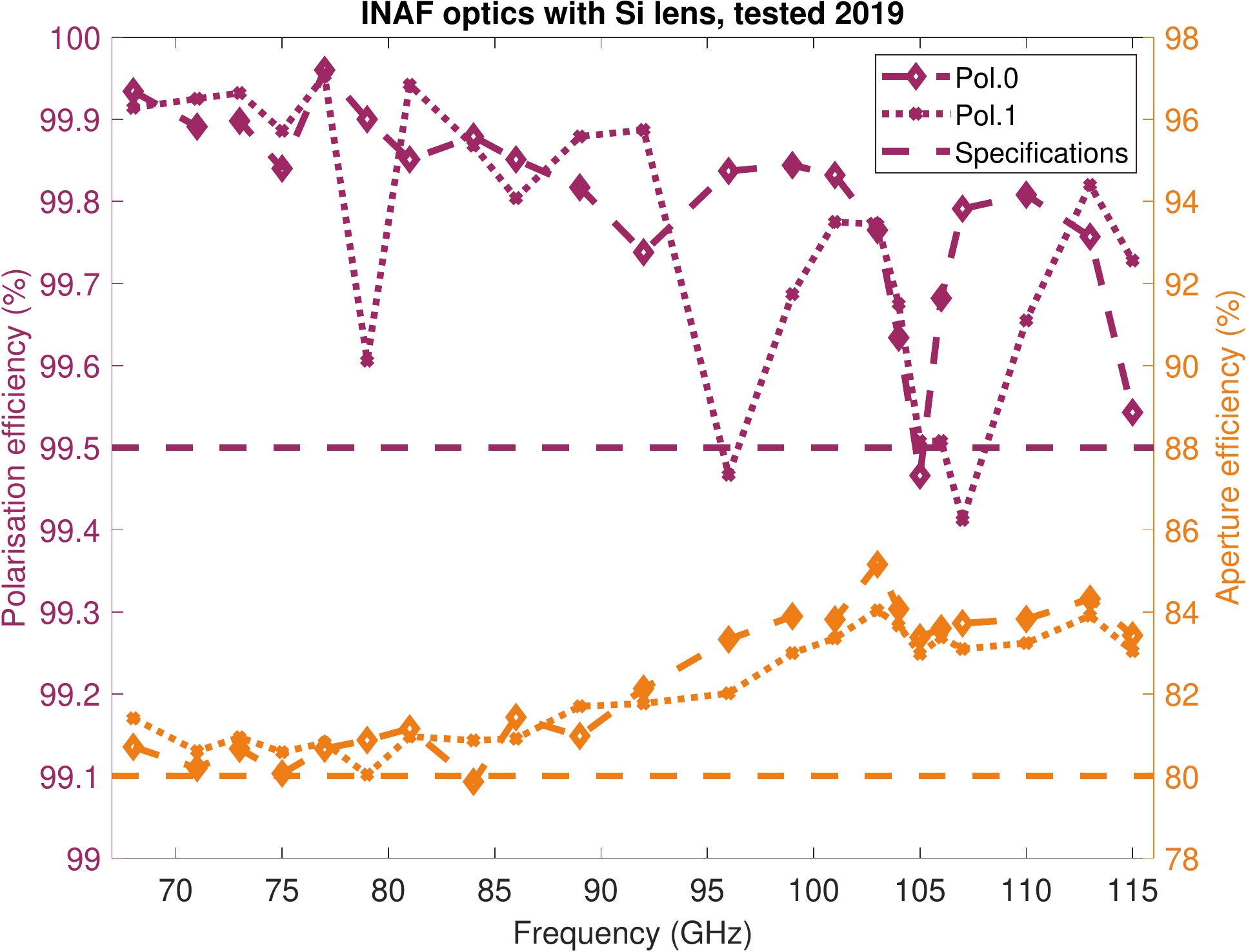}
\includegraphics[width=0.94\columnwidth]{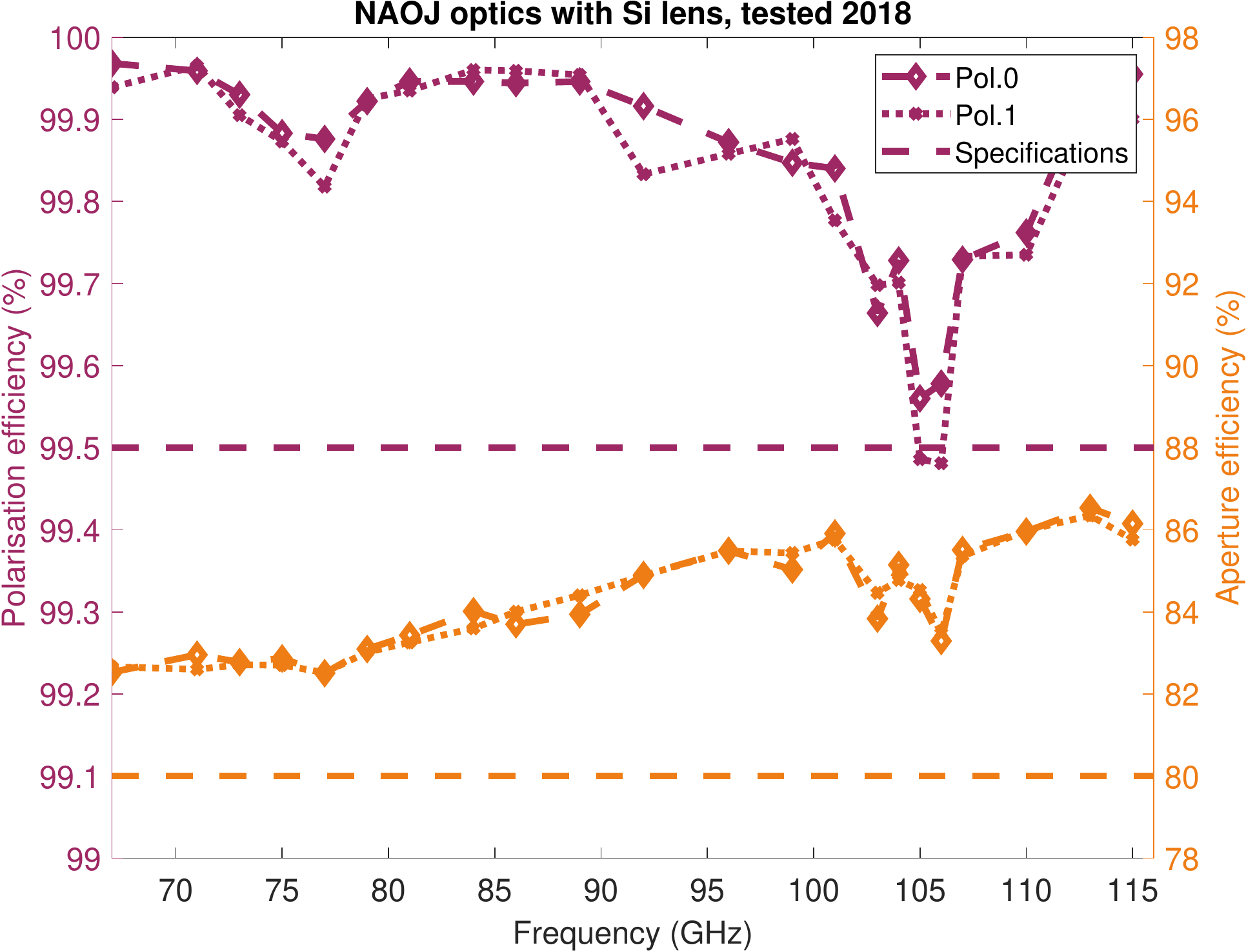}
\caption{Measured polarisation (left y-axis) and aperture (right y-axis) efficiencies of the full UdC, INAF, and NAOJ optics including the Si lens (upper, middle, and lower panels respectively). Aperture efficiencies (orange) for all designs are compliant with the ALMA specification of $>80\%$ across full frequency range. Polarisation efficiencies (purple) are mostly compliant with the ALMA specification of $>99.5\%$, with some degradation between 103-112~GHz.
}\label{fig:optical_effs}
\end{figure}
%--------------------------------------------------------------------

Figure~\ref{fig:optical_effs} reports the aperture and polarisation efficiencies, derived from the measured beam patterns, for the three different optical assemblies we designed, prototyped, and tested using the test setup described in Appendix \ref{sec:optics-test-setup}. These are the key metrics for the performance of receiver optics.
The aperture efficiency (orange curves in Fig.~\ref{fig:optical_effs}) is defined in terms of coupling to the secondary mirror of the telescope, and is a product of the spillover, amplitude, phase, focus and polarisation efficiencies for each assembly, respectively. 
We note that for all of the measurements reported here, the silicon lens prototype was used. 

Our requirement for the aperture efficiency is that it must be $>80\%$ over the full band.
Overall, the measured aperture efficiencies of all three optical designs are compliant with the ALMA specifications (denoted by dashed red lines).
Though still compliant with the requirements, the slightly reduced aperture efficiency of the INAF optics is explained by the non-optimum coupling to the silicon lens, which features a design optimised specifically for use with the UdC optics.  
In practice, the lens works almost equally well with the NAOJ optics, which have a similar beam pattern to the UdC optics, but it does not perfectly match the INAF optics, resulting in about 2\% lower measured aperture efficiency (as expected from the design differences).

The polarisation efficiencies (purple curves in Fig.~\ref{fig:optical_effs}) are also largely compliant with the $>95.5\%$ ALMA requirement, with some degradation between 103-112~GHz for all three types of optics. 
We believe this degradation is related to the spurious mode excitation at the feedhorn OMT interface due to possible misalignment. 
We conclude that the three designs demonstrate excellent and similar optical performances, largely compliant with the ALMA requirements. The marginal non-compliance with the polarisation efficiency requirement, common to all three designs, can be improved by tightening the manufacturing and alignment tolerances.
We are currently working to implement more accurate alignment of these elements, and expect improvements in the performance as a result.

%--------------------------------------------------------------------
\begin{figure}
\centering
\includegraphics[width=\columnwidth]{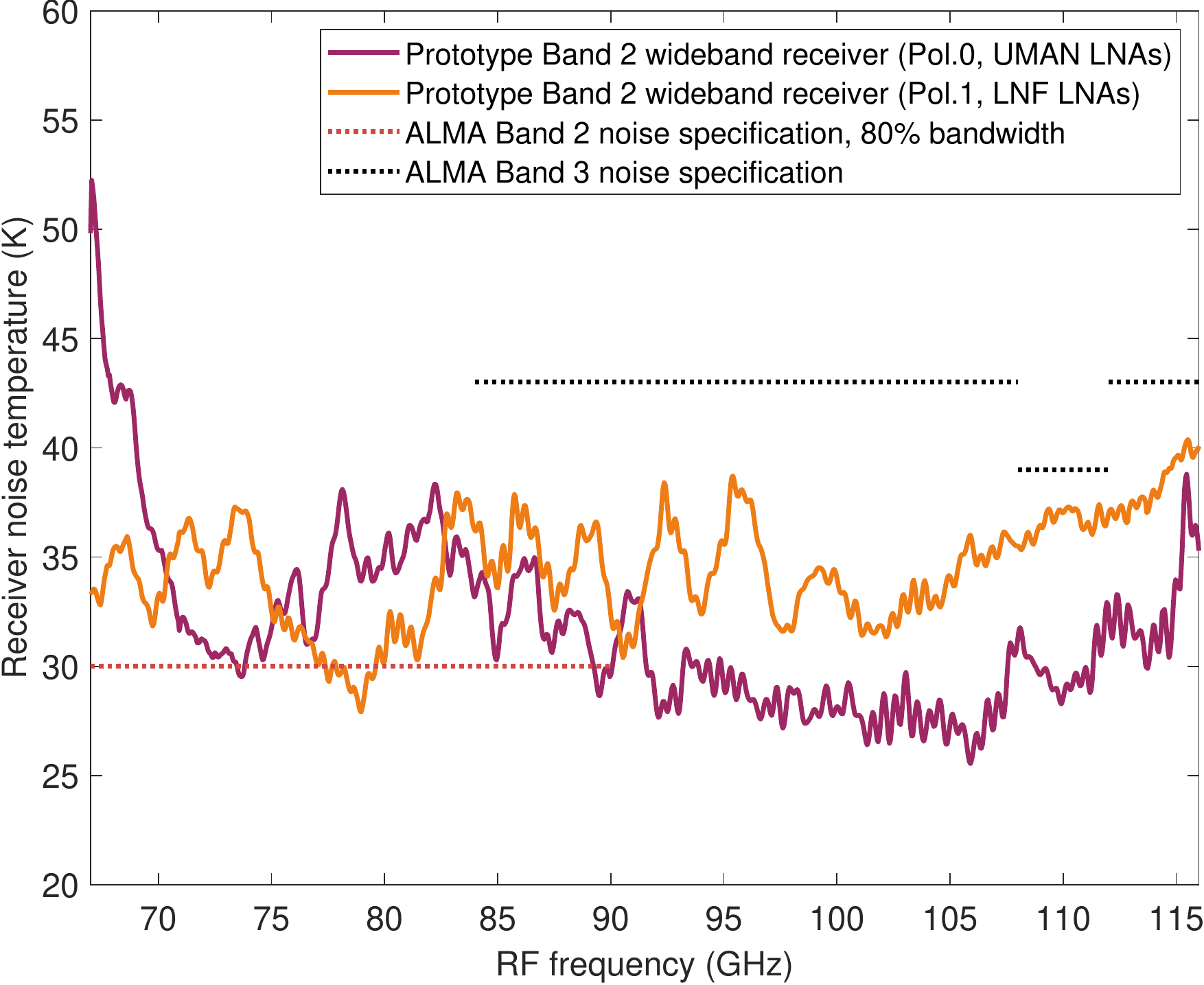}
\caption{Measured receiver noise temperature for two polarisation channels of the prototype receiver. The channel labelled `Pol.~0' is based on an LNA from UoM, and that labelled `Pol.~1' is based an LNA from LNF. 
}\label{fig:noise_meas2}
\end{figure}
%--------------------------------------------------------------------

The prototype cryogenic receiver was designed to accommodate each version of the optics described above. All three feedhorn and OMT assemblies were installed and fully characterised in the receiver prototype using the setup described in Appendix \ref{sec:rx-test-setup}. 
Figure~\ref{fig:noise_meas2} shows the uncorrected receiver noise temperature of the receiver equipped with cascaded LNAs from UoM, installed in the Pol.~0 polarisation channel, and cascaded LNAs from LNF installed in the Pol.~1 polarisation channel.  
These particular tests have been performed with the cartridge assembly using the NAOJ feedhorn and OMT, which have lower insertion loss and thus allow for the better noise performance, currently the most challenging requirement for this receiver.
We tested the prototype receiver with the silicon and UHMW-PE lenses and confirmed very similar noise performance (within $<1$~K on average).

While the achieved performance is already on a state-of-the-art level, the noise performance below 90~GHz does not yet meet the ALMA requirements. We believe, however, that there is still room for further optimisation and improvement. The new generation LNAs under development aim to improve both the noise and the gain, specifically for the frequency range 70-90 GHz. Our simulations also show that below 90 GHz there is considerable beam truncation at the cryostat  top plate and at the lens aperture. This issue is being addressed by modifying the optical layout and optimising the lens size.

%--------------------------------------------------------------------
\begin{figure}
\centering
\includegraphics[width=\columnwidth]{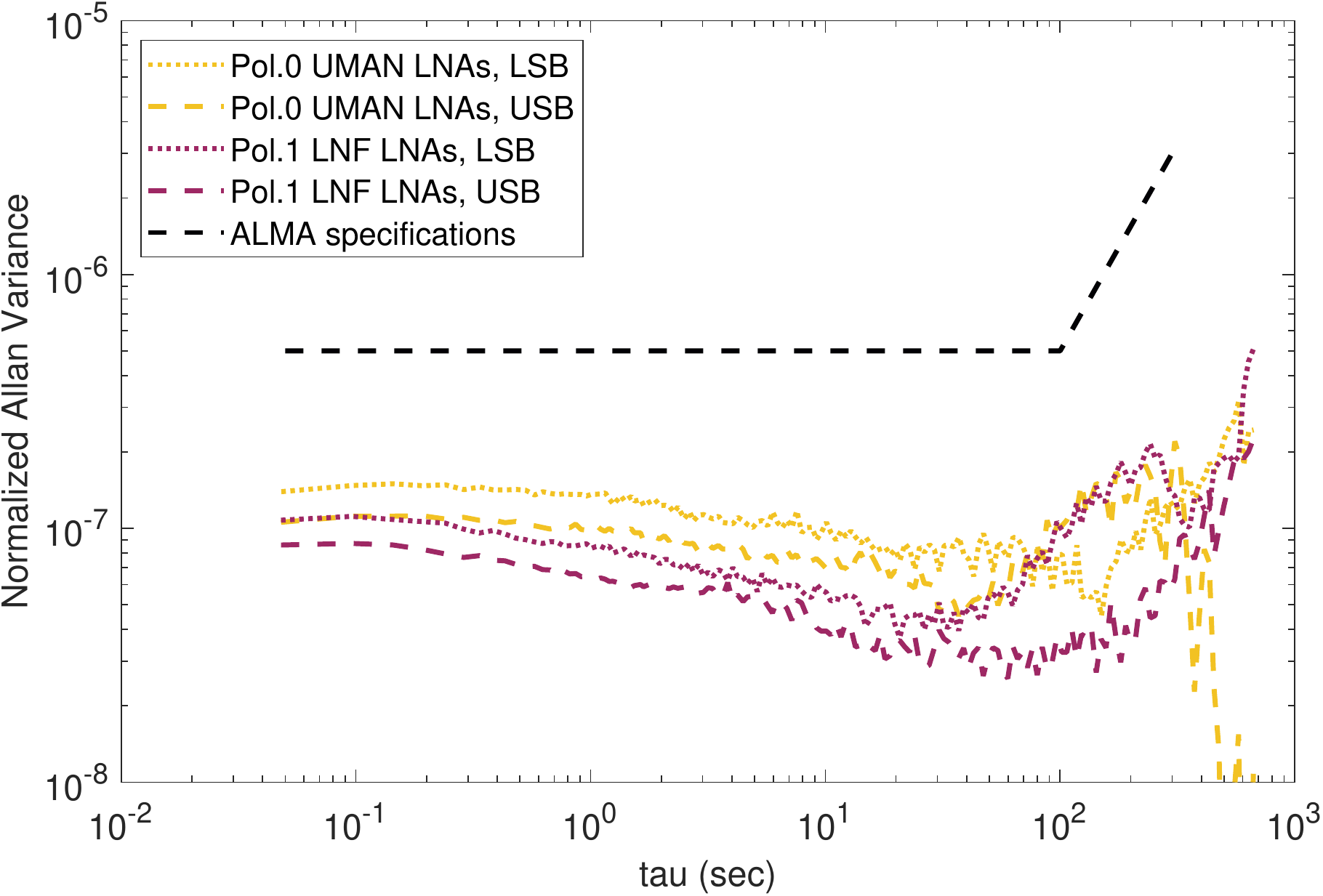}
\caption{Total power amplitude stability of the receiver prototype at an LO setting of 92~GHz for the two polarisation channels, which are based on LNF  and UMAN LNAs. Dashed green line indicates the ALMA requirements.
}\label{fig:stability}
\end{figure}
%--------------------------------------------------------------------

The stability over time of the InP MMICs may become an issue given their large bandwidth and high operating frequency, if not addressed properly. Other active components may also potentially contribute to the system' stability. In order to ensure the absence of oscillations in the LNAs, and to confirm compatibility of the selected design and components selection with ALMA requirements, total power amplitude stability tests were performed. The normalised Allan variance is calculated from the 30 minutes time series measurements with the resolution bandwidth of 8~GHz, and the results are shown in Fig.~\ref{fig:stability}. The ALMA requirements are shown in the plot as guidelines. The results suggest absence of oscillations, or other types of instabilities, over both the short and long term.

%--------------------------------------------------------------------
\section{Summary and Discussion}\label{sec:conclusions}

In this paper, we present the science case and an overview of the technical development for an ALMA Band 2  receiver covering the full 67-116 GHz atmospheric window, originally defined as ALMA Bands 2 and 3.  
A number of challenges have been met in realising this wideband receiver design and we have developed multiple technological solutions to produce critical elements capable of meeting most of the ALMA requirements. 
We consider each set of solutions for the passive and active components technologically mature and a viable approach for the final receiver to be installed in ALMA.
However, although the receiver represents state-of-the-art technology, it does not meet all the ALMA specifications over the entire 67-116~GHz band, specifically the noise temperature specification at RF frequencies $< 90$~GHz. We will, therefore, continue to optimise the receiver, and the cryogenic low noise amplifiers in particular, with the goal of meeting  ALMA's requirements.

This wideband design will open a new window for ALMA, which will make use of ALMA’s unparalleled collecting area, resolution, and low noise, to probe astrophysics in this band to much greater sensitivities than possible to date.  
We expect the new receiver system to be operational starting 2024.  This wide RF Band 2 will improve observational efficiency for a number of key science cases crucial to the ALMA 2030 Development Roadmap \citep{alma_roadmap}, laying the foundations for future wide bandwidth receiver systems that have the capacity to more than double ALMA’s on-sky bandwidth. 

\section*{Acknowledgements}
This work was supported by the EU-ALMA Development Study Programme; by JSPS KAKENHI Grant Number 15H02074; by ESO-INAF collaboration agreements N. 64776/15/69626/HNE, N. 74288/16/78223/OSZ, N. 81231/18/88652/ADI and by the Italian Ministero dell\'\,Istruzione, Universit\`a e Ricerca through the grant Progetti Premiali 2012 -- iALMA (CUP C52I13000140001); by CONICYT through projects Basal AFB-170002 and Quimal 14002.
G.E.C. was supported by National Science Foundation Graduate Research Fellowship 1650114.

%--------------------------------------------------------------------
\bibliographystyle{aa} % style aa.bst
\bibliography{references} % your references Yourfile.bib
%--------------------------------------------------------------------

\appendix

\section{Test systems}\label{sec:testsystems}

\begin{figure}
\centering
\includegraphics[width=\columnwidth]{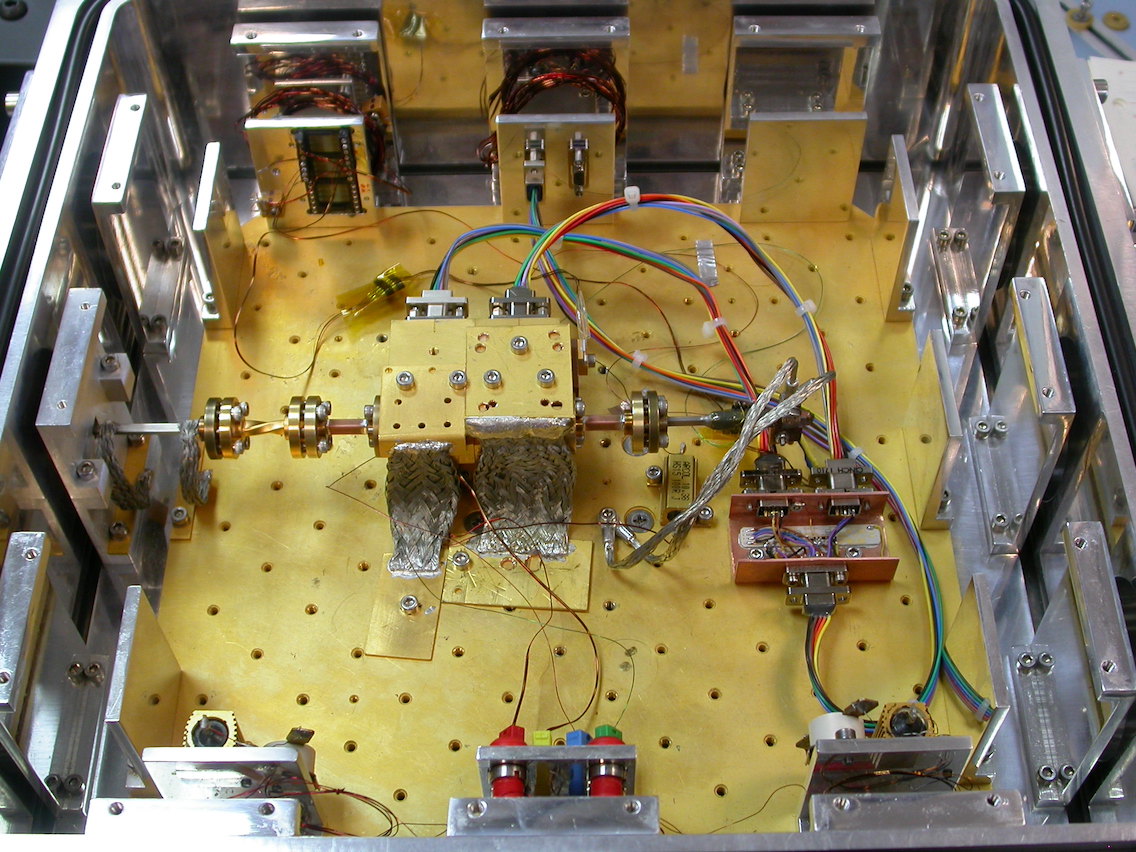}
\caption{View inside the Yebes Observatory cryostat with the configuration used for the noise measurement of two cascaded amplifiers using the heated load method.}
\label{fig:yebes_cryo}
\end{figure}

%----------------------------------------------------------------

\subsection{Low noise amplifier test system}\label{sec:lna-test-setup}

The system used for the comparison of the cryogenic LNAs, which is at located at Yebes Observatory, Spain, is shown in Fig.~\ref{fig:yebes_cryo}. This uses a 4~K Sumitomo pulse tube cooler and the temperature of the cold stage is actively stabilised with a Lake shore 336 cryogenic temperature controller. The measurement was obtained by the heated load method.  A special, low thermal capacity, low-reflection cryogenic waveguide termination fitted with a calibrated temperature sensor and nichrome wire heater was used to present different noise temperatures at the input of the amplifier. The load was built in a section of high thermal conductivity copper waveguide which was directly soldered to a short section of thin-wall stainless steel waveguide which supports the thermal gradient between the input termination and the amplifier. The effect of the loss and temperature gradient in the stainless steel waveguide section was carefully calculated to correct the equivalent noise temperature presented at the input flange of the amplifier (the correction was approximately only -0.8~K for a 50~K physical temperature of the termination). The temperatures of the input load for the hot and cold states used for the Y factor calculation were 50 K and 20 K. Two independent PID control loops were used to maintain constant the physical temperature of the amplifier (at 15 K) while the input load was set either to 50 or 20 K.

Outside the cryostat, a downconverter (Virginia Diodes Inc, USA) and a noise figure meter were used to measure the noise power in the hot and cold states. The frequency could be selected in a synthesised local oscillator in the 70-116 GHz range.  A four port WR10 mechanical waveguide switch at the output of the cryostat allowed the connection of an avalanche noise diode to the input of the downconverter to calibrate the gain and noise of the receiver or the direct connection of the output of the amplifier to a spectrum analyser harmonic mixer to monitor the presence of undesired self-oscillations of the amplifier. The calibration of the receiver system allows obtaining the gain of the amplifier as well as correcting for the noise contribution of the downconverter.  The noise temperature data presented for the amplifier is the corrected value. The absolute accuracy of the noise temperature measurements was estimated by Monte Carlo calculations, obtaining a value of $\pm 0.8~$K (at 2$\sigma$).

\subsection{Optical test setup}\label{sec:optics-test-setup}

%----------------------------------------------------------------
\begin{figure*}
\centering
\includegraphics[width=\textwidth]{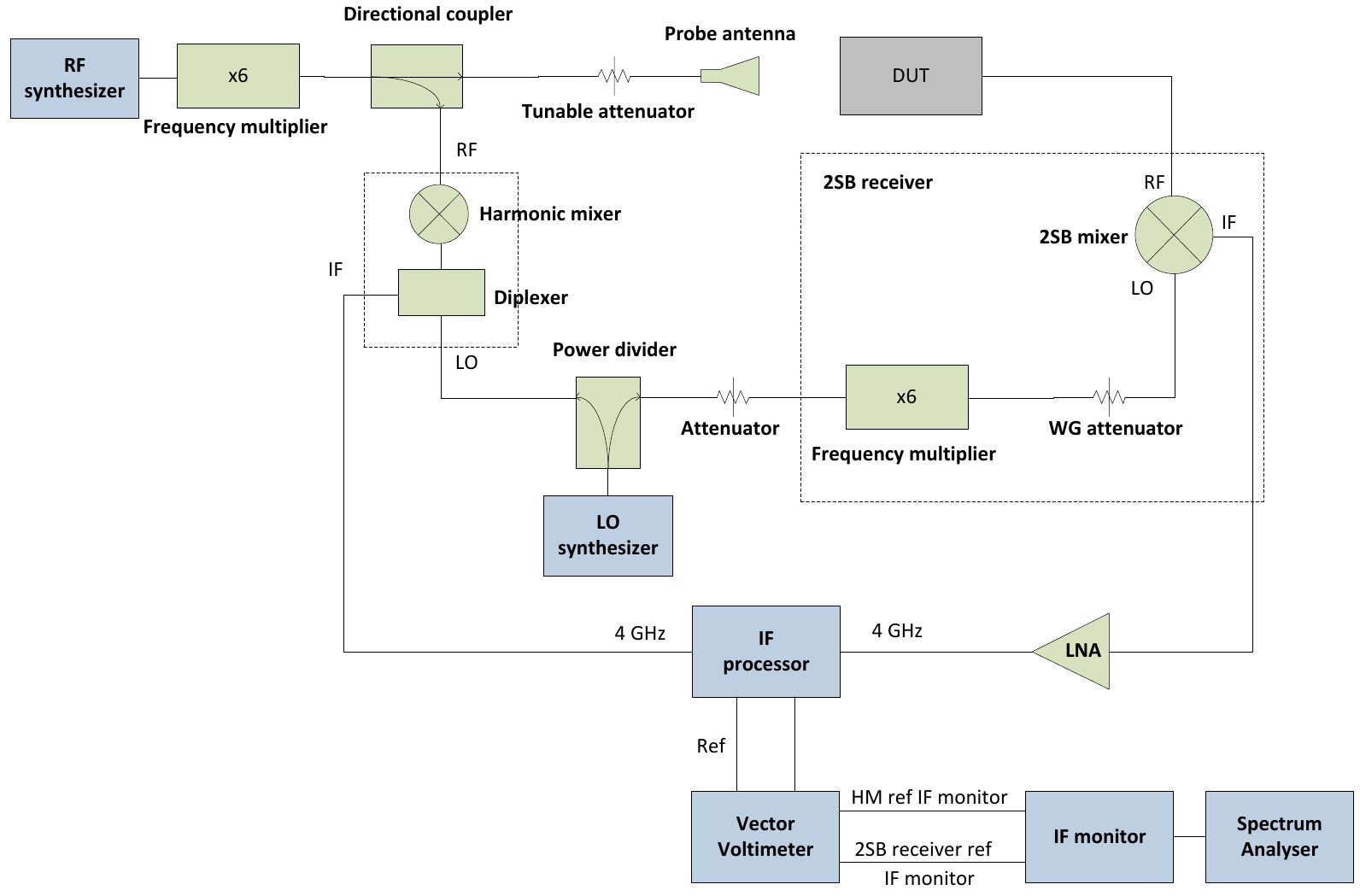}
\caption{Block diagram for the optical test setup used for measurements of the feedhorn near-field response. Figure adapted from \cite{Yagoubov2016}.}
\label{fig:optical_test_setup}
\end{figure*}
%----------------------------------------------------------------

The optical performances of all three types of optics were measured at ESO using a near-field beam measurement system, the block diagram for which is shown in Figure \ref{fig:optical_test_setup} (adapted from \citealt{Yagoubov2016}). This system has been designed to accept any combination of optical components and allows testing the feed horns alone ask well as the complete optical configuration including the lens and infrared filters. The system consists of a transmitting open-ended waveguide probe antenna attached to an (X,Y)-scanner equipped with Z- rotation stage for co- and cross-polarisation measurements, a room-temperature 2SB receiver, an electronics rack and a control PC. The maximum scannable plane is 150 mm $\times$ 150 mm. The sampling rate in the XY plane is greater than 0.5$\lambda$ in order to meet the Nyquist criterion (i.e. samples are $< 0.5\lambda$ apart). The signal of an RF synthesizer is frequency multiplied by a factor of 6 and fed to the probe antenna. The radiated RF signal is received by the device under test (DUT) and down-converted to an IF frequency of 4~GHz with the 2SB receiver. The LO signal for the down-conversion is obtained by frequency multiplication, also by a factor of 6, of the signal provided by a second synthesizer. A reference signal with the phase information of the RF and LO signals is created by mixing samples of the RF signal and the low frequency LO signal, with a harmonic mixer. The resulting reference signal at the IF frequency is compared in amplitude and phase with the IF signal from the 2SB receiver, and the result is read with the vector voltmeter. The far-field beam pattern is computed from the near-field data by Fourier transform and the optical efficiencies are then calculated.

%----------------------------------------------------------------

\subsection{Receiver test system}\label{sec:rx-test-setup}

Key performance metrics of the receiver, that is, noise, passband variations, stability, etc, were determined through several cryogenic measurement campaigns at the INAF-OAS premises in Bologna.
The cryogenic receiver is integrated inside a cartridge test cryostat which is composed of three stages, connected with corresponding stages of a GM cryocooler (Sumitomo RDK 3ST), providing references at: $\sim$80~K, $\sim$15~K, and $\sim$4~K (not used).

We apply the standard Y-factor method, based on the use of hot and cold thermal loads, to measure the receiver noise. The setup reuses the microwave calibrator produced for the {\it Planck} Low Frequency Instrument (LFI; \citealt{Mennella2010,Sandri2010}). It consists of a bed of Eccosorb CR110 pyramids and has an emissivity $>0.9999$ over the frequency range of interest. The load is situated in a large, liquid-nitrogen-filled polystyrene box to provide the cold reference, optically coupled to the receiver through the aluminium mirror oriented at 45$^\circ$ with respect to\ the beam direction.
A fan continuously dries the box and the lens to prevent ice formation on tips of the cold load and on the vacuum lens.  A panel of Eccosorb CV3 foam absorber (emissivity $>$ 0.999), backed by aluminium and thermalised at room temperature, provides the hot reference: it directly faces the window (lens) of the cryostat, at a distance of a few centimetres, and is connected to a step motor allowing the signal to chop between the hot and cold reference.

\section{Acronyms and abbreviations used in the work}\label{sec:appendix_acroy}
We list the acronyms used throughout this work in Table~\ref{tab:jargon}.  

\begin{table*}
    \caption{Acronyms and abbreviations used throughout the text.}
    \label{tab:jargon}
    \centering
    \begin{tabular}{l l}
        \hline\hline%\noalign{\smallskip}
        Acronym & Meaning \\%\noalign{\smallskip}
        \hline\noalign{\smallskip}
        2SB  & two sideband (i.e.\ sideband separating)\\
        ACA  & Atacama Compact Array (Morita Array)\\
        ADS  & (Keysight) Advanced Design System \\
        ALMA & Atacama Large Millimeter/Submillimeter Array\\
        AME  & anomalous microwave emission\\
        AR   & anti-refective / anti-reflection\\
        ASIAA & Academia Sinica, Institute of Astronomy \& Astrophysics\\
        C2R  & circular-to-rectangular \\      
        CAD  & computer aided design\\
        CCA  & cold cartridge assembly\\
        CMB  & cosmic microwave background\\
        CNC  & computer numerical control\\
        CO   & carbon monoxide\\
        CLP  & Chilean Peso\\
        CRAL & Cahill Radio Astronomy Laboratory\\
        DSB  & dual-sideband\\
        DUT  & device under test\\
        EM   & electromagnetic\\
        ESO  & European Southern Observatory\\
        FBM  & fundamental balanced mixers\\
        FE   & front end (receiver cryostat)\\
        FWHM & full width at half maximum\\
        FZ   & float zone\\
        GaAs & Gallium Arsenide\\
        GARD & Group for Advanced Receiver Development\\
        HEMT & high electron mobility transistor\\
        HFSS & (ANSYS) High-Frequency Structure Simulator\\
        HDPE & high-density polyethylene\\
        IF   & intermediate frequency\\
        IR   & infra-red\\
        INAF & Istituto Nazionale di Astrofisica\\
        InP  & Indium Phosphide\\
        IRA  & Istituto di Radioastronomia\\
        LFI  & Low Frequency Instrument (on {\it Planck})\\
        LNA  & low noise amplifier\\
        LNF  & Low Noise Factory\\
        LO   & local oscillator\\
	MMFE & mode-matching/finite-elements\\
        MMIC & monolithic microwave integrated circuit\\
        NAOJ & National Astronomical Observatory of Japan\\
        NGC  & Northrop Grumman Corporation\\
        NRAO & National Radio Astronomical Observatory\\
        OAA  & Osservatorio Astrofisico di Arcetri\\
        OAS  & Osservatorio di Astrofisica e Scienza dello Spazio di Bologna\\
        OMT  & orthomode transducer\\
        PCL  & phase centre location\\
        PLL  & phase locked loop\\
        PWV  & preciptable water vapour\\
        RF   & radio frequency (i.e. on-sky frequency)\\
        RPG  & Radiometer Physics GmbH\\
        SIS  & superconductor-insulator-superconductor\\
        SSB  & single-sideband\\
        SZ   & Sunyaev-Zeldovich (effect)\\
        UdC  & Universidad de Chile\\
        UHMW-PE & Ultra-high-molecular-weight polyethylene\\
        UoM  & University of Manchester\\
        VNA  & vector network analyser\\
        WCA  & warm cartridge assembly\\
        YIG  & yttrium iron garnet\\%\noalign{\smallskip}
        \hline
    \end{tabular}
\end{table*}

%----------------------------------------------------------------
%Here we include a list of figures and tables:
%\listoffigures
%\listoftables
\end{document}